\documentclass[5p,10pt]{elsarticle}

\journal{Nature Energy}

\biboptions{numbers,sort&compress,super}

\usepackage{libertine}
\usepackage{libertinust1math}
\usepackage{geometry}
\usepackage{amsmath}
\usepackage{bbold}
\usepackage{graphicx}
\usepackage{eurosym}
\usepackage{mathtools}
\usepackage{url}
\usepackage{booktabs}
\usepackage{epstopdf}
\usepackage{xfrac}
\usepackage{tabularx}
\usepackage{bm}
\usepackage{subcaption}
\usepackage{longtable}
\usepackage{multirow}
\usepackage{threeparttable}
\usepackage[export]{adjustbox}
\usepackage[version=4]{mhchem}
\usepackage[colorlinks]{hyperref}
\usepackage[nameinlink,sort&compress,capitalise]{cleveref}
\usepackage[leftcaption,raggedright]{sidecap}
\usepackage[prependcaption,textsize=footnotesize]{todonotes}
\usepackage{blindtext}

\usepackage{siunitx}
\DeclareSIUnit\year{a}
\DeclareSIUnit{\tco}{t_{\ce{CO2}}}
\DeclareSIUnit{\sieuro}{\mbox{\euro}}
\DeclareSIUnit{\twh}{{\tera\watt\hour}}
\DeclareSIUnit{\mwh}{{\mega\watt\hour}}
\DeclareSIUnit{\kwh}{{\kilo\watt\hour}}

\newcommand{\bneuro}[1]{#1\,bn\euro{}/a}

\renewcommand{\ttdefault}{\sfdefault}

\begin{document}

\begin{frontmatter}

	\title{Energy Imports and Infrastructure in a Carbon-Neutral European Energy System}

	\author[tub]{Fabian Neumann\corref{correspondingauthor}}
	\ead{f.neumann@tu-berlin.de}
	\author[pik]{Johannes Hampp}
	\author[tub]{Tom Brown}

	\address[tub]{Department of Digital Transformation in Energy Systems, Institute of Energy Technology,\\Technische Universität Berlin, Fakultät III, Einsteinufer 25 (TA 8), 10587 Berlin, Germany}
	\address[pik]{Potsdam Institute for Climate Impact Research (PIK), Member of the Leibniz Association, P.O.~Box 60 12 03, 14412 Potsdam, Germany}

	\begin{abstract}
		Importing renewable energy to Europe offers many potential benefits, including reduced energy costs, lower pressure on infrastructure development, and less land-use within Europe. However, there remain many open questions: on the achievable cost reductions, how much should be imported, whether the energy vector should be electricity, hydrogen or hydrogen derivatives like ammonia or steel, and their impact on Europe's domestic energy infrastructure needs. This study integrates the TRACE global energy supply chain model with the sector-coupled energy system model for Europe PyPSA-Eur to explore scenarios with varying import volumes, costs, and vectors. We find system cost reductions of 1-14\%, depending on assumed import costs, with diminishing returns for larger import volumes and a preference for methanol, steel and hydrogen imports. Keeping some domestic power-to-X production is beneficial for integrating variable renewables, utilising waste heat from fuel synthesis and leveraging local sustainable carbon sources. Our findings highlight the need for coordinating import strategies with infrastructure policy and reveal maneuvering space for incorporating non-cost decision factors.

	\end{abstract}

\end{frontmatter}

%%% promise of imports %%%

Importing renewable energy to Europe promises several advantages for achieving a
swift energy transition. It could lower costs, help circumvent the slow domestic
deployment of renewable energy infrastructure and reduce pressure on land usage
in Europe. Many parts of the world have cheap and abundant renewable energy
supply potentials that they could offer to existing or emerging global energy
markets.\cite{irenaGlobalHydrogen2022,luxSupplyCurves2021,vanderzwaanTimmermansDream2021,fasihiLongTermHydrocarbon2017,reichenbergDeepDecarbonization2022,galvanExportingSunshine2022,armijoFlexibleProduction2020,pfennigGlobalGISbasedPotential2023}
Partnering with these regions could help Europe reach its carbon neutrality
goals while stimulating economic development in exporting countries.

%%% dangers of imports %%%

However, even if energy imports are economically attractive for Europe, a strong
reliance may not be desirable because of energy security concerns. Awareness of
energy security has risen since Russia throttled fossil gas supplies to Europe
in 2022,\cite{pedersenLongtermImplications2022} at a time when the EU27 imported
around two-thirds of its fossil energy needs.\cite{eurostatCompleteEnergy2023}
Europe must take care to avoid repeating the mistakes of previous decades when
it became dependent on a small number of exporters with market power and reliant
on rigid pipeline infrastructure.

%%% dependence of energy imports on energy infrastructure %%%

Europe's strategy for clean energy imports will also strongly affect the
requirements for domestic energy infrastructure. Previous research found many
ways to develop a self-sufficient energy
system.\cite{pickeringDiversityOptions2022,trondleHomemadeImported2019,brownSynergiesSector2018}
To support such scenarios without energy imports into Europe, reinforcing the
European power grid or building a hydrogen network was often identified as
beneficial.\cite{neumannPotentialRole2023,victoriaSpeedTechnological2022}
However, depending on the volumes of imports and the energy vectors
(electricity, hydrogen or hydrogen derivatives), Europe might not need to expand
its hydrogen transport infrastructure. Most hydrogen is used to make derivative
products (e.g.~, ammonia for fertilisers or Fischer-Tropsch fuels for aviation
and shipping).\cite{neumannPotentialRole2023} If Europe imported these products
at scale, much of the hydrogen demand would fall away. In consequence, this
would reduce the need for hydrogen transport. However, if hydrogen itself is
imported, this would require a pipeline topology tailored to accommodate
hydrogen arriving from North Africa or shipping routes to Northern
Europe.

%%% review of policy strategies %%%

Policy has reflected these different visions for imports in various ways. In
particular, hydrogen imports have recently attracted considerable interest, with
plans of the European Commission\cite{europeancommissionRepowerEUPlan} to import
10~Mt (333~TWh\footnote{All mass-energy conversion is based on the lower heating
value (LHV). Steel is included in energy terms applying 2.1 kWh/kg as released
by the oxidation of iron.}) hydrogen and derivatives by 2030. Desire to import
hydrogen and derivative products is also present in various national
strategies.\cite{corbeauNationalHydrogenStrategies2024} In particular, Germany
seeks to cover up to 70\% of its hydrogen consumption through imports by 2030
and pursues bilateral partnerships to accomplish
this.\cite{bundesministeriumfuerwirtschaftundklimaschutzFortschreibungNationalenWasserstoffstrategie2023}
Conversely, hydrogen roadmaps of
Denmark,\cite{danishministryofclimateenergyandutilitiesRegeringensStrategiPowertoX2021}
Ireland,\cite{departmentoftheenvironmentclimateandcommunicationsgovernmentofirelandNationalHydrogenStrategy2023}
Spain,\cite{marcoestrategicodeenergiayclimaRutaHidrogenoApuesta2020} and the
United
Kingdom,\cite{ukdepartmentforenergysecurity&netzeroHydrogenStrategyUpdate2023}
recognise these countries' potential to become major exporters of renewable
energy, whereas France's strategy focuses on local hydrogen production to meet
domestic needs.\cite{frenchgovernmentStrategieNationalePour2023} Additionally,
in recent plans by transmission system
operators,\cite{entso-eTYNDP2024Project2024}  European grid development plans
reveal renewed enthusiasm for electricity imports via ultra-long HVDC cables,
evolving from early
DESERTEC\cite{desertecfoundationDESERTECSustainableWealth2024} ideas to
contemporary proposals like the Morocco-UK Xlinks
project.\cite{xlinksMoroccoUKPower2023}

%%% literature review %%%

While many previous academic studies have evaluated the cost of `green'
renewable energy and material imports in the form of
electricity,\cite{lilliestamEnergySecurity2011,triebSolarElectricity2012,lilliestamVulnerabilityTerrorist2014,bogdanovNorthEastAsian2016,benaslaTransitionSustainable2019,reichenbergDeepDecarbonization2022}% (some with reference to the DESERTEC idea),
hydrogen,\cite{timmerbergHydrogenRenewables2019,ishimotoLargescaleProduction2020,brandleEstimatingLongterm2021,luxSupplyCurves2021,galvanExportingSunshine2022,collisDeterminingProduction2022,galimovaImpactInternational2023,schmitzImplicationsHydrogenImport2024}
ammonia,\cite{nayak-lukeTechnoeconomicViability2020,armijoFlexibleProduction2020,galimovaFeasibilityGreen2023}
methane,\cite{luxSupplyCurves2021,agoraenergiewendeHydrogenImport2022}
steel,\cite{trollipHowGreen2022a,devlinRegionalSupply2022,lopezDefossilisedSteel2023}
carbon-based
fuels,\cite{fasihiLongTermHydrocarbon2017,sherwinElectrofuelSynthesis2021} or a
broader variety of power-to-X
fuels,\cite{vanderzwaanTimmermansDream2021,pfennigGlobalGISbased2022,irenaGlobalHydrogen2022,solerEFuelsTechno2022,hamppImportOptions2023,gengeSupplyCosts2023,galimovaGlobalTrading2023a}
these do not address the interactions of imports with European energy
infrastructure requirements. On the other hand, among studies dealing with the
detailed planning of net-zero energy systems in Europe, some do not consider
energy
imports,\cite{pickeringDiversityOptions2022,brownSynergiesSector2018,victoriaSpeedTechnological2022}
while others only consider hydrogen imports or a limited set of alternative
import
vectors.\cite{gilsInteractionHydrogen2021,seckHydrogenDecarbonization2022,wetzelGreenEnergy2023,kountourisUnifiedEuropean2023,neumannPotentialRole2023}
Only a few consider at least elementary cost
uncertainties,\cite{frischmuthHydrogenSourcing2022,schmitzImplicationsHydrogenImport2024} and none investigate a
larger range of potential import volumes across subsets of available import
vectors.

%%% main paper idea - scenarios %%%

In this study, we explore the full range between the two poles of complete
self-sufficiency and wide-ranging renewable energy imports into Europe in
scenarios with high shares of wind and solar electricity and net-zero carbon
emissions. We investigate how the infrastructure requirements of a
self-sufficient European energy system that exclusively leverages domestic
resources from the continent may differ from a system that relies on energy
imports from outside of Europe. For our analysis, we integrate an open model of
global energy supply chains, TRACE,\cite{hamppImportOptions2023} with a
spatially and temporally resolved sector-coupled open-source energy system model
for Europe, PyPSA-Eur,\cite{PyPSAEurSecSectorCoupled} to investigate the impact
of imports on European energy infrastructure needs. We evaluate potential import
locations and costs for different supply vectors, by how much system costs can
be reduced through imports, and how their inclusion affects deployed transport
networks and storage. For this purpose, we perform sensitivity analyses
interpolating between very high levels of imports and no imports at all,
exploring low and high costs for imports to account for associated
uncertainties, and system responses to the exclusion of subsets of import
vectors, in order to probe the flatness of the solution space. This allows us to
draw robust policy conclusions from our results.

\begin{figure*}
    \begin{subfigure}[t]{\textwidth}
        \caption{Global perspective for energy imports into Europe}
        \label{fig:options:global}
        \includegraphics[width=\textwidth]{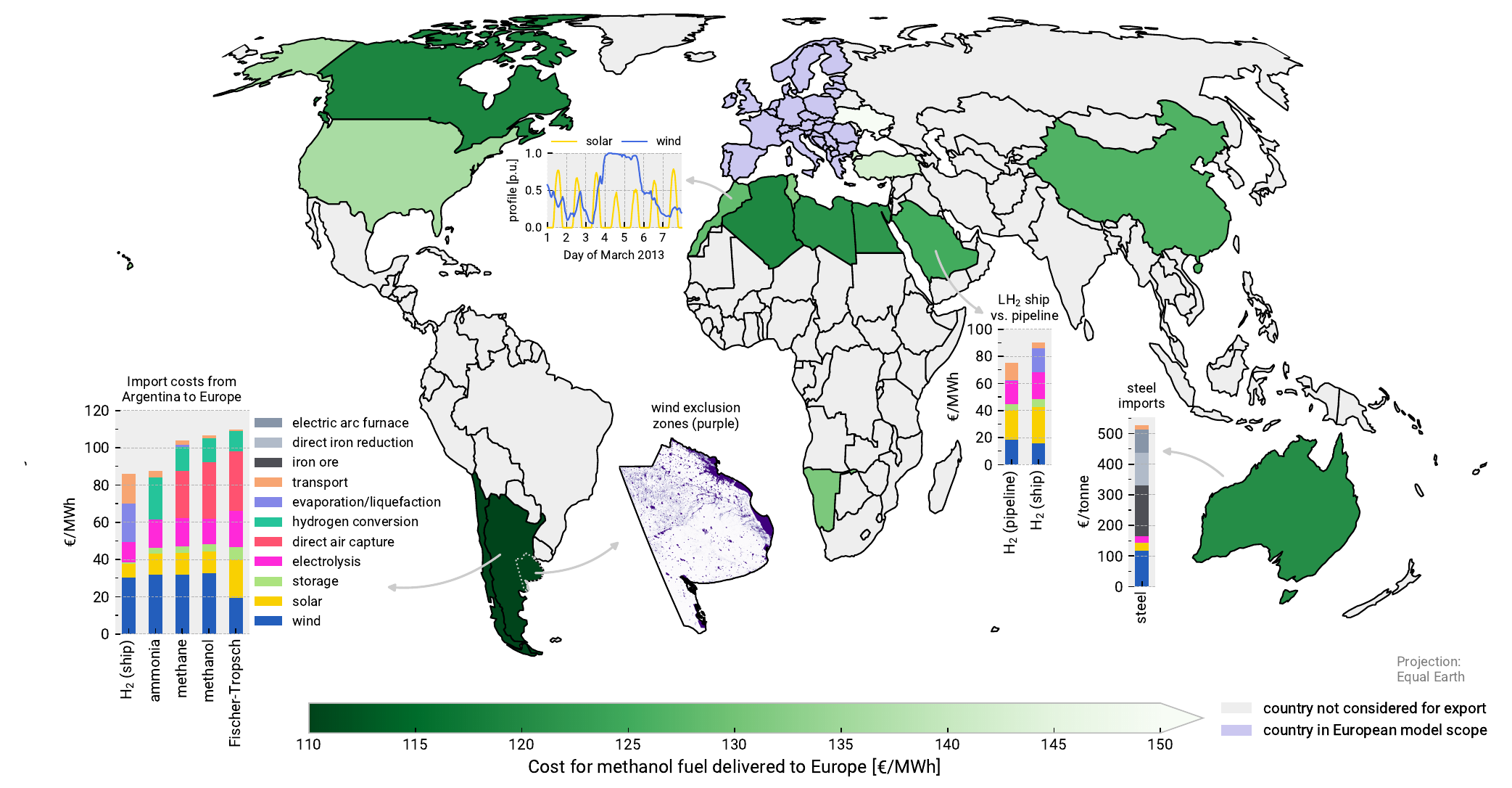}
    \end{subfigure}
    \begin{subfigure}[t]{0.7\textwidth}
        \caption{European perspective for inbound energy imports}
        \label{fig:options:europe}
        \centering
        \includegraphics[width=\textwidth]{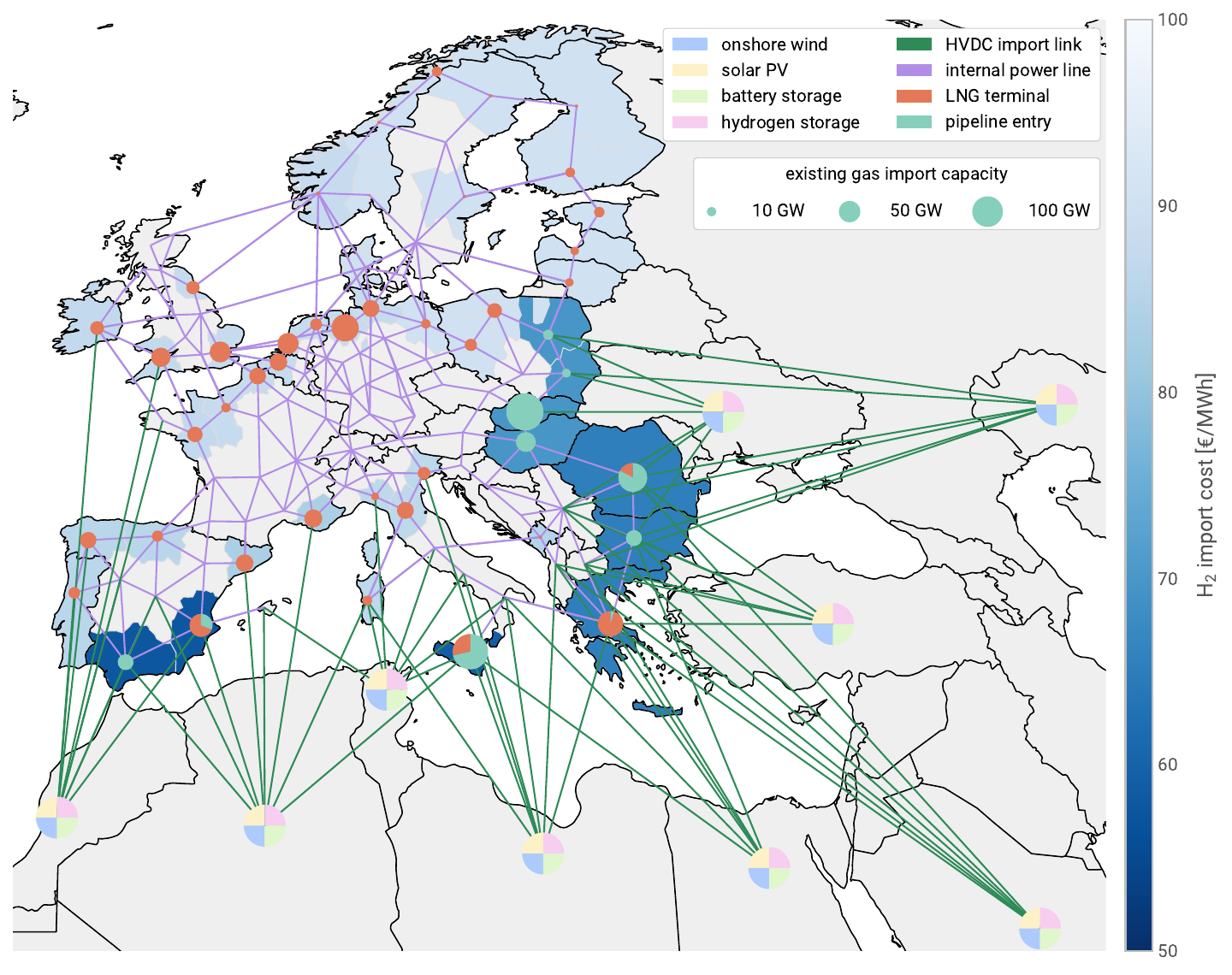}
    \end{subfigure}
    \begin{subfigure}[t]{0.3\textwidth}
        \caption{Cost distribution by origin and vector}
        \label{fig:options:distribution}
        \centering
        \includegraphics[width=\textwidth]{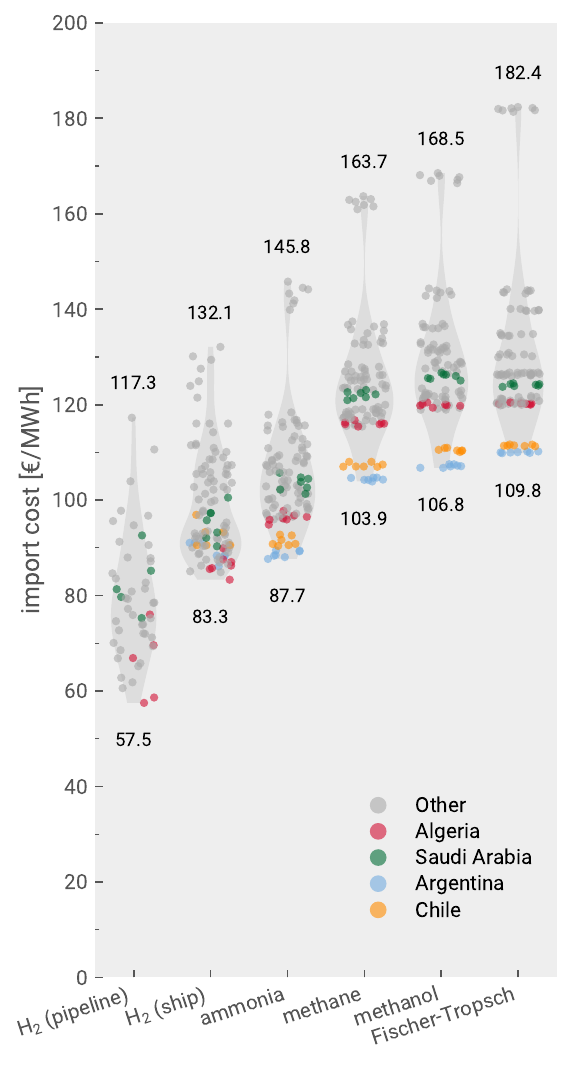}
    \end{subfigure}
    \caption{\textbf{Overview of considered import options.}
        \textit{Panel (a)} shows the regional differences in the cost to deliver
        green methanol to Europe (choropleth layer), the cost composition of
        different import vectors (bar charts), an illustration of the wind and
        solar availability in Morocco, and an illustration of the land
        eligibility analysis for wind turbine placement in the region of Buenos
        Aires in Argentina. \textit{Panel (b)} depicts considered potential
        entry points for energy imports into Europe like the location of
        existing and planned LNG terminals and gas pipeline entry points, the
        lowest costs of hydrogen imports in different European regions
        (choropleth layer), and the considered connections for long-distance
        HVDC import links and hydrogen pipelines from the MENA region,
        Kazakhstan, Turkey and Ukraine. \textit{Panel (c)} displays the
        distribution and range of import costs for different energy carriers and
        entry points with indications for selected countries of origin from the
        TRACE model (violin charts), i.e.~differences in identically coloured
        markers are due to regional differences in the transport costs to
        entrypoints.}
    \label{fig:options}
\end{figure*}

%%% technical discussion of import vectors %%%

As possible import options, we consider electricity by transmission line,
hydrogen as gas by pipeline and liquid by ship, methane as liquid by ship,
liquid ammonia, steel, methanol and Fischer-Tropsch fuels by ship. Each energy
vector has unique characteristics with regards to its production, transport and
consumption (\cref{fig:si:balances-a,fig:si:balances-b}). Electricity offers
the most flexible usage but is challenging to store and requires variability
management if sourced from wind or solar energy. Hydrogen is easier to store and
transport in large quantities but at the expense of conversion losses and less
versatile applications. Large quantities could be used for backup power and
heat, steel production, and the domestic synthesis of shipping and aviation
fuels. On the other hand, imported synthetic carbonaceous fuels like methane,
methanol and Fischer-Tropsch fuels could largely substitute the need for
domestic synthesis. There is much more experience with storing and transporting
these fuels and part of the existing infrastructure could potentially be reused
or repurposed. However, they require a sustainable carbon source and,
particularly for methane, effective carbon management and leakage
prevention.\cite{shirizadehImpactMethaneLeakage2023} Ammonia is similarly easier
to handle than hydrogen but does not require a carbon source. However, it faces
safety and acceptance concerns due to its toxicity and potentially adverse
effects on the global nitrogen
cycle.\cite{bertagniMinimizingImpactsAmmonia2023,wolframUsingAmmoniaShipping2022}
Its demand in Europe is mostly driven by fertiliser usage. Steel represents the
import of energy-intensive materials and offers low transport costs.

Further conversion of imported fuels is also possible once they have arrived in
Europe, e.g.~hydrogen could be used to synthesise carbon-based fuels, ammonia
could be cracked to hydrogen, methane and methanol could be reformed to hydrogen
or combusted for power generation with or without carbon capture. However,
conversion losses can make it less attractive economically to use a high-value
hydrogen derivative merely as a transport and storage vessel only to reconvert
it back to hydrogen or electricity.

%%% brief methodology PyPSA-Eur %%%

The PyPSA-Eur\cite{PyPSAEurSecSectorCoupled} model co-optimises the investment
and operation of generation, storage, conversion and transmission
infrastructures in a single linear optimisation problem. The model is further
given the opportunity to relocate some energy-intensive industries within Europe
capturing a potential renewables pull
effect.\cite{verpoortEstimatingRenewables2023,samadiRenewablesPull2023} We
resolve 110 regions comprising the European Union without Cyprus and Malta as
well as the United Kingdom, Norway, Switzerland, Albania, Bosnia and
Herzegovina, Montenegro, North Macedonia, Serbia, and Kosovo. In combination
with a 4-hourly equivalent time resolution for one year, grid bottlenecks,
renewable variability, and seasonal storage requirements are efficiently
captured. Weather variations between years are not considered for computational
reasons. The model includes regional demands from the electricity, industry,
buildings, agriculture and transport sectors, international shipping and
aviation, and non-energy feedstock demands in the chemicals industry.
Transmission infrastructure for electricity, gas and hydrogen, and candidate
entry points like existing and prospective LNG terminals and cross-continental
pipelines are also represented. We utilize techno-economic assumptions for
2030\cite{dea2019}, reflecting that infrastructure required for achieving carbon
neutrality must be built well in advance of reaching this goal. While enforcing
net-zero emissions for carbon dioxide, we also limit the annual carbon
sequestration potential to 200~Mt$_{\text{CO}_2}$/a. This suffices to offset
unabatable industrial process emissions of around 140~Mt$_{\text{CO}_2}$/a and
limited use of fossil fuels beyond that, whose emissions are compensated either
through capturing emissions at source with a capture rate of 90\% or via carbon
dioxide removal.

More details are included in the \nameref{sec:methods} section.

\section*{Cost assessment of energy and material import vectors}

%%% brief methodology TRACE %%%

Green fuel and steel import costs seen by the model are based on an extension of
recent research by Hampp et al.,\cite{hamppImportOptions2023} who assessed the
levelised cost of energy exports for different green energy and material supply
chains in various world regions (\cref{fig:options:global}). Our selection of
exporting countries comprises Australia, Argentina, Chile, Kazakhstan, Namibia,
Turkey, Ukraine, the Eastern United States and Canada, mainland China, and the
MENA region. Regional supply cost curves for these countries are developed based
on renewable resources, land availability and prioritised domestic demand.
Unlike domestic electrofuel synthesis in Europe, which could use captured CO$_2$
from point sources, direct air capture is assumed to be the only carbon source
of imported fuels. Concepts involving the shipment of captured CO$_2$
from Europe to exporting countries for carbonaceous fuel synthesis are not
considered.\cite{treeenergysolutionsGreenCycle2024,fonderSyntheticMethaneClosing2024}

%%% brief methodology TRACE-PyPSA-Eur scoupling %%%

We use these supply curves to determine the region-specific lowest import cost
for each carrier, thus incorporating the potential trade-off between import cost
and import location (\cref{fig:options:europe}). For hydrogen derivatives, the
lowest-cost suppliers are Argentina and Chile for all entry points into Europe.
Electricity imports are endogenously optimised, meaning that the capacities and
operation of wind and solar generation as well as storage in the respective
exporting countries and the HVDC transmission lines, are co-planned with the
rest of the system. Hydrogen and methane can be imported where there are
existing or planned LNG terminals or pipeline entry-points (excluding
connections through Russia). This results in lower hydrogen import costs, where
it can be imported by pipeline. Ammonia, carbonaceous fuels and steel are not
spatially resolved in the model, assuming they can be transported within Europe
at negligible additional cost.

\begin{figure*}
    \includegraphics[width=\textwidth]{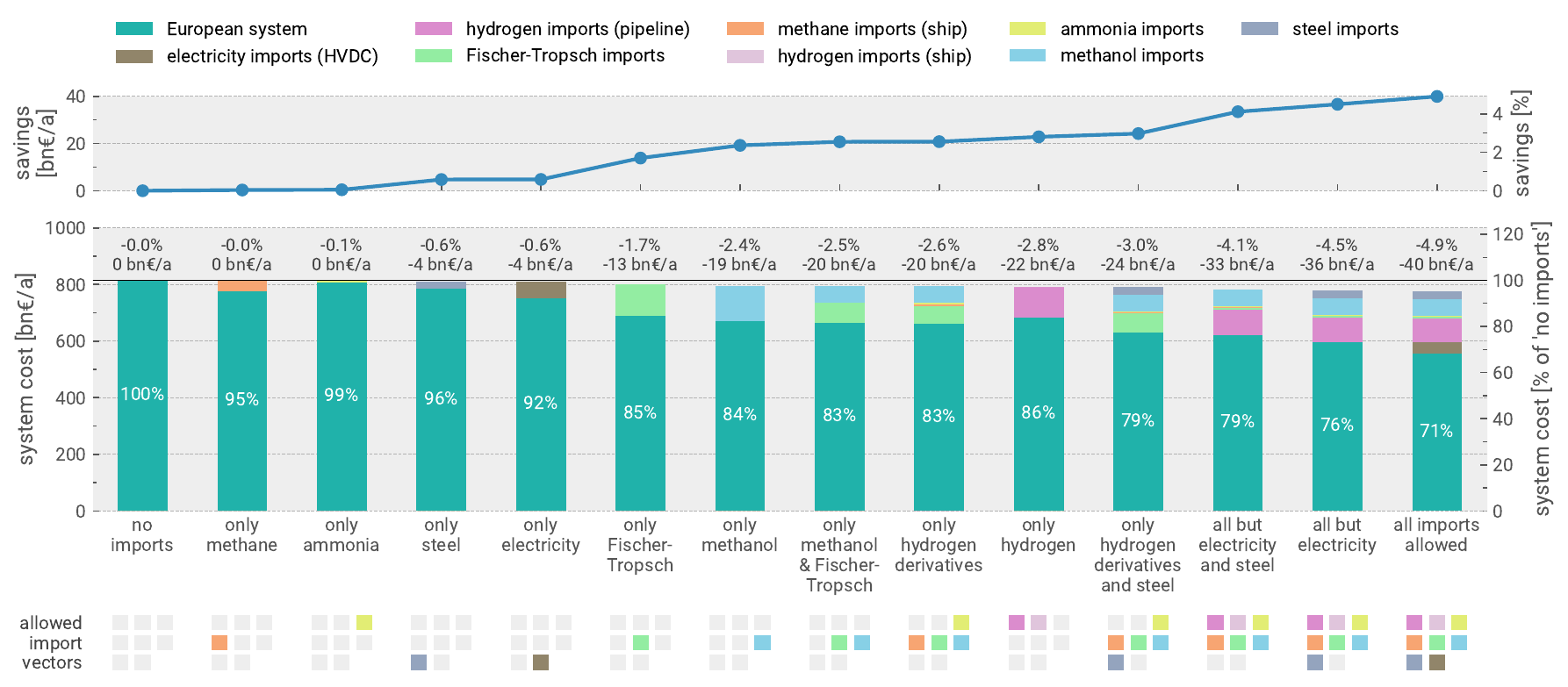}
    \caption{\textbf{Potential for cost reductions with reduced sets of import options.}
        Subsets of available import options are sorted by ascending cost
        reduction potential. Top panel shows profile of total cost savings.
        Bottom panel shows composition and extent of imports in relation to
        total energy system costs. Percentage numbers in bar plot indicate the
        share of total system costs spent on domestic energy infrastructure.
        Alternative scenarios of this figure with higher and lower import cost
        assumptions are included in the supplementary material. }
    \label{fig:sensitivity-bars}
\end{figure*}

\section*{Cost savings for fuel and material import combinations}

\begin{figure}
    \includegraphics[width=\columnwidth]{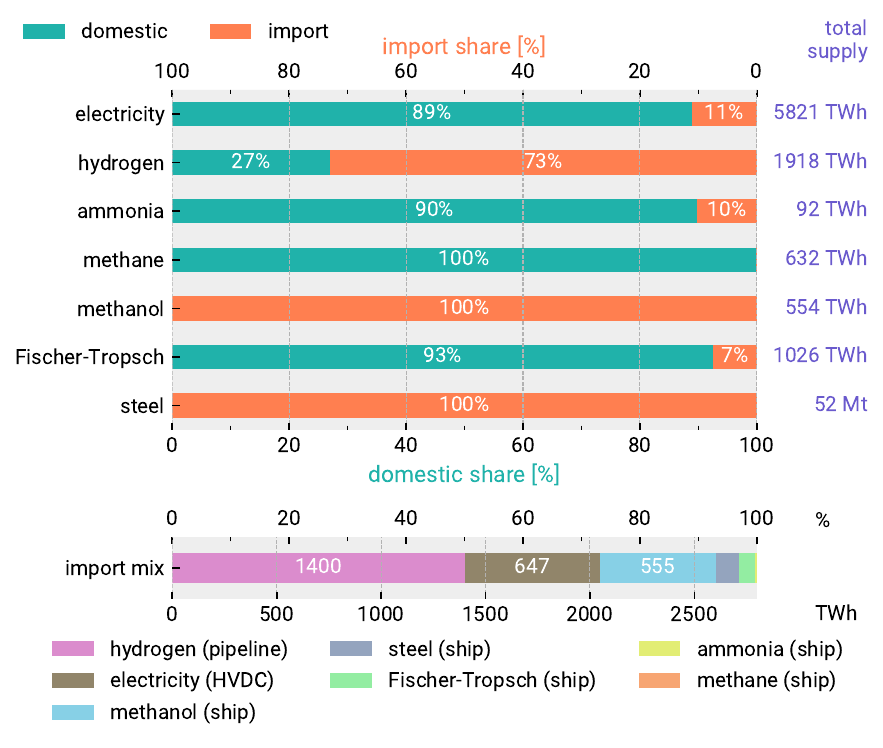}
    \caption{\textbf{Shares of imports and domestic production by carrier and optimised import carrier mix for import scenario with flexible carrier choice.}
        The figure also shows total supply for each carrier. Import shares for
        further import scenarios are included in the supplementary material.
        Steel is included in energy terms applying 2.1 kWh/kg as released by the
        oxidation of iron. }
    \label{fig:import-shares}
\end{figure}

In \cref{fig:sensitivity-bars}, we first explore the cost reduction potential of
various energy and material import options. In the absence of energy imports,
total energy system costs add up to \bneuro{815}\footnote{All currency values
are given in \euro{}$_{2020}$.}. By enabling imports from outside of Europe and
considering all import vectors, we find a potential reduction of total energy
system costs by up to \bneuro{39}. This corresponds to a relative reduction of
4.9\%. For cost-optimal imports, around 71\% of these costs are used to develop
domestic energy infrastructure. The remaining 29\% are spent on importing a
volume of 52~Mt of green steel and around 2700 TWh of green energy, which is
almost a quarter of the system's total energy supply (\cref{fig:import-shares}).

Next, we investigate the impact of restricting the available import options to
subsets of import vectors. We find that if only hydrogen can be imported, cost
savings are reduced to \bneuro{22} (2.8\%). This is because by using hydrogen as
an intermediary carrier, low-cost renewable electricity from abroad can still be
leveraged for the synthesis of derivative products in Europe. For this purpose,
pipeline-based hydrogen imports are preferred to ship-based imports as liquid.
When direct hydrogen imports are excluded from the available import options,
cost savings are similar with \bneuro{24} (3\%). Focusing imports exclusively on
liquid carbonaceous fuels derived from hydrogen, i.e.~methanol or Fischer-Tropsch
fuels, consistently achieves cost savings of \bneuro{13-20} (1.7-2.5\%).

On the contrary, restricting options to only ammonia or methane imports yields
negligible cost savings. Small savings below \bneuro{5} (0.6\%) can be reached
if only electricity or steel can be imported. This is due to the lower volume
and variety of usage options for ammonia, methane and steel compared to
hydrogen, methanol and Fischer-Tropsch fuels. Furthermore, the direct import of
electricity poses more challenges for system integration. Generally, our results
indicate a preference for methanol, hydrogen and steel imports over electricity
imports, with a mix emerging as the most cost-effective approach.
\cref{fig:si:subsets} show additional insights into how varying import costs
affect these findings.

\begin{figure*}
    \centering
    \footnotesize
    (a) no imports allowed \\
    \includegraphics[width=\textwidth]{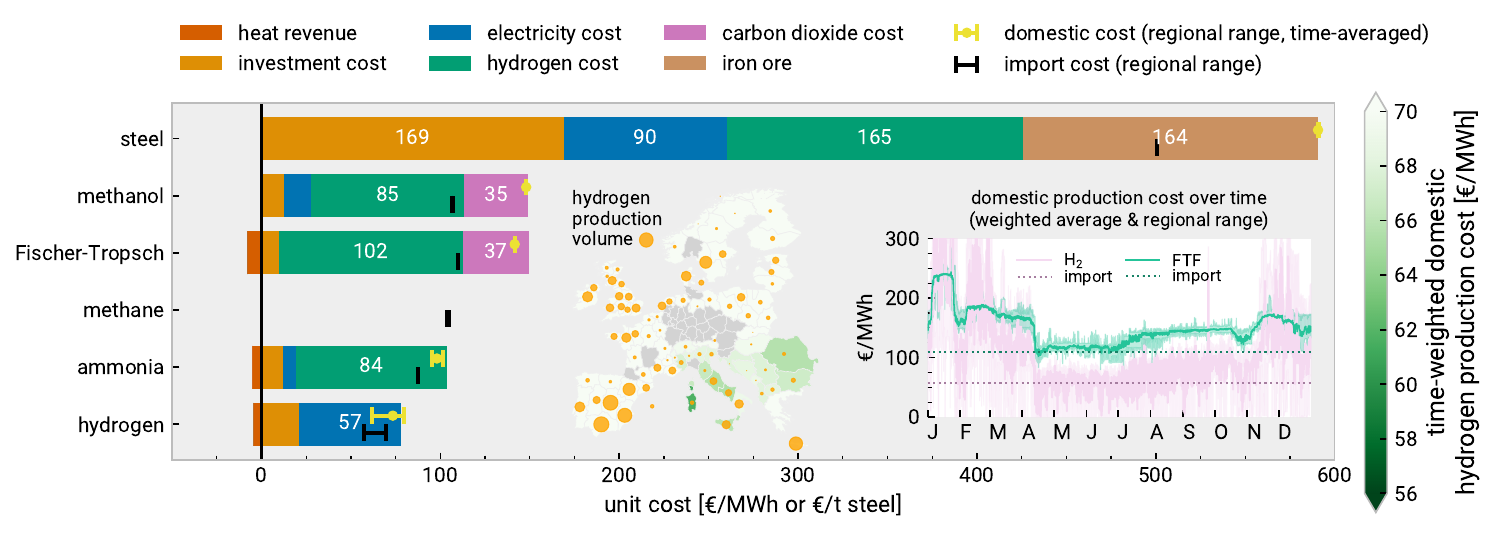} \\
    (b) only hydrogen imports allowed \\
    \includegraphics[width=\textwidth, trim=0cm 0cm 0cm 1.5cm,clip]{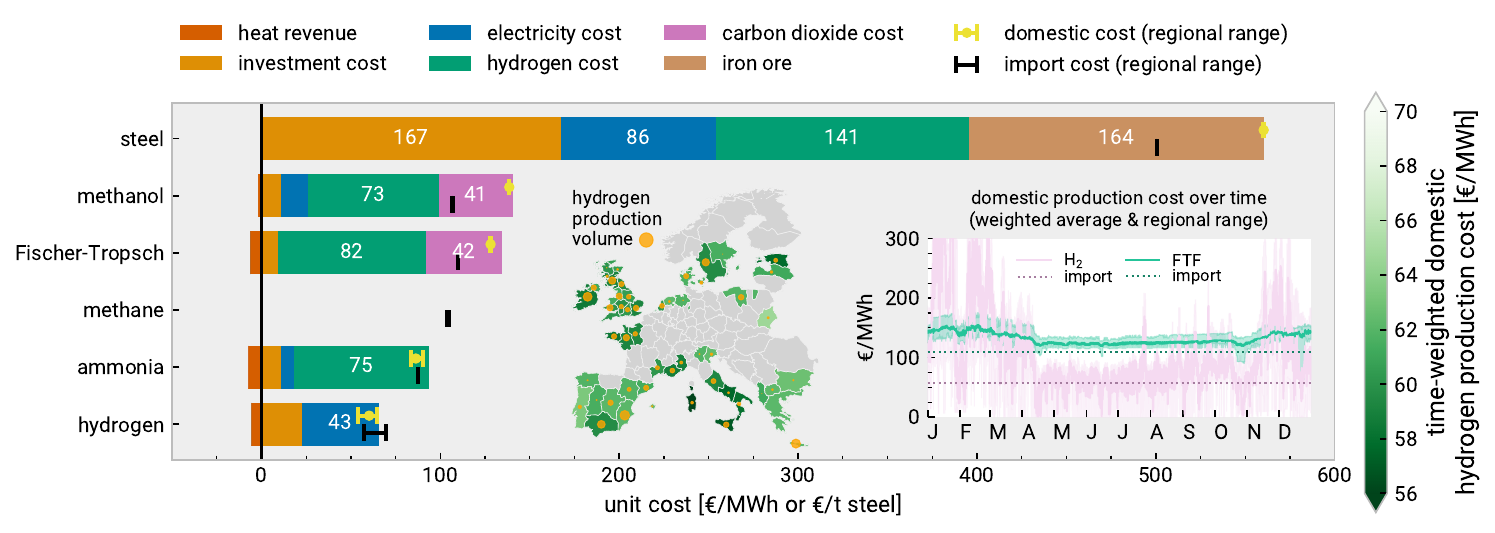} \\
    (c) all imports allowed \\
    \includegraphics[width=\textwidth, trim=0cm 0cm 0cm 1.5cm, clip]{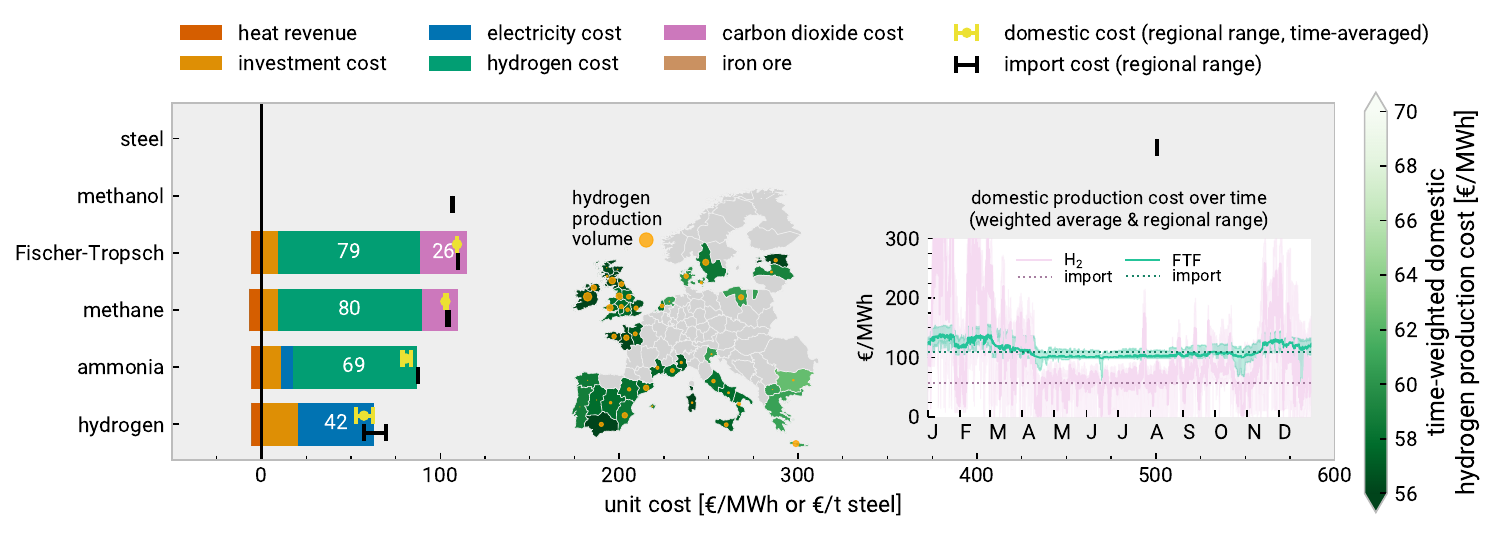}
    \caption{\textbf{Comparison of domestic synthetic production costs and import costs for varying import scenarios.}
        The three panels (a), (b), and (c) refer to different import scenarios.
        In each panel, the \textit{bar charts} show the production-weighted
        average costs of domestic production of steel, hydrogen and its
        derivatives split into its cost and revenue components. These have been
        computed using the marginal prices of the respective inputs and outputs
        for the production volume of each region and snapshot. Capital
        expenditures are distributed to hours in proportion to the production
        volume. Missing bars indicate that no domestic production occured in the
        scenario, e.g.~for the case of methane where all demand is met by
        biogenic and fossil methane and no synthetic production occured
        (cf.~energy balances in supplementary material). All hydrogen is
        produced from electrolysis; i.e.~the model did not choose to produce
        hydrogen via steam methane reforming with or without carbon capture. For
        each bar, the yellow errorbars show the range of time-averaged domestic
        production costs across all regions. The black error bars show the range
        of import costs across all regions. The \textit{maps} relate the
        hydrogen production volume to the weighted cost of domestic hydrogen
        production (left colorbar). The \textit{time series} indicate the
        variance of the domestic production cost over time for hydrogen and
        Fischer-Tropsch fuel (FTF) including the regional spectrum as shaded
        area.}
    \label{fig:market-values}
\end{figure*}

\section*{Import dynamics for different energy carriers}

%%% state results %%%

\cref{fig:import-shares} outlines which carriers are imported in which
quantities in relation to their total supply under default assumptions when the
vector and volume can be flexibly chosen (``all imports allowed'' in
\cref{fig:sensitivity-bars}). In energy terms, cost-optimal imports comprise
around 50\% hydrogen, more than 20\% electricity, and around 20\% of
carbonaceous fuels. Noticeably, all crude steel and methanol for shipping and
industry is imported. Also, around three-quarters of the total hydrogen supply
is imported. Hydrogen is imported so that it can be processed into derivative
products domestically rather than direct applications for hydrogen. Smaller
import shares are observed for electricity, Fischer-Tropsch fuels, and ammonia,
which are mostly domestically produced.

%%% explanation through market values %%%

To explain the import shares in \cref{fig:import-shares} in more detail, we
compare import costs with average domestic production cost split by cost and
revenue components in \cref{fig:market-values}. First, for the scenario without
imports, imported fuel appear to be substantially cheaper than domestic
production. The high demand for hydrogen and derivative products
(\cref{fig:si:balances-a,fig:si:balances-b}) means that the most attractive
domestic potentials for renewable electricity and carbon dioxide have been
exhausted. Power from wind and solar needs to be produced in regions with worse
capacity factors and direct air capture becomes the price-setting technology for
\ce{CO2} as biogenic and industrial sources ($\approx$600~Mt$_\text{\ce{CO2}}$)
are depleted.

%%% hydrogen imports lower pressure on domestic supply chain %%%

Part of this gap is closed when hydrogen imports are allowed. By sourcing
cheaper hydrogen from outside Europe, the domestic costs of derivative fuel
synthesis are reduced. This hybrid approach has the largest effect on
Fischer-Tropsch production due to its higher hydrogen demand compared to
methanolisation and the Haber-Bosch process. Hydrogen imports also decouple the
synthesis from the seasonal variation of domestic hydrogen production costs.

%%% waste heat integration %%%

The potential for waste heat utilisation from fuel synthesis within Europe adds
further appeal to this hybrid approach. By importing hydrogen rather than the
derivative product, heat supply into district heating networks from synthesis
processes can create an additional revenue stream of up to 10 \euro{}/MWh,
depending on the process.

% Taking ammonia as example, the levelised cost is 73
% \euro{}/MWh for domestic production compared to 88 \euro{}/MWh for imported
% ammonia.

The waste heat integration is also the reason why in \cref{fig:import-shares},
with all import vectors allowed, all methanol is imported, whereas
Fischer-Tropsch fuels and ammonia are produced mainly domestically using high
shares of imported hydrogen. Because the thermal discharge from the methanol
synthesis is primarily used for the distillation of the methanol-water output
mix, its waste heat potential is considered much lower compared to
Fischer-Tropsch, Haber-Bosch and Sabatier processes. Therefore, it is less
attractive to retain this part of the value chain within Europe.

%%% explanations of low cost differences if all imports allowed %%%

With all import vectors allowed, we see minimal cost differences between
domestic production and imports as the supply curves reach equilibrium. This is
because imports of hydrogen and derivative products lower the strain on the
domestic supply chain. Thereby, domestic production would only be ramped up
where it competes with imports and associated infrastructure costs. Such regions
exist in Southern Europe or the British Isles and, therefore, not all hydrogen
is imported (\cref{fig:si:cost-supply-curves}).

\begin{figure*}
    \begin{subfigure}[t]{\columnwidth}
        \caption{cost variations applied to all carriers but electricity}
        \label{fig:sensitivity-costs:A}
        \includegraphics[width=\columnwidth]{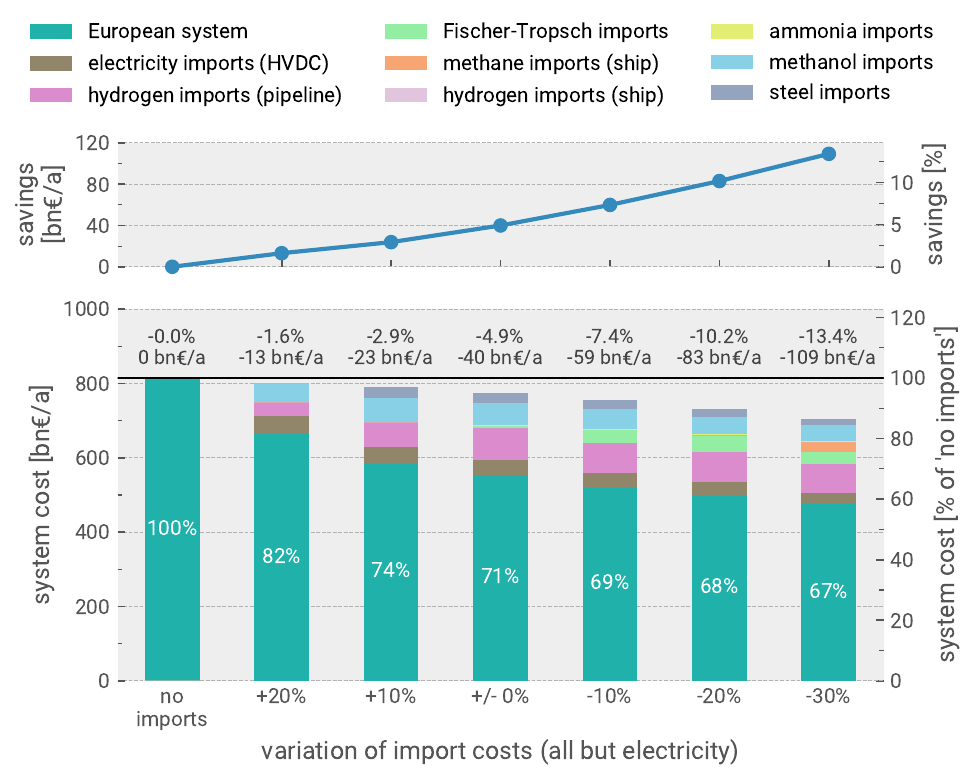}
    \end{subfigure}
    \begin{subfigure}[t]{\columnwidth}
        \caption{cost variations only applied to carbonaceous fuels}
        \label{fig:sensitivity-costs:B}
        \includegraphics[width=\columnwidth]{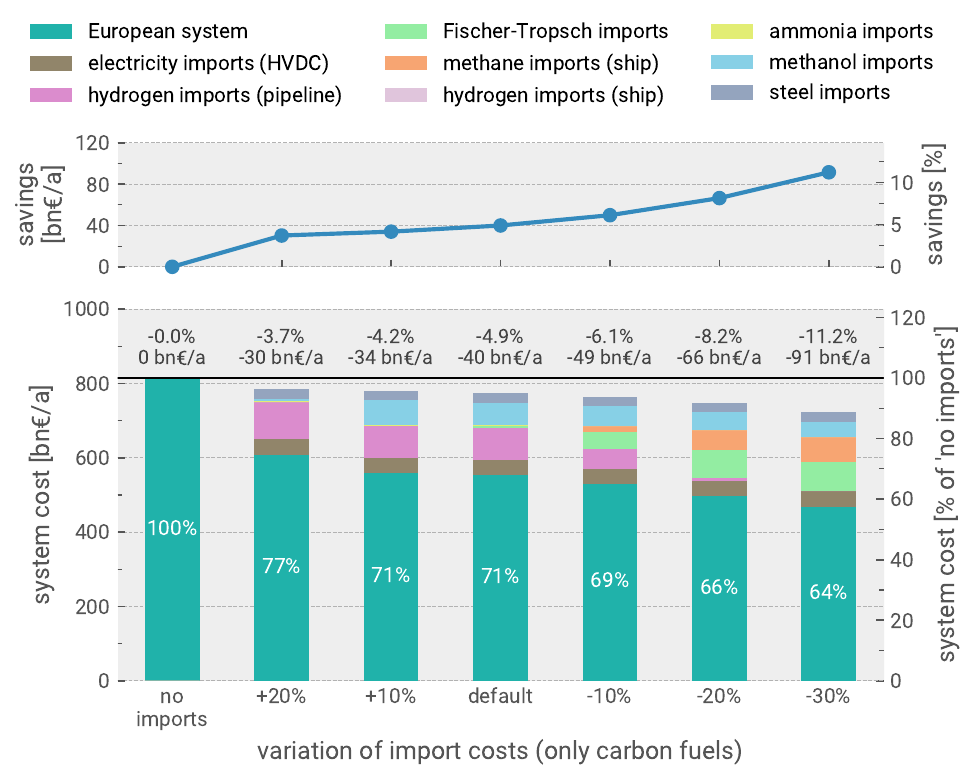}
    \end{subfigure}
    \caption{\textbf{Effect of import cost variations on cost savings and import shares with all vectors allowed.}
    In panel (a), indicated relative cost changes are applied uniformly to all
    vectors but electricity imports. In panel (b), cost changes are only applied
    to carbonaceous fuels (methane, methanol and Fischer-Tropsch). Top subpanels
    show potential cost savings compared to the scenario without imports. Bottom
    subpanels show the share and composition of different import vectors in
    relation to total energy system costs. The information is shown both in
    absolute terms and relative terms compared to the scenario without imports.
    }
    \label{fig:sensitivity-costs}
\end{figure*}

\section*{Sensitivity of potential cost savings to import costs}

It should be noted, however, that the cost-optimal import mix also strongly
depends on the assumed import costs. This uncertainty is addressed in
\cref{fig:sensitivity-costs}. \cref{fig:sensitivity-costs:A} highlights the
extensive range in potential cost reductions if higher or lower import costs
could be attained and underlines the resulting variance in cost-effective import
mixes. Within $\pm 20\%$ of the default import costs applied to all carriers but
electricity, total cost savings vary between \bneuro{13} (1.6\%) and \bneuro{83}
(10.2\%). Within this range, import volumes vary between 1700 and 3800~TWh.
Across most scenarios, there is a stable role for methanol, steel, electricity
and hydrogen imports. One significant difference, however, are Fischer-Tropsch
fuel imports starting from cost reductions of 10\% and their absence at cost
increases beyond 10\%.

A breakdown of potential causes for such cost variations such as cost of
capital, cost of carbon dioxide and investment costs are presented in
\cref{tab:cost-uncertainty}. Some of these only affect the cost of
carbonaceous fuels. One central assumption for carbon-based fuels is that
imported fuels rely exclusively on direct air capture (DAC) as a carbon source.
Arguments for this assumption relate to the potential remoteness of the ideal
locations for renewable fuel production or the absence of industrial point
sources in the exporting country. On the other hand, domestic electrofuels can
mostly use less expensive captured carbon dioxide from industrial point sources
or biogenic origin. Therefore, the higher cost for DAC partially cancels out the
savings from utilising better renewable resources abroad. This is one of the
reasons why there is substantial power-to-X production in Europe, even with
corresponding import options. The availability of cheaper biogenic CO$_2$ in the
exporting country, for instance, would make importing carbonaceous fuels more
attractive (\cref{tab:cost-uncertainty}).

When the relative cost variation of $\pm 20\%$ is only applied to carbon-based
fuels (\cref{fig:sensitivity-costs:B}), hydrogen imports are increasingly
displaced by methane and Fischer-Tropsch imports with falling costs. However, it
takes a cost increase of more than 10\% for domestic methanol production to
become more cost-effective than methanol imports. This underlines the robust
benefit of importing methanol.

% FN import volumes in cost optimum +20%: 1705 TWh +10%: 2391 TWh +-0%: 2800 TWh
% -10%: 3121 TWh -20%: 3412 TWh -30%: 3766 TWh

\begin{figure*} 
    \includegraphics[width=\textwidth]{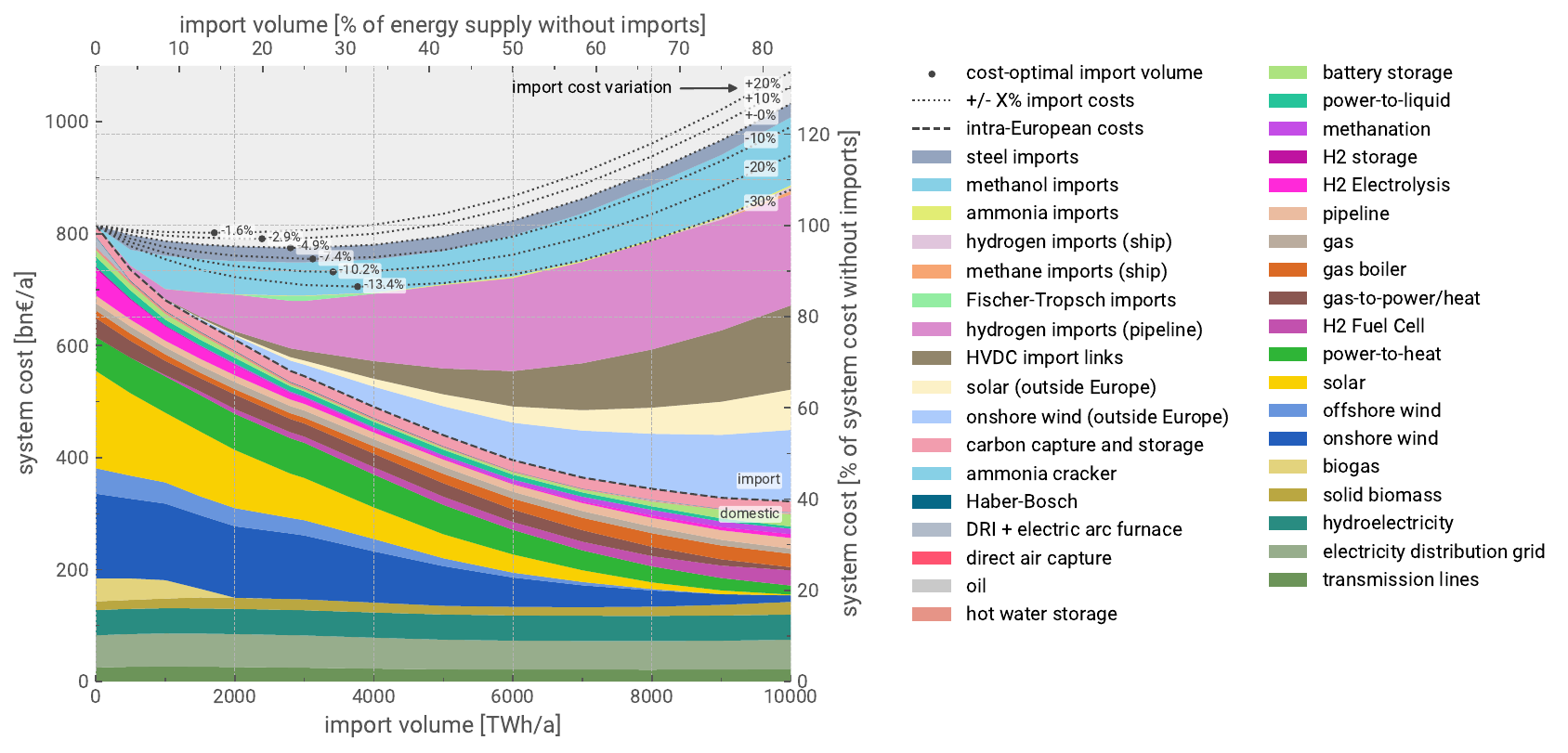}
    \caption{\textbf{Sensitivity of import volume on total system cost and composition.}
        The dashed line splits total system cost into domestic and foreign cost.
        Dotted lines represent import cost variations, indicating the respective
        altered profile of total system cost for given prescribed import
        volumes. The black markers denote the maximum cost reductions and
        cost-optimal import volume for a given import cost level (extreme points
        of the profiles). Steel is included in energy terms applying 2.1 kWh/kg
        as released by the oxidation of iron. Cost alterations are uniformly
        applied to all imports opotions but direct electricity imports. }
    \label{fig:sensitivity-volume}
\end{figure*}

\section*{Attainable cost savings for varying import volumes} 

What is consistent across all import cost variations is the flat solution space
around the respective cost-optimal import volumes. Increasing or decreasing the
total amount of imports barely affects system costs within $\pm 1000$ TWh. This
is illustrated in \cref{fig:sensitivity-volume}, which shows the system cost as
a function of enforced import volumes and different import costs. A wide range
scenarios with import volumes below 5600~TWh (4000~TWh with +20\% and 7500~TWh
with -20\% import costs) have lower total energy system costs than the scenario
without any imports. These import volumes are roughly twice the cost-optimal
import volumes, which are indicated by the black markers in
\cref{fig:sensitivity-volume} and correspond to the bars previously shown in
\cref{fig:sensitivity-costs:A}. Naturally, the cost-optimal volume of imports
increases as their costs decrease, but the response weakens with lower import
costs.

As we explore the effect of increasing import volumes on system costs, we find
that already 43\% (36-61\% within $\pm$20\% import costs) of the 4.9\%
(1.6-10.2\%) total cost benefit (\bneuro{17}) can be achieved with the first 500
TWh of imports. This corresponds to only 18\% (15-29\%) of the cost-optimal
import volumes, highlighting the diminishing return of large amounts of energy
imports in Europe. While importing 1000 TWh already realises 70\% of the maximum
cost savings with our default assumptions, this maximum is only obtained for
2800~TWh of imports. For these initial 1000~TWh, primary crude steel and
methanol imports are prioritised, followed by hydrogen and, subsequently,
electricity beyond 2000 TWh. Once more than 5000~TWh are imported, less than
half the total system cost would be spent on domestic energy infrastructure.

As imports increase, there is a corresponding decrease in the need for domestic
power-to-X (PtX) production and renewable capacities. A large share of the
hydrogen, methanol and raw steel production is outsourced from Europe, reducing
the need for domestic wind and solar capacities. This trend is further
characterised by the displacement of biogas usage in favour of hydrogen imports
around the 2000~TWh mark as demand for domestic \ce{CO2} utilisation drops. An
increase in the amount of hydrogen imported coincides with an increasing use of
hydrogen fuel cells for electricity and central heat supply in district heating
networks, partially displacing the use of methane. Regarding electricity imports
from the MENA region, \cref{fig:sensitivity-volume} reveals a mix of wind and
solar power to establish favourable feed-in profiles for the European system
integration and higher utilisation rates for the long-distance HVDC links with a
capacity-weighted average of 72\%. Utilisation rates are high because a
considerable share of the import costs of electricity can be attributed to power
transmission.

% FN LCOE for solar and wind in MENA are mostly similar potentials are not
% constraining DZ: solar 34 €/MWh, wind 32 €/MWh TN: solar 30 €/MWh, wind 38
% €/MWh LY: solar 24 €/MWh, wind 37 €/MWh

As import costs are varied, the composition of the domestic system and import
mix is primarily similar (\cref{fig:si:volume}). The main differences are a
more prominent role for Fischer-Tropsch fuel imports with lower import costs and
green methane for high import volumes. It should also be noted that the windows
for cost savings are much smaller if only subsets of import options are
available (\cref{fig:si:volume-subsets}). However, up to an import volume of
2000~TWh, excluding electricity imports would not diminish the cost-saving
potential substantially.

\begin{figure*}
    \includegraphics[width=\textwidth]{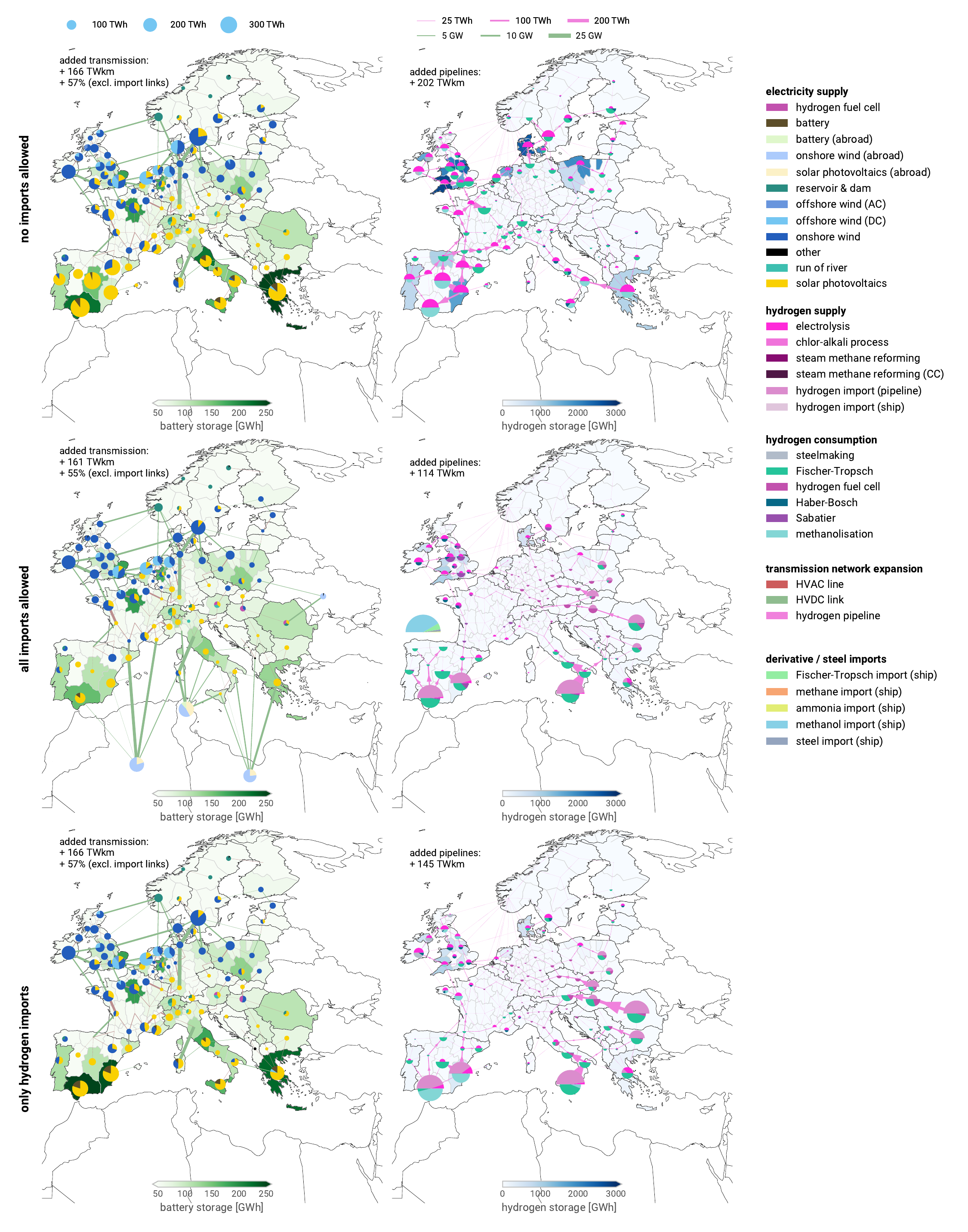}
    \caption{\textbf{Layout of European energy infrastructure for different import scenarios.}
        Left column shows the regional electricity supply mix (pies), added HVDC
        and HVAC transmission capacity (lines), and the siting of battery
        storage (choropleth). Right column shows the hydrogen supply (top half
        of pies) and consumption (bottom half of pies), net flow and direction
        of hydrogen in newly built pipelines (lines), and the siting of hydrogen
        storage subject to geological potentials (choropleth). Total volumes of
        transmission expansion are given in TWkm, which is the sum product of
        the capacity and length of individual connections. The half circle in
        the Bay of Biscay indicates the imports of carriers that are not
        spatially resolved: ammonia, steel, methanol, Fischer-Tropsch fuels.
        Hydrogen imports are shown at the entry points. Hydrogen imports in
        Bulgaria and Romania originate from Algeria and Egypt. Maps for more
        scenarios are included in the supplementary material. }
    \label{fig:import-infrastructure}
\end{figure*}

\section*{Interactions of import strategy \& domestic infrastructure}

Across the range of import scenarios analysed, we find that the decision which
import vectors are used strongly affects domestic energy infrastructure needs
(\cref{fig:import-infrastructure}).

%%% self-sufficiency %%%

In the fully self-sufficient European energy supply scenario, we see large
\mbox{power-to-X} production within Europe to cover the demand for hydrogen and
hydrogen derivatives in steelmaking, high-value chemicals, green shipping and
aviation fuels. Production sites are concentrated in Southern Europe for
solar-based electrolysis and the broader North Sea region for wind-based
electrolysis. The steel and ammonia industries relocate to the periphery of
Europe in Spain and Scotland, where hydrogen is cheap and abundant. Electricity
grid reinforcements are focused in Northwestern Europe and long-distance HVDC
connections but are broadly distributed overall. Hydrogen pipeline build-out is
strongest in Spain and France to transport hydrogen from the Southern production
hubs to fuel synthesis sites. Most of these pipelines are used unidirectionally,
with bidirectional usage where pipelines link hydrogen production and low-cost
geological storage sites (for instance, between Greece and Italy and Southern
Spain).

%%% flexible imports %%%

Considering imports of renewable electricity, green hydrogen, and electrofuels
substantially alters the infrastructure buildout in Europe. Imports displace
much of the European power-to-X production capacities and, particularly,
domestic solar energy generation in Southern Europe. In contrast, the British
Isles retain some domestic electrolyser capacities to produce synthetic methane
locally, also leveraging the Sabatier process's waste heat. The electricity
imports are distributed evenly between the North African countries Algeria,
Libya, and Tunisia and across multiple entry points in Spain, France, Italy and
Greece. This facilitates grid integration without strong reinforcement needs.
Electricity imports are also optimised to achieve higher utilisation rates above
70\% for the HVDC import connections. This is realised by mixing wind and solar
generation for seasonal balancing and using some batteries for short-term
storage (\cref{fig:si:import-operation}).

While the amount and locations of domestic power grid reinforcements are not
significantly affected by the import of electricity and other fuels, the extent
of the hydrogen network is halved and its routing is significantly altered.
Compared to the self-sufficiency scenario, the cost-benefit of the hydrogen
network shrinks from \bneuro{11} (1.3\%) to \bneuro{3} (0.4\%). This is caused
by substantial amounts of methanol imports that diminish the demand for hydrogen
in Europe and, hence, the need to transport it. In combination with the steel
and ammonia industry relocation, longer hydrogen pipeline connections are then
predominantly built to meet hydrogen CHP demands to bring electricity and heat
to renewables-poor and grid-poor regions in Eastern Europe and Germany.
Moreover, the hydrogen network helps absorb inbound hydrogen in South and
Southeast Europe, transporting some hydrogen, which is not directly used for
fuel synthesis at the entry points, to neighbouring regions.

%%% overarching trends %%%

A further observation is the high value of power-to-X production for system
integration and the role of waste heat in the siting of fuel synthesis plants
(\cref{fig:si:infra-b}). Using the process waste heat in district heating
networks with seasonal thermal storage generates notable cost savings of
\bneuro{11-21} (1.3-2.6\%). Consequently, savings are lower when imports
displace domestic PtX infrastructure. To realise these benefits, PtX facilities
tend to be located in densely populated areas (e.g.~Paris or Hamburg), which
drives part of the the hydrogen network. Notably, because of the waste heat
produced in Fischer-Tropsch and Sabatier plants, these tend to locate where
space heating demand is high. This is not the case for methanolisation plants,
which have lower waste heat potential. Alongside the flexible operation of
electrolysis to integrate variable wind and solar feed-in and the broad
availability of industrial and biogenic carbon sources in Europe, waste heat
usage is a key factor that makes electricity and hydrogen imports with
subsequent domestic conversion more attractive relative to the direct import of
derivative products. Infrastructure layouts for further import scenarios are
presented in \cref{fig:si:infra-b} to \cref{fig:si:infra-d}.

\section*{Discussion and conclusions}
\label{sec:discussion}

Our analysis offers insights into how renewable energy imports might reduce
overall systems costs and interact with European energy infrastructure. Our
results show that imports of green energy reduce costs of a carbon-neutral
European energy system by \bneuro{39} (5\%), noting, however, that the
uncertainty range is considerable. While we find that some imports are robustly
beneficial, system cost savings range between 1\% and 14\% depending on the
import costs. What is consistent, however, are the diminishing return of energy
imports for larger quantities, with peak cost savings below imports of
4000~TWh/a. We also find that there is value in pursuing some \mbox{power-to-X}
production in Europe as a source of flexibility for wind and solar integration
and as a source of waste heat for district heating networks. Another location
factor in favour of European \mbox{power-to-X} is the  wide availability of
sustainable biogenic and industrial carbon sources, which helps to reduce
reliance on costly direct air capture.

Overall, we find that the import vectors used strongly affect domestic
infrastructure needs. For example, only a smaller hydrogen network would be
required if hydrogen derivatives were largely imported. This underscores the
importance of coordination between energy import strategies and infrastructure
policy decisions. Our results present a quantitative basis for further
discussions about the trade-offs between system cost, carbon neutrality, public
acceptance, energy security, infrastructure buildout and imports.

The small differences in cost between some scenarios is particularly relevant
because factors other than pure costs might then drive the designs of import
strategies. The relatively limited cost benefit of imports and value chain
reordering, may speak against pursuing this avenue. A desire for energy
sovereignty would motivate more domestic supply and diversified imports. For
instance, focusing on ship-bourne imports would reduce pipeline lock-in and
mitigate the risks of sudden supply disruptions and abuse of market power.
Focussing on carriers that are already a globally traded commodity may also be
more appealing. Producing renewable energy locally would bring value creation
and jobs to Europe that are currently outsourced to fossil fuel producers
abroad. For similar reasons, the import of intermediary raw products like sponge
iron, which represents the most energy-intensive part of the steel value chain,
could also be a relevant option.

There is also a social dimension to the import strategy and the question of how
fast the associated infrastructure can get built. Policymaking in Europe might
prefer easy-to-implement systems featuring, for instance, lower domestic
infrastructure requirements, reuse of existing infrastructure, lower technology
risk, and reduced land usage for broader public support than the most
cost-effective solution. However, in outsourcing potential land use and
infrastructure conflicts to abroad, potential exporting countries must weigh the
prospect of economic development against internal social and environmental
issues. Ultimately, Europe's energy strategy must balance cost savings from
green energy and material imports with broader concerns like geopolitics,
economic development, public opinion and the willingness of potential exporting
countries in order to ensure a swift, secure and sustainable energy future. Our
research shows that there is maneuvering space to accommodate such non-cost
concerns.

\section*{Methods}
\label{sec:methods}

\subsection*{Modelling of the European energy system}

For our analysis, we use the European sector-coupled high-resolution energy
system model PyPSA-Eur\cite{horschPyPSAEurOpen2018a} based on the open-source
modelling framework PyPSA\cite{brownPyPSAPython2018} (Python for Power System
Analysis) in a setup similar to Neumann et al.~\cite{neumannPotentialRole2023},
covering the energy demands of all sectors including electricity, heat,
transport, industry, agriculture, as well as international shipping and
aviation.

The model simultaneously optimises spatially explicit investments and the
operation of generation, storage, conversion and transmission assets to minimise
total system costs in a linear optimisation problem, which is solved with
\textit{Gurobi}.\cite{gurobi} The capacity expansion is based on technology cost
and efficiency projections for the year 2030, many of which are taken from the
technology catalogue of the Danish Energy Agency.\cite{DEA} Choosing projections
for the year 2030 for a net-zero carbon emission scenarios more likely to be
reached by mid-century acknowledges that much of the required infrastructure
must be constructed well in advance of reaching this goal. Spatially, the model
resolves 110 regions in Europe,\cite{frysztackiStrongEffect2021} covering the
European Union, the United Kingdom, Norway, Switzerland and the Balkan countries
without Malta and Cyprus. Temporally, the model is solved with an uninterrupted
4-hourly equivalent resolution for the weather year 2013, using a segmentation
clustering approach implemented in the \textit{tsam}
toolbox.\cite{hoffmannParetooptimalTemporal2022} In terms of investment periods,
no pathway optimisation is conducted, but a greenfield approach is pursued
except for existing hydro-electricity and transmission infrastructure targeting
net-zero CO$_2$ emissions.

Networks are considered for electricity, methane and hydrogen
transport.\cite{ENTSOE,plutaSciGRIDGas2022a} However, different to Neumann et
al.,\cite{neumannPotentialRole2023} pipeline retrofitting to hydrogen is
disabled for computational reasons such that all hydrogen pipelines are assumed
to be newly built. Data on the gas transmission system is further supplemented
by the locations of fossil gas extraction sites and gas storage facilities based
on SciGRID\_gas,\cite{plutaSciGRIDGas2022a} as well as investment costs and
capacities of existing and planned LNG
terminals\cite{instituteforenergyeconomicsandfinancialanalysisEuropeanLNG2023}
Moreover, a carbon dioxide network is not explicitly co-optimised since CO$_2$
is not spatially resolved in this model version.\cite{hofmannDesigningCO22023}

The overall annual sequestration of CO$_2$ is limited to 200
Mt$_{\text{CO}_2}$/a. This number allows for sequestering the industry's
unabated fossil emissions (e.g. in the cement industry) while minimising
reliance on carbon removal technologies. The carbon management features of the
model trace the carbon cycles through various conversion stages: industrial
emissions, biomass and gas combustion, carbon capture, storage or long-term
sequestration, direct air capture, electrofuels, recycling, and waste-to-energy
plants.

Renewable potentials and time series for wind and solar electricity generation
are calculated with \textit{atlite},\cite{hofmannAtliteLightweight2021}
considering land eligibility constraints like nature reserves or distance
criteria to settlements. Given low onshore wind expansion in many European
countries in recent years,\cite{ourworldindataInstalledWind2023} restrictive
onshore wind expansion potentials are applied, using a 1.5 MW/km$^2$ factor for
the eligible land area. Geological potentials for hydrogen storage are taken
from Caglayan et al.\cite{caglayanTechnicalPotential2020} Biomass potentials are
restricted to residues from agriculture and forestry, as well as waste and
manure, based on the medium potentials specified for 2030 in the JRC-ENSPRESO
database.\cite{ruizENSPRESOOpen2019} The finite biomass resource can be employed
for low-temperature heat provision in industrial applications, biomass boilers
and CHPs, and biofuel production for use in aviation, shipping and the chemicals
industry. Additionally, we allow biogas upgrading, including the capture of the
CO$_2$ contained in biogas. The total assumed bioenergy potentials are 1569~TWh
with a carbon content corresponding to 546~Mt$_{\text{CO}_2}$/a, which is not
fully available as a feedstock for fuel synthesis due to imperfect capture
rates of up to 90\%.

Heating supply technologies like heat pumps, electric boilers, gas boilers, and
combined heat and power (CHP) plants are endogenously optimised separately for
decentral use and central district heating. District heating networks can
further be supplemented with waste heat from various power-to-X processes
(electrolysis, methanation, methanolisation, ammonia synthesis, Fischer-Tropsch
fuel synthesis).

While the shipping sector is assumed to use methanol as fuel, land-based
transport, including heavy-duty vehicles, is deemed fully electrified in the
presented scenarios. Aviation can decide to use green kerosene derived from
Fischer-Tropsch fuels or methanol. Besides methanol-to-kerosene, further
usage options for methanol have been added.
These include
methanol-to-olefins/aromatics for the production of green plastics,
methanol-to-power\cite{brownUltralongdurationEnergyStorage2023} in open-cycle gas
turbines or Allam cycle turbines, and steam reforming of methanol with or
without carbon capture. For the synthesis of electrofuels, we also account for
potential operational restrictions by considering a minimum part load of 30\%
for methanolisation and methanation compared to 70\% for Fischer-Tropsch
synthesis, both within Europe and abroad.

A further core improvement of the model regards the physical representation of
energy transport over long distances. For gas and hydrogen pipelines, we
incorporate electricity demands for compression of 1\% and 2\% per 1000km
of the transported energy, respectively.\cite{gasforclimateEuropeanHydrogen2021}
For HVDC transmission lines, we assume 2\% static losses at the substations and
additional losses of 3\% per 1000km. The losses of high-voltage AC transmission
lines are estimated using a piecewise linear approximation as proposed in
Neumann et al.,\cite{neumannAssessmentsLinear2022} in addition to the linearised
power flow equations.\cite{horschLinearOptimal2018} Up to a maximum capacity
increase of 30\%, we consider dynamic line rating (DLR), leveraging the cooling
effect of wind and low ambient temperatures to exploit existing transmission
assets fully.\cite{glaumLeveragingExisting2023} To approximate N-1 resilience,
transmission lines may only be used up to 70\% of their rated dynamic capacity.
To prevent excessive expansion of single connections, the expansion of power
transmission lines between two regions is limited to 15 GW for HVAC and 25 GW
for HVDC lines, while a similar constraint of 50.7 GW is placed on hydrogen
pipelines, which corresponds to three parallel 48-inch
pipelines.\cite{gasforclimateEuropeanHydrogen2021}

Finally, we also developed the possibility for the model to relocate the steel
and ammonia industry within Europe, mainly to level the playing field between
non-European green steel imports and domestic production. This is achieved by
explicitly modelling the cost, efficiency and operation of hydrogen direct iron
reduction (H2-DRI) and electric arc furnaces (EAF), which can be sited all over
Europe, and the cost to procure iron ore. We further allow the oversizing of
steelmaking plants to allow flexible production in response to the renewables
supply conditions.

\subsection*{Modelling of import supply chains and costs}

\begin{figure*}
    \centering
    \includegraphics[width=.82\textwidth]{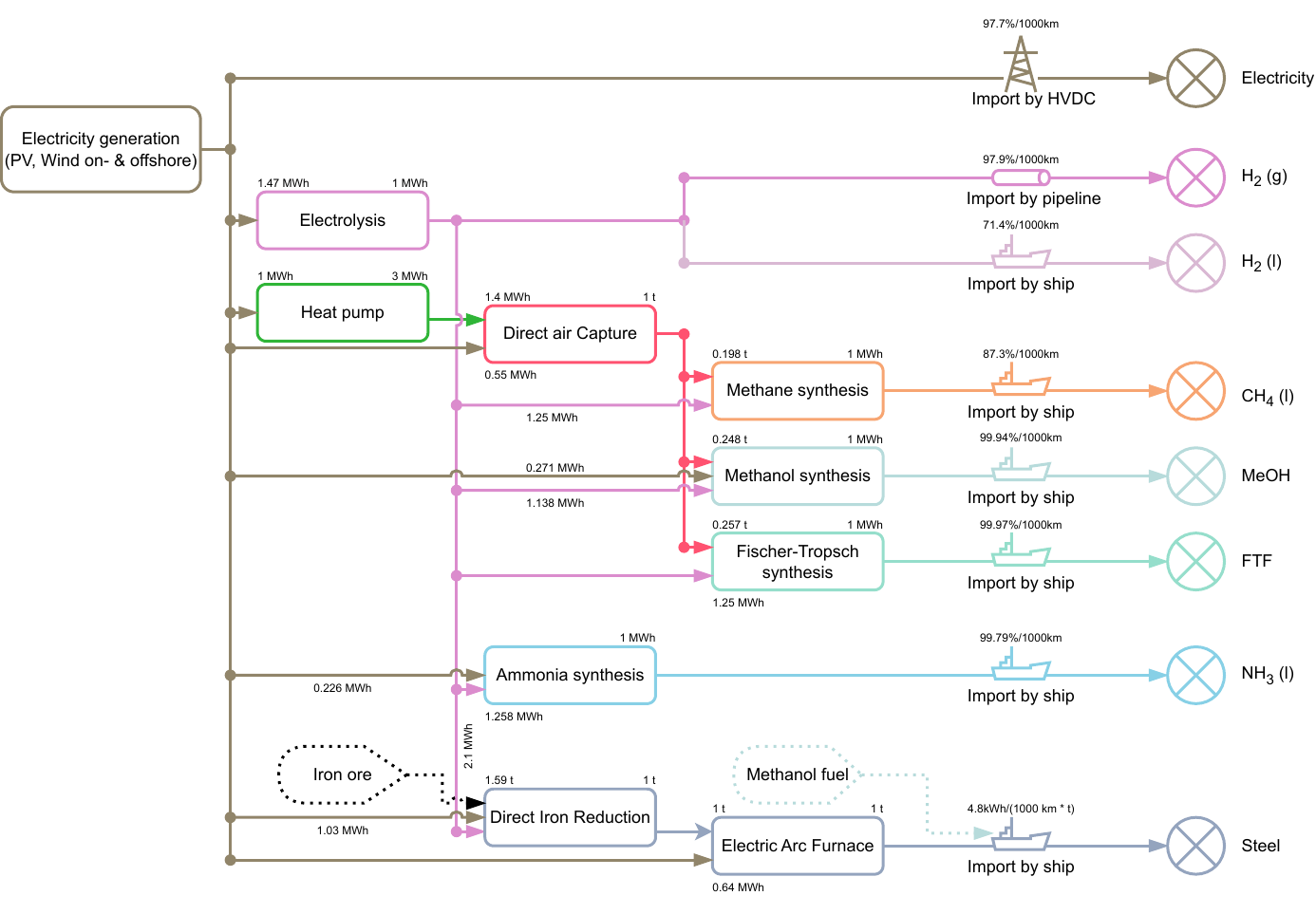}
    \caption{\textbf{Schematic overview of the import supply chains.} The
    illustration includes key input-output ratios of the different conversion
    processes and the transport efficiencies for the different import vectors.}
    \label{fig:import-esc-scheme}
\end{figure*}

The European energy system model is extended with data from the TRACE model
\cite{hamppImportOptions2023} to assess the costs of different vectors for
importing green energy and material into Europe from various world regions. For
each vector, we identify locations with existing or planned import
infrastructure where the respective carrier may enter the European energy
system.

Starting from the methodology by Hampp et al.\cite{hamppImportOptions2023}, some
adjustments were made to the original TRACE model. Namely, land availability and
wind and solar time-series are determined using
\textit{atlite}\cite{hofmannAtliteLightweight2021} instead of
\textit{GEGIS}.\cite{mattssonAutopilotEnergy2021} Techno-economic assumptions
were aligned with those used in the European model, and steel was included as an
energy-intensive material import vector. The exporting countries comprise
Australia, Argentina, Chile, Kazakhstan, Namibia, Turkey, Ukraine, the Eastern
United States and Canada, mainland China, and counties in the MENA region.

To determine the levelised cost of energy for exports, the methodology first
assesses the regional potentials for wind and solar energy. A regional
electricity cost supply curve is determined, from which projected future local
energy demand is subtracted. Thereby, domestic consumption is prioritised and
supplied by the countries' best renewable resources even though we do not model
the energy transition in exporting countries in detail. The remaining wind and
solar electricity supply can then be used to produce the specific energy or
material vector using water electrolysis for \ce{H2}, direct air capture (DAC)
for \ce{CO2}, air separation units (ASU) for \ce{N2}, synthesis of methane,
methanol, ammonia or Fischer-Tropsch fuels from \ce{H2} with \ce{CO2} or
\ce{N2}, and \ce{H2} direct iron reduction (DRI) with subsequent processing in
electric arc furnaces (EAF) for the processing of iron ore (103.7 \euro{}/t)
into green steel. Other CO$_2$ sources than DAC are not considered in the
exporting countries, a notable difference from the European model. Liquid
organic hydrogen carriers (LOHC) are not considered as export vector due to
their lower technology readiness level (TRL) compared to other
vectors.\cite{irenaGlobalHydrogen2022} Further details on the energy and
feedstock flow and process efficiencies are outlined in
\cref{fig:import-esc-scheme}.

For each vector, an annual reference export demand of 500~TWh (lower heating
value, LHV) or 100~Mt of steel is assumed, mirroring large-scale energy and
material infrastructures and export volumes, corresponding to approximately 40\%
of current LNG
imports\cite{instituteforenergyeconomicsandfinancialanalysisEuropeanLNG2023} and
66\% of European steel
production.\cite{eurofer-theeuropeansteelassociationEuropeanSteel2023}

%%% capacity expansion %%%

Based on these supply chain definitions, a capacity expansion optimisation is
performed to determine the cost-optimal combination of infrastructure and
process capacities for all intermediary products and delivering the final
carrier either through pipelines (\ce{H2(g)}, \ce{CH4(g)}) or by ship
(\ce{H2(l)}, \ce{CH4(l)}, \ce{NH3(l)}, \ce{MeOH}, Fischer-Tropsch fuel,
and steel). Exports from each of the regions shown in \cref{fig:options:global}
are modelled to each of twelve European import locations based on large port
locations, determining the levelised costs of energy or steel the European entry
point will see for each supply chain. All energy supply chains are assumed to
consume their energy vector as fuel for transport to Europe, except for steel,
which uses externally bought methanol as shipping fuel.

%%% import constraints in European model %%%

The resulting levelised cost of exported energy specific to the respective
importing regions is added as a constant marginal import cost for all
chemical energy carriers and steel. For the import of hydrogen and methane,
candidate entry points are identified based on where existing and prospective
LNG terminals and cross-continental pipelines are located. This includes new LNG
import terminals in Europe in response to ambitions to phase out Russian gas
supply in 2022.
\cite{instituteforenergyeconomicsandfinancialanalysisEuropeanLNG2023} To achieve
regional diversity in potential gas and hydrogen imports and avoid vulnerable
singular import locations, we allow the expansion beyond the reported capacities
only up to a factor of 2.5, taking the median value of reported investment costs
for LNG terminals.\cite{GlobalGas2022} A surcharge of 20\% is added for hydrogen
import terminals due to the lack of practical experience. Carbonaceous fuels,
ammonia, and steel imports are not spatially resolved due to their low transport
costs and, therefore, are not constrained by the availability of suitable entry
points. To present energy and material imports in a common unit, the embodied
energy in steel is approximated with the 2.1 kWh/kg released in iron oxide
reduction, i.e.~energy released by combustion.\cite{kuhnIronRecyclable2022}

%%% special handling of electricity imports %%%

Owing to the variability of wind and solar electricity, the supply chain of
electricity imports is endogenously optimised with the rest of the European
system rather than using a constant levelised cost of exported electricity. This
comprises the optimisation of wind and solar capacities, batteries and hydrogen
storage in steel tanks, and the size and operation of HVDC link connection into
Europe based on the availability time series in neighbouring countries as
illustrated in \cref{fig:options:europe}. Underground hydrogen storage options
are not considered due to the limited availability of salt caverns in many of
the best renewable resource regions in the countries that are considered
exporting.\cite{hevinUndergroundStorage2019} We also assume that the energy
supply chains dedicated to exports will be islanded from the rest of the local
energy system. Europe's connection options with exporting countries are confined
to the 5\% nearest regions, with additional ultra-long distance connection
options to Ireland, Cornwall and Brittany following the vision of the Xlinks
project between Morocco and the United Kingdom.\cite{xlinksMoroccoUKPower2023}
Connections through Russia or Belarus are excluded, and thus, some connections
are affected by additional detours beyond the regularly applied detour factor of
125\% of the as-the-crow-flies distance. Similar to intra-European HVDC
transmission, a 3\% loss per 1000km and a 2\% converter station loss are assumed.

%%% more detailed results %%%

As illustrated in \cref{fig:options}, for imports of hydrogen by pipeline,
nearby countries like Algeria and Egypt emerged as lowest cost exporters (ca.~57
\euro{}/MWh). Importing hydrogen by ship is substantially more expensive due to
liquefaction and evaporation losses. Algeria could offer supply through this
vector at 84 \euro{}/MWh. For all other hydrogen derivatives, Argentina and
Chile offer the lowest cost imports between 88 and 110~\euro{}/MWh or
501~\euro{}/t for steel. Methanol is found to be cheaper than the
Fischer-Tropsch route because it is assumed to be more flexible (30\% minimum
part load compared to 70\% for
Fischer-Tropsch).\cite{brownUltralongdurationEnergy2023} The lower process
flexibility shifts the energy mix towards solar electricity and causes higher
levels of curtailment, increasing costs. The transport costs of \ce{CH4(l)} are
lower than for \ce{H2(l)} since the liquefaction consumes less energy and
individual ships can carry more energy with \ce{CH4(l)}. Pipeline imports of
\ce{CH4(g)} were also considered, but costs were higher than for \ce{CH4(l)}
shipping under the assumption that new pipelines would have to be built.
Consequently, the model preferred LNG imports over pipeline imports.

\section*{Acknowledgements}

J.H. gratefully acknowledges funding from the Kopernikus-Ariadne project (FKZ
03SFK5A and 03SFK5A0-2) by the German Federal Ministry of Education and
Research (\textit{Bundesministerium für Bildung und Forschung, BMBF}).

\section*{Author contributions}

% following https://casrai.org/credit/

\textbf{F.N.}:
Conceptualization --
Data curation --
Formal Analysis --
% Funding acquisition --
Investigation --
Methodology --
% Project administration --
% Resources --
Software --
% Supervision --
Validation --
Visualization --
Writing - original draft --
% Writing - review \& editing
\textbf{J.H.}:
Conceptualization --
Data curation --
Formal Analysis --
% Funding acquisition --
Investigation --
Methodology --
% Project administration --
% Resources --
Software --
% Supervision --
Validation --
Visualization --
% Writing - original draft --
Writing - review \& editing
\textbf{T.B.}:
Conceptualization --
% Data curation --
Formal Analysis --
Funding acquisition --
Investigation --
Methodology --
Project administration --
% Resources --
Software --
Supervision --
Validation --
% Visualization --
% Writing - original draft --
Writing - review \& editing

\section*{Declaration of interests}

The authors declare no competing interests.

\section*{Data and code availability}

A dataset of the model results will be made available on \url{zenodo} after peer-review.
The code to reproduce the experiments is available at \url{https://github.com/fneum/import-benefits}.

\renewcommand{\ttdefault}{\sfdefault}
\bibliography{/home/fneum/zotero-bibtex.bib}

\appendix

\onecolumn

\section{Causes and severity of import cost uncertainty}

\begin{table*}[!htb]
    \small
    \centering
    \begin{tabular}{lrrrr}
        \toprule
        Factor & Change & Unit & Change & Unit\\
        \midrule
        higher WACC of 12\% (e.g.~high project risk) & +43.1 & \euro{}/MWh  &
        +39.3 & \% \\
        higher WACC of 10\% (e.g.~high project risk) & +25.3 & \euro{}/MWh  &
        +23.0 & \% \\
        higher WACC of 8\% (e.g.~high project risk) & +8.2 & \euro{}/MWh  & +7.4
        & \% \\
        higher direct air capture investment cost (+200\%) & +55.8 & \euro{}/MWh
        & +50.8 & \% \\
        higher direct air capture investment cost (+100\%) & +28.1 & \euro{}/MWh
        & +25.6 & \% \\
        higher direct air capture investment cost (+50\%) & +14.1 & \euro{}/MWh
        & +12.9 & \% \\
        higher direct air capture investment cost (+25\%) & +7.1 & \euro{}/MWh &
        +6.5 & \% \\
        higher electrolysis investment cost (+200\%) & +29.2 & \euro{}/MWh  &
        +26.6 & \% \\
        higher electrolysis investment cost (+100\%) & +16.7 & \euro{}/MWh  &
        +15.2 & \% \\
        higher electrolysis investment cost (+50\%) & +9.0 & \euro{}/MWh  & +8.2
        & \% \\
        higher electrolysis investment cost (+25\%) & +4.7 & \euro{}/MWh  & +4.3
        & \% \\
        Argentina and Chile not available for export & +10.1 & \euro{}/MWh  &
        +9.2 & \% \\
        \midrule
        lower WACC of 3\% (e.g.~government guarantees) & -29.5 & \euro{}/MWh  &
        -26.8 & \% \\
        lower WACC of 5\% (e.g.~government guarantees) & -15.5 & \euro{}/MWh  &
        -14.1 & \% \\
        lower WACC of 6\% (e.g.~government guarantees) & -8.0 & \euro{}/MWh  &
        -7.2 & \% \\
        sell excess curtailed electricity at 40 \euro{}/MWh & -24.7 & \euro{}/MWh  &
        -22.6 & \% \\
        sell excess curtailed electricity at 30 \euro{}/MWh & -15.6 & \euro{}/MWh  &
        -14.2 & \% \\
        sell excess curtailed electricity at 20 \euro{}/MWh & -8.0 & \euro{}/MWh  &
        -7.2 & \% \\
        option to use available biogenic or cycled \ce{CO2} for 60 \euro{}/t & -21.7 &
        \euro{}/MWh  & -19.7 & \% \\
        option to use available biogenic or cycled \ce{CO2} for 80 \euro{}/t & -16.1 &
        \euro{}/MWh  & -14.7 & \% \\
        option to use available biogenic or cycled \ce{CO2} for 100 \euro{}/t & -10.6 &
        \euro{}/MWh  & -9.7 & \% \\
        option to build geological hydrogen storage at 2.4 \euro{}/kWh
        (reduction by 95\%) & -8.2 & \euro{}/MWh  & -7.4 & \% \\
        option to use power-to-X waste heat streams for direct air capture &
        -3.8 & \euro{}/MWh  & -3.4 & \% \\
        highly flexible operation of fuel synthesis plant (20\% minimum
        part-load instead of 70\%) & -5.4 & \euro{}/MWh  & -4.9 & \% \\
        \bottomrule
    \end{tabular}
    \caption{\textbf{Examples for potential import cost increases or decreases.}
    The table presents cost sensitivities in absolute and relative terms based
    on the supply chain for producing Fischer-Tropsch fuels in Argentina for
    export to Europe. The reference fuel import cost for this case is 109.8
    \euro{}/MWh. Responses to changes in the input assumptions are not
    additive.}
    \label{tab:cost-uncertainty}
\end{table*}

In \cref{tab:cost-uncertainty}, we vary some of the techno-economic assumptions
for evaluating green fuel supply chains in the exporting countries to justify
the range of import cost deviations from the defaults. These relate to
technology costs, financing costs, excess power and heat handling, fuel
synthesis flexibility, and the availability of geological hydrogen storage and
alternative sources of CO$_2$. For all the following sensitivities, it should be
noted that they are not additive.

A higher weighted average cost of capital (WACC) than the uniformly applied 7\%,
e.g.~due to higher project risks, and lower WACC, e.g.~due to the
government-backing of a project, have a substantial effect on the import cost
calculations; an increase or decrease by just one percentage point already
alters the costs per unit of energy by more than
7\%.
 
Likewise, a failure to achieve the anticipated cost reductions for electrolysers
and DAC systems would also result in far-reaching cost increases for green
energy imports, especially if the fuel contains carbon. The availability of
biogenic CO$_2$ (or fossil CO$_2$ from industrial processes that is largely
cycled between use and synthesis and, hence, not emitted to the atmosphere) can
reduce the green fuel cost by 20\% if it can be provided for 60 \euro{}/t and by
10\% if made available for 100 \euro{}/t.

The default assumptions for export supply chains assume islanded fuel synthesis
sites that are not connected to the local electricity system. The isolation
drives the system into high curtailment rates of 8\%. If surplus electricity
production could be sold and absorbed by the local power grid, considerable cost
reductions could be achieved (refer to \cref{tab:cost-uncertainty}).

Besides integration with the local energy system, process integration using
waste heat streams from power-to-X plants for direct air capture and flexible
Fischer-Tropsch synthesis similar to methanolisation can also reduce fuel cost
by 3-5\% each. Conditions that would allow for geological hydrogen storage
reduce the need for flexible synthesis plant operation and could reduce import
costs by more than 7\%. However, even though many potential export countries
possess geological hydrogen storage potential, suitable storage sites are not
always co-located with the countries' best renewable potentials.

Finally, cost rises can also be expected if the most competitive exporting
countries are not offering to export green energy. Argentina and Chile have a
margin of 10 \euro{}/MWh over the next cheapest exporting country
(i.e.~Australia, Algeria and Libya with 120 \euro{}/MWh). If these countries
were unavailable for import, costs would rise by almost 10\%.

\newpage

\section{Limitations}

Several limitations of the study should be noted. The model is a simplified
representation of the energy system and does not capture all aspects of the
real-world transition. First of all, the optimization results represent a
long-term equilibrium that does not account for the transition process from the
mostly fossil-based energy system to a carbon-neutral one. This means that the
model does not capture constraints regards neither the mid-term ramp-up of green
energy export capacities nor the speed at which supporting infrastructure could
scale up. Furthermore, global competition betweeen potential exporting
countries, as well as between importing countries, make import volumes and
prices more difficult to predict than our purely cost-based analysis can
provide. Besides unclear market developments, local issues in exporting regions
such as potential water scarcity to produce large amounts of hydrogen in
renewable-rich but arid countries is not addressed. Moreover, the fact that we
assume that import costs are not time-dependent and available on demand may
underestimate some intermediate storage requirements for imports at
entry-points, especially for hydrogen.

In terms of industry relocation within Europe, our modelling is constrained to
the migration of steel, ammonia and chemicals. Other sectors like concrete and
alumina production is not considered for relocation. We make this choice so that
relocation within Europe can compete with imports from abroad, which equates a
migration of the industry branch out of the European value chain. Potential
repurcussions on local jobs and relocatoin costs are not captured by the model.
Finally, it should be acknowledged that our assumptions about the broad
availability of district heating networks across Europe to absorb
\mbox{power-to-X} waste heat are relatively progressive as it is not certain
that central heating will become available in regions which have not based their
heating supply on district heating so far.

\section{Supplementary Figures}

\begin{figure}[!htb]
    \footnotesize
    (a) 10\% higher import costs \\
    \includegraphics[width=\textwidth]{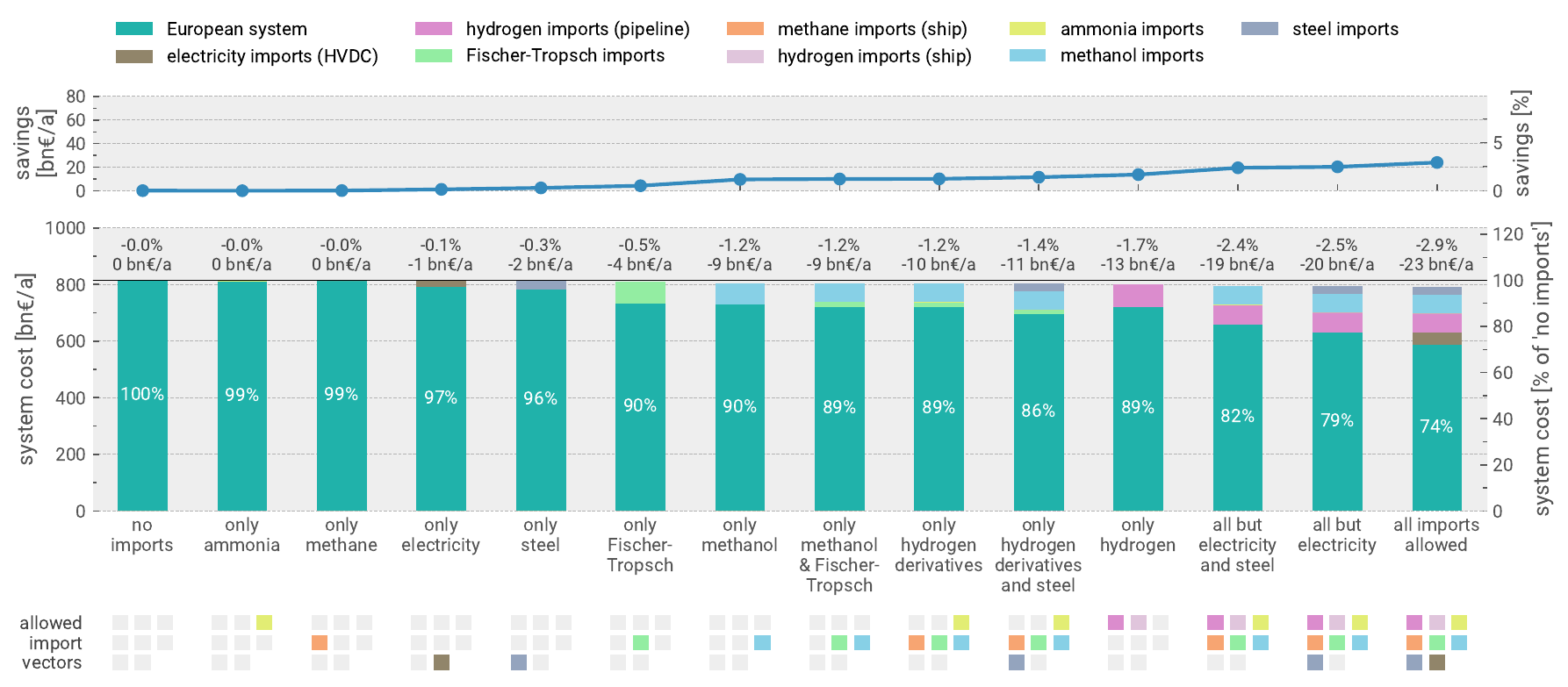} \\
    (b) 10\% lower import costs \\
    \includegraphics[width=\textwidth]{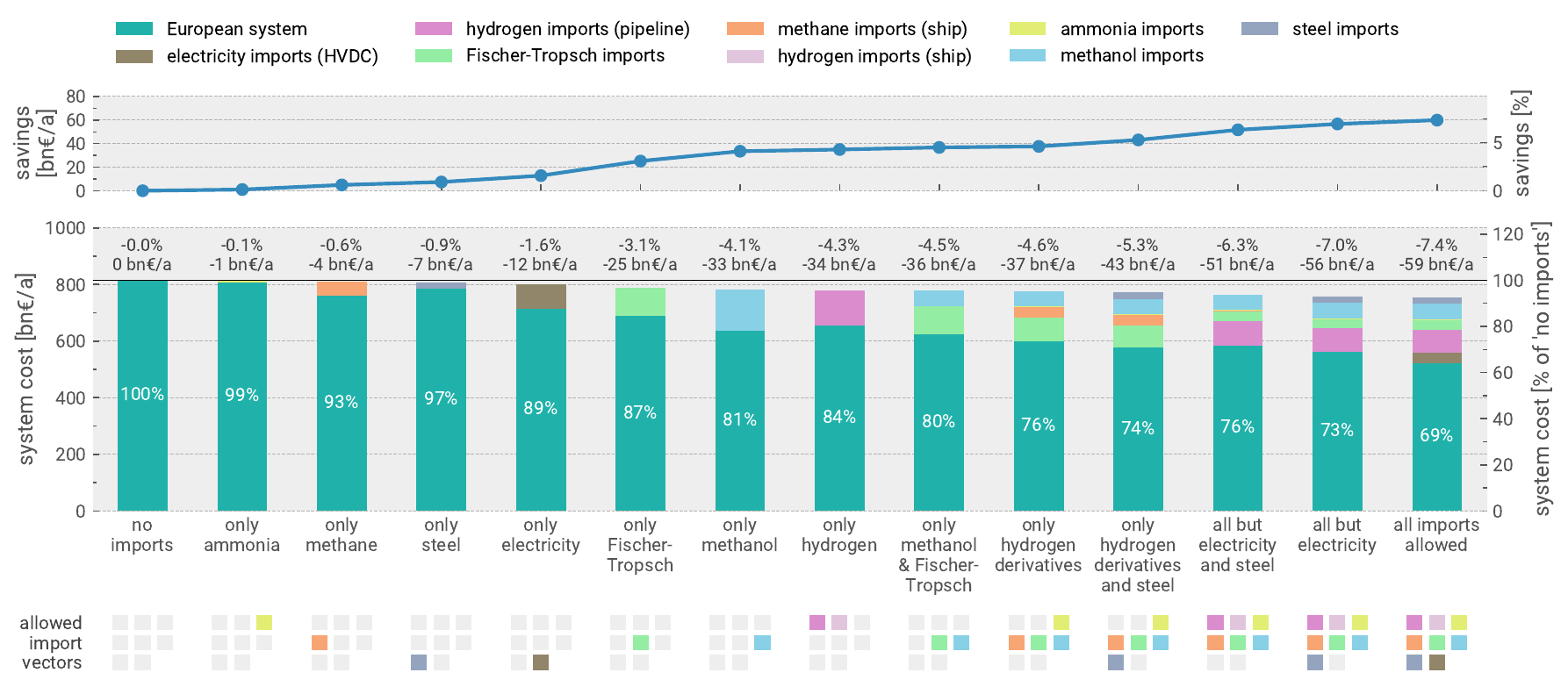} \\
    (c) 20\% lower import costs \\
    \includegraphics[width=\textwidth]{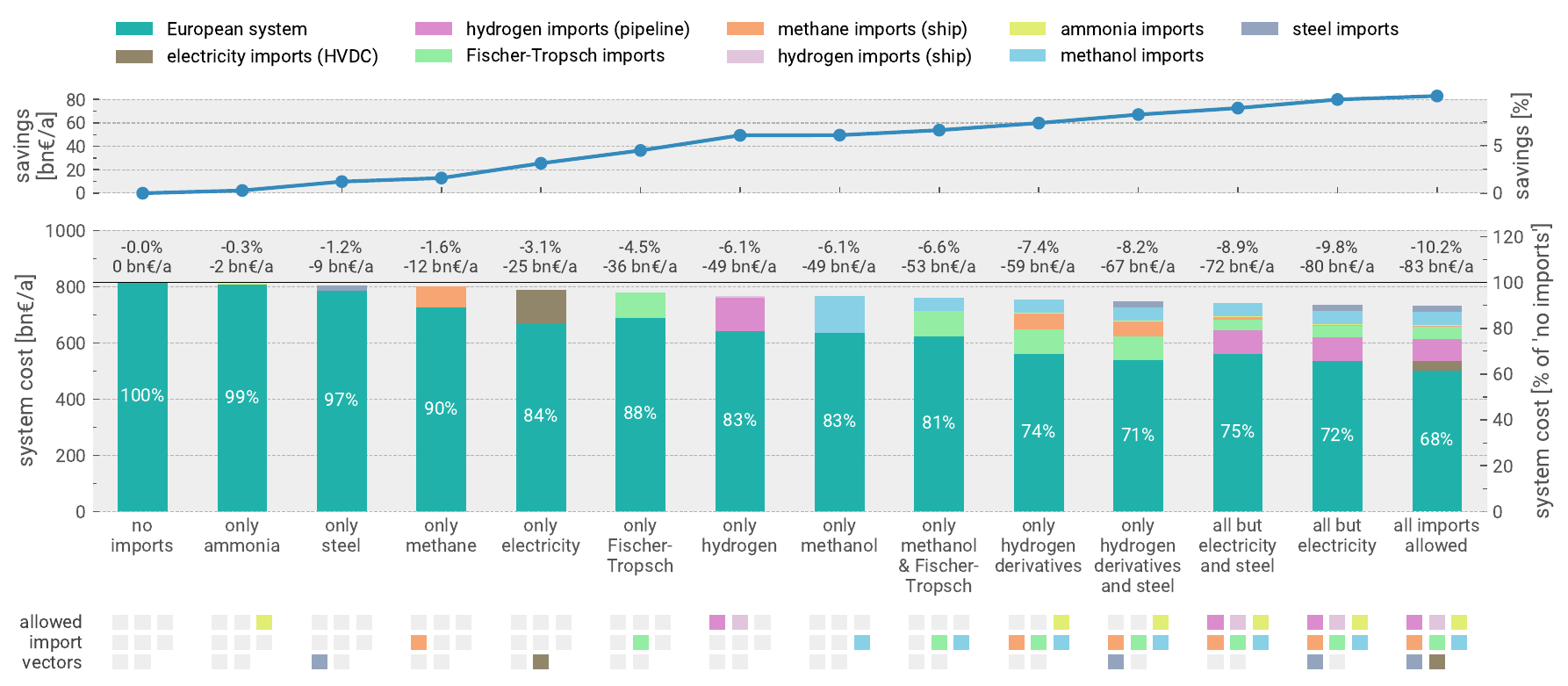} \\
    \caption{\textbf{Potential for cost reductions with reduced sets of import options for varying import costs.}}
    \label{fig:si:subsets}
\end{figure}

\begin{figure*}
    \small
    (a) import costs +20\% \\
    \includegraphics[width=\textwidth]{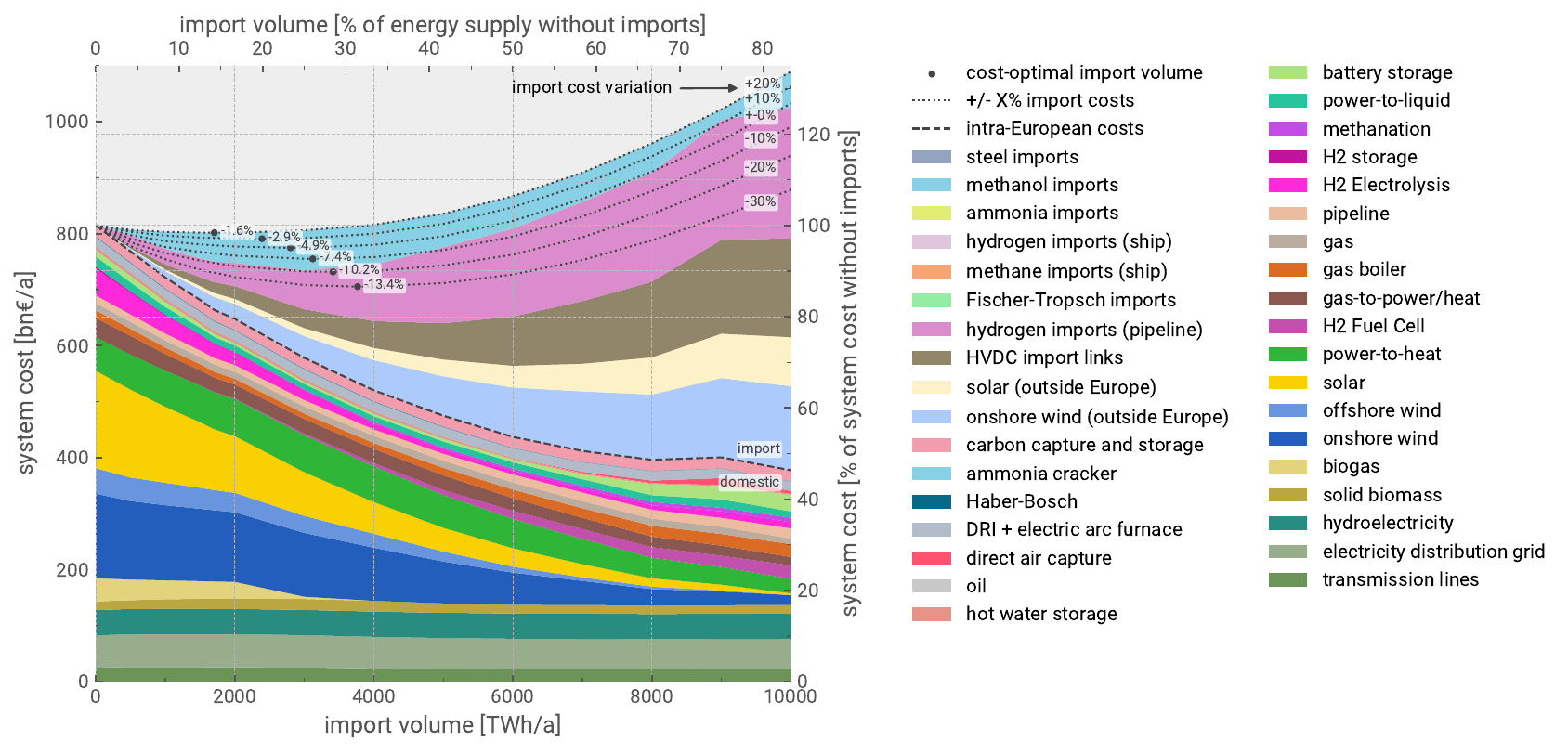}
    \begin{tabular}{ll}
        (b) import costs +10\% & (c) import costs -10\% \\
        \includegraphics[width=0.495\textwidth, trim=0cm 0cm 12.8cm 0cm, clip]{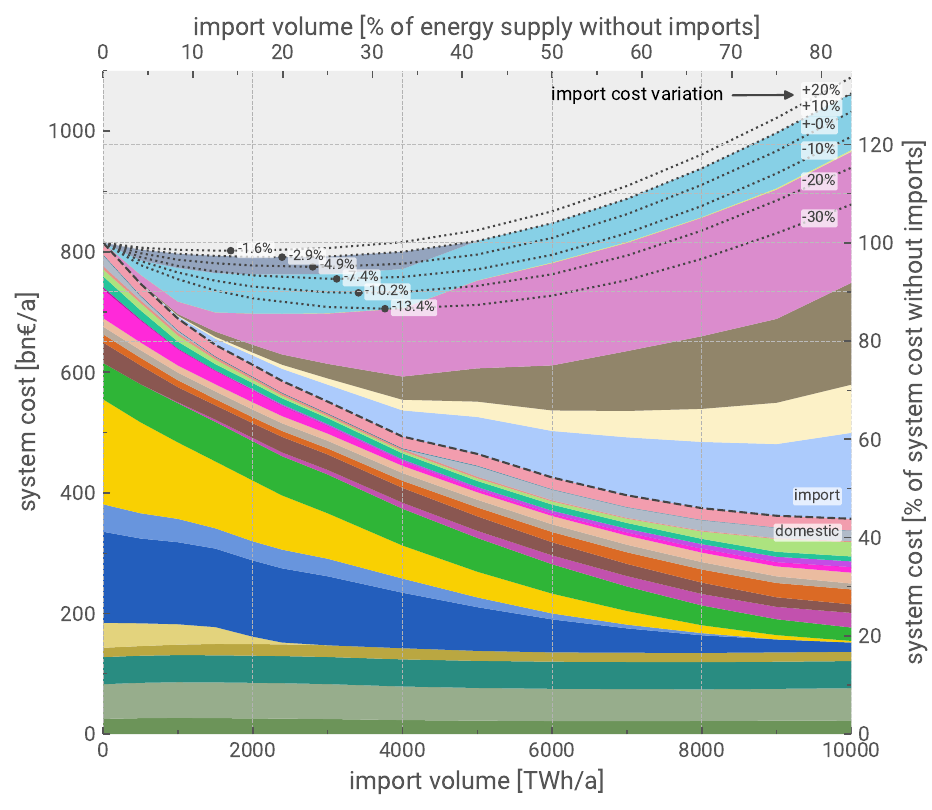} &
        \includegraphics[width=0.495\textwidth, trim=0cm 0cm 12.8cm 0cm, clip]{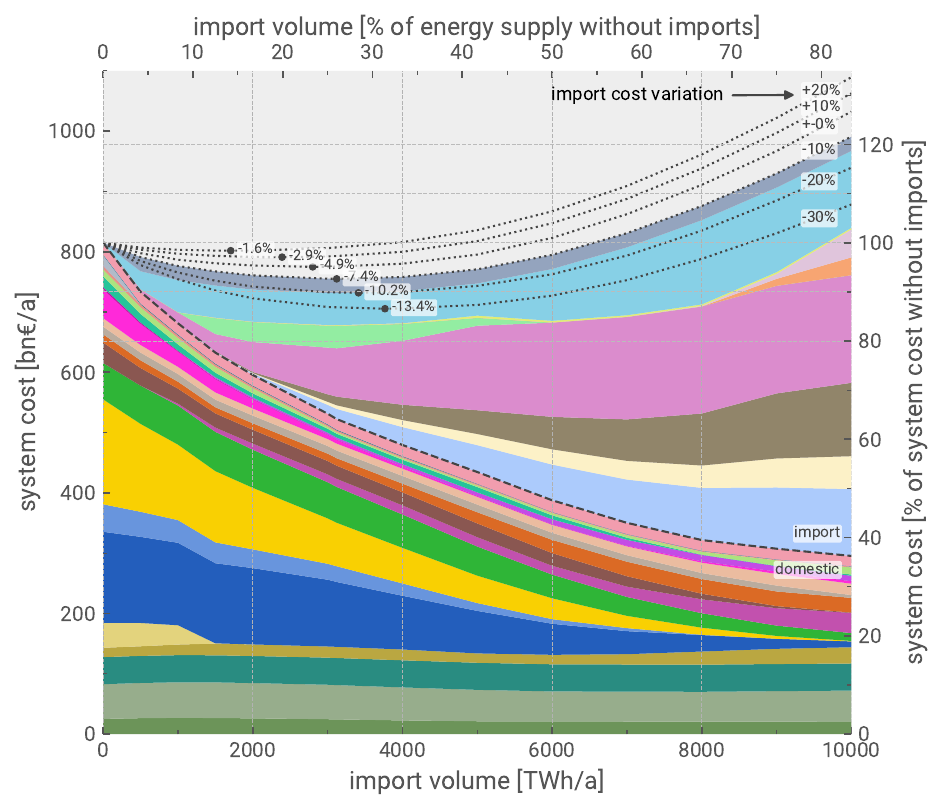} \\
        (d) import costs -20\% & (e) import costs -30\% \\
        \includegraphics[width=0.495\textwidth, trim=0cm 0cm 12.8cm 0cm, clip]{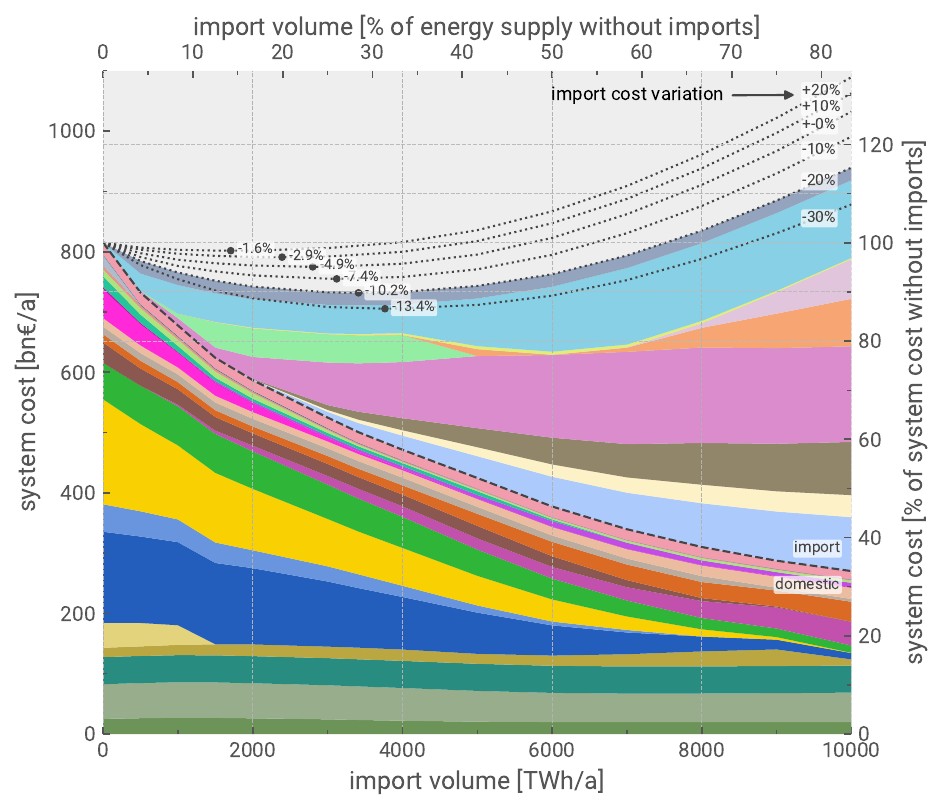} &
        \includegraphics[width=0.495\textwidth, trim=0cm 0cm 12.8cm 0cm, clip]{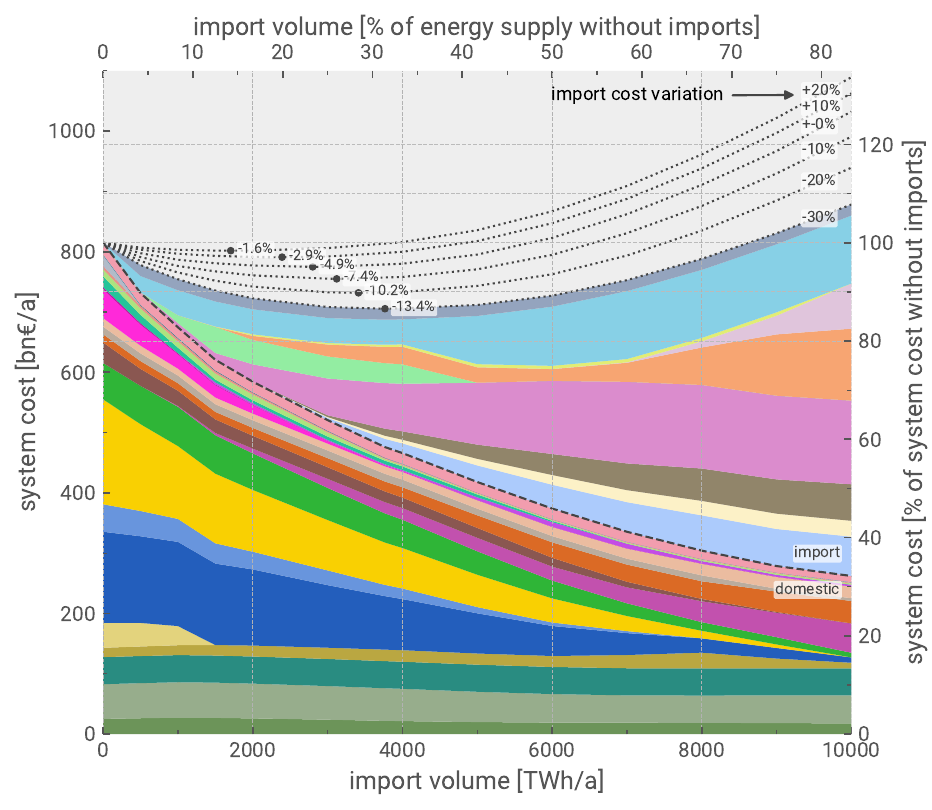}
    \end{tabular}
    \caption{\textbf{Sensitivity of import volume on total system cost and composition for varying import costs.}}
    \label{fig:si:volume}
\end{figure*}

\begin{figure*}
    \includegraphics[width=\textwidth]{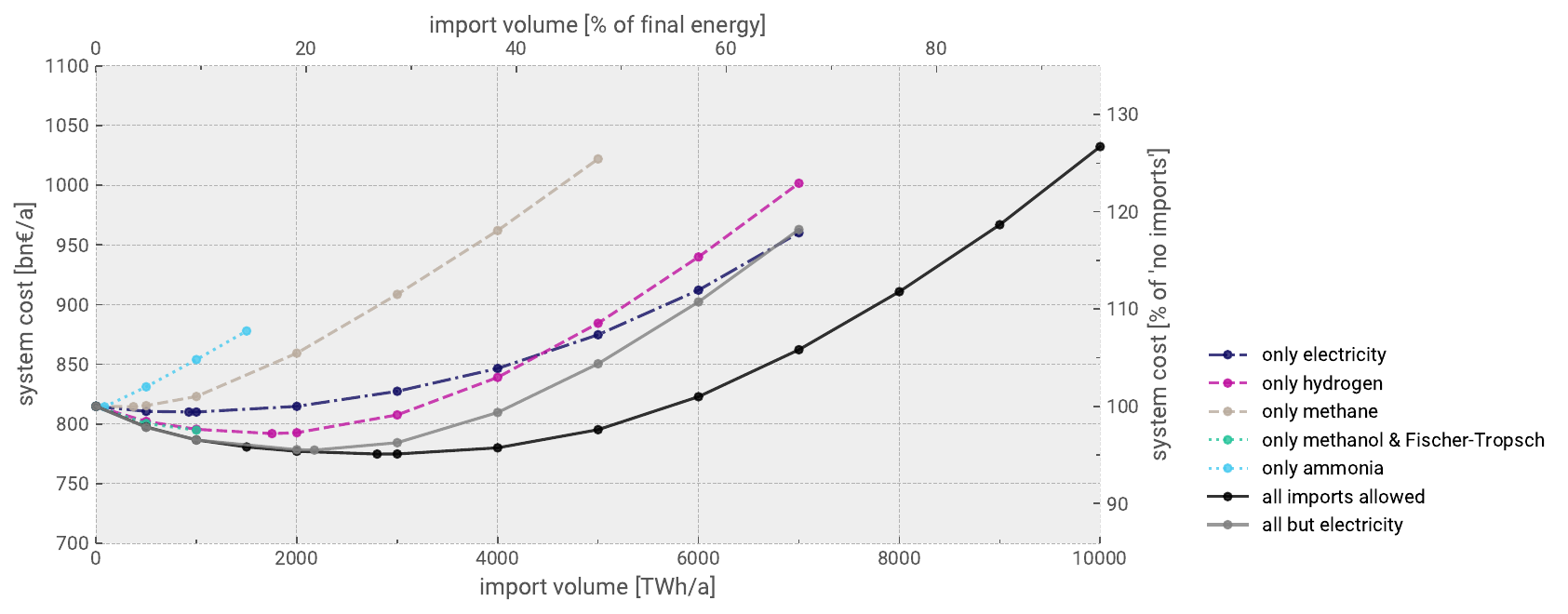}
    \caption{\textbf{Sensitivity of import volume on total system cost with subsets of import vectors available.}}
    \label{fig:si:volume-subsets}
\end{figure*}

\begin{figure*}
    \footnotesize
    \begin{tabular}{cc}
        (a) only electricity imports & (b) only hydrogen imports \\
        \includegraphics[width=0.49\textwidth]{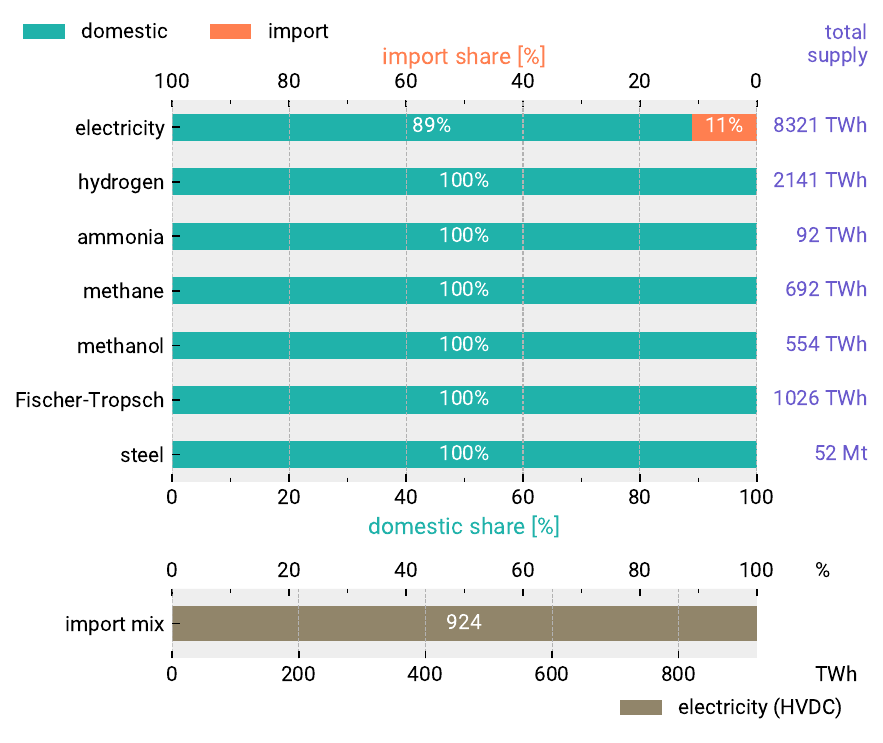} &
        \includegraphics[width=0.49\textwidth]{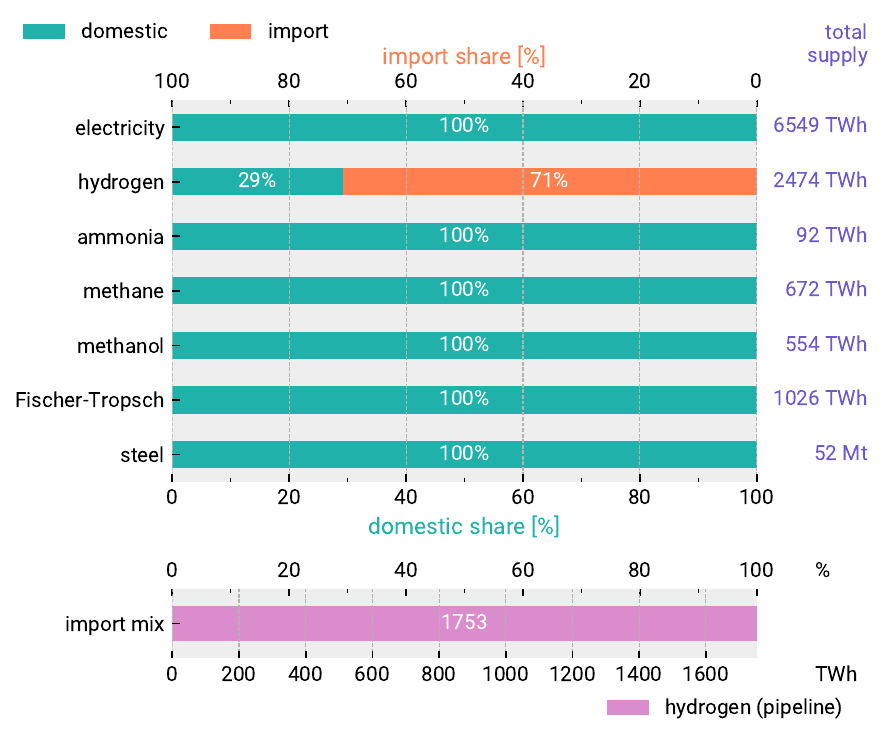} \\
        (c) -10\% import costs (all carriers but electricity) & (d) -10\% import cost (only carbonaceous fuels) \\
        \includegraphics[width=0.49\textwidth]{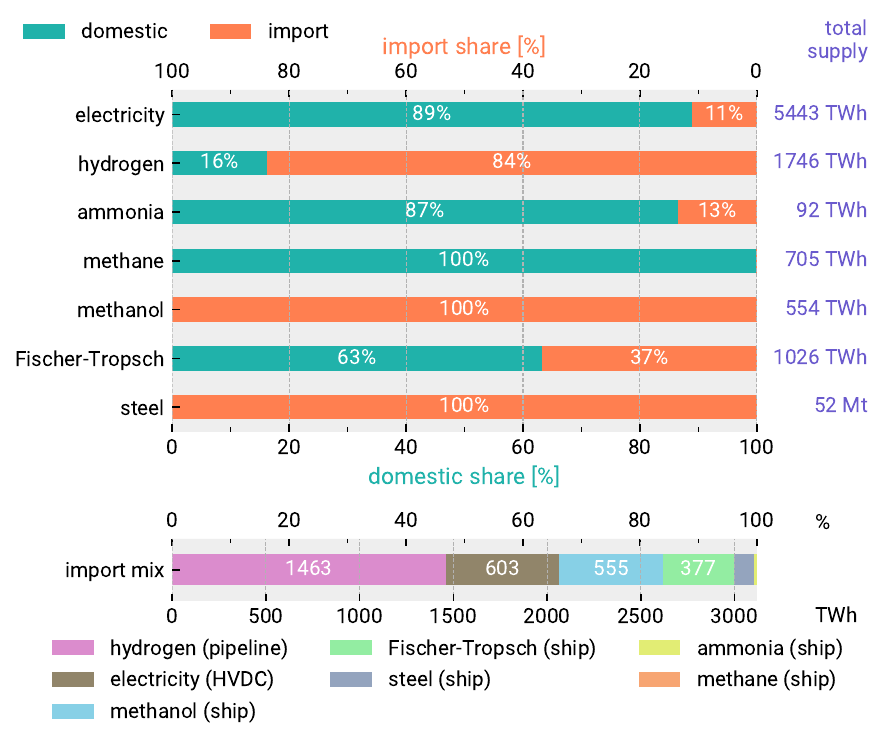} &
        \includegraphics[width=0.49\textwidth]{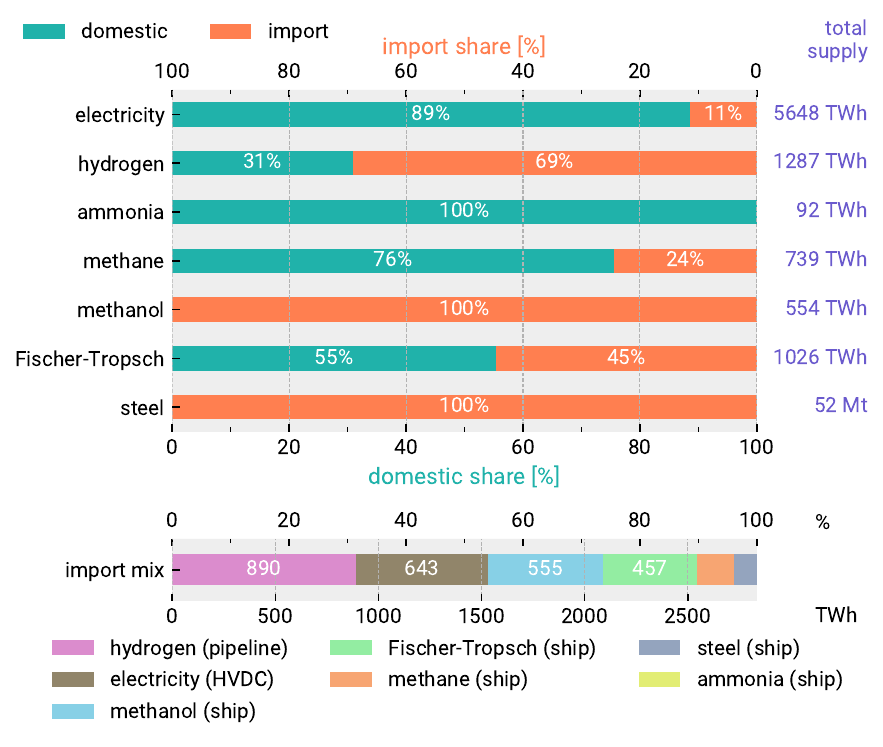} \\
    \end{tabular}
    \caption{\textbf{Import shares and mix for different import scenarios.}}
    \label{fig:si:import-shares}
\end{figure*}

\begin{figure*}
    \footnotesize
    \begin{tabular}{ccc}
        (a) H$_2$ / no imports allowed & (b) H$_2$ / only hydrogen imports & (c) H$_2$ / all imports allowed \\
        \includegraphics[width=0.325\textwidth]{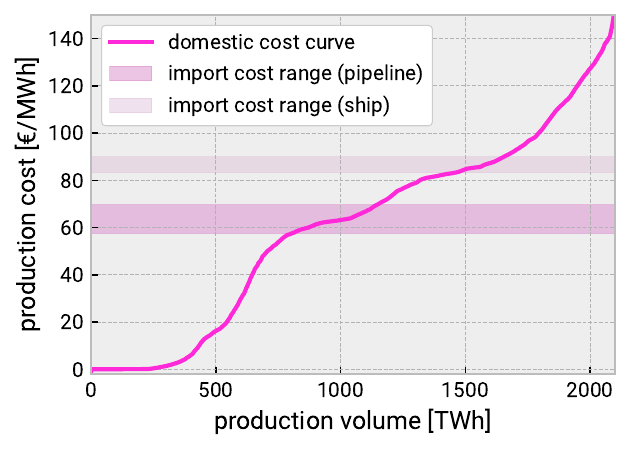} &
        \includegraphics[width=0.325\textwidth]{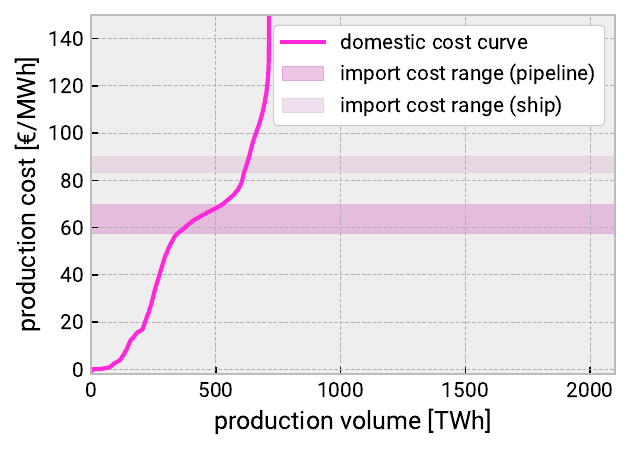} &
        \includegraphics[width=0.325\textwidth]{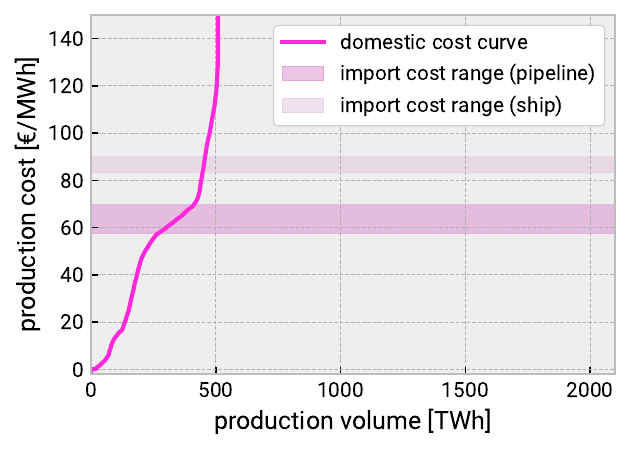} \\
        (d) MeOH / no imports allowed & (e) MeOH / only hydrogen imports & (f) MeOH / all imports allowed \\
        \includegraphics[width=0.325\textwidth]{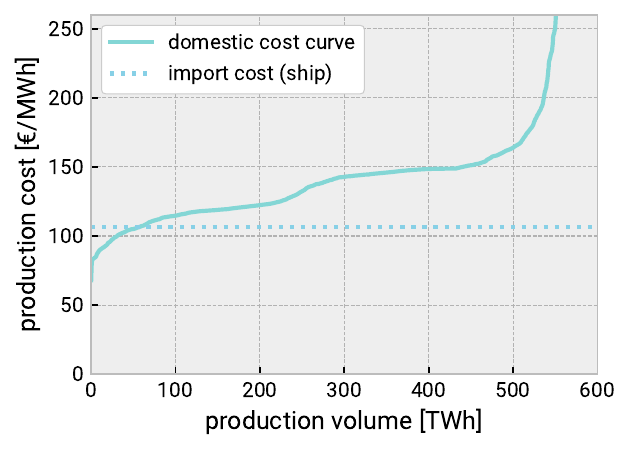} &
        \includegraphics[width=0.325\textwidth]{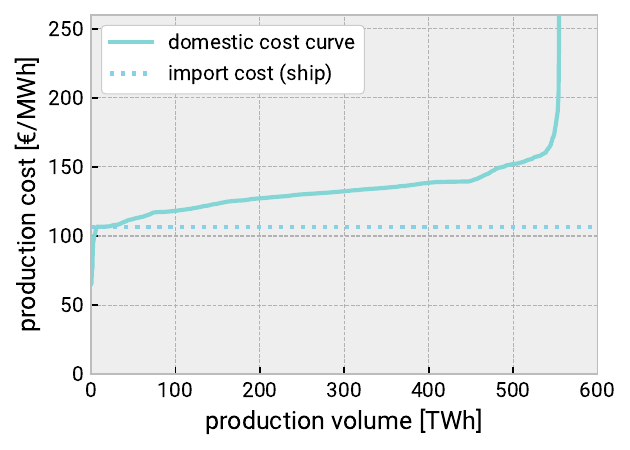} &
        \includegraphics[width=0.325\textwidth]{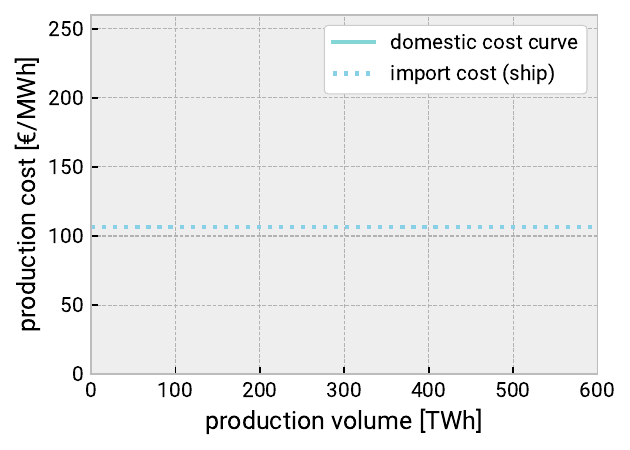} \\
        (g) Fischer-Tropsch / no imports allowed & (h) Fischer-Tropsch / only hydrogen imports & (i) Fischer-Tropsch / all imports allowed \\
        \includegraphics[width=0.325\textwidth]{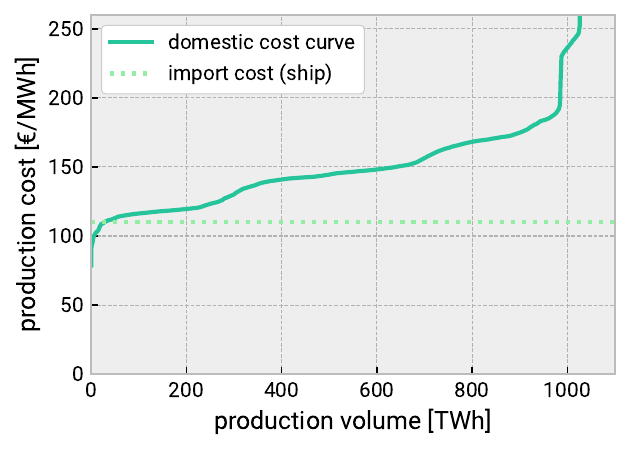} &
        \includegraphics[width=0.325\textwidth]{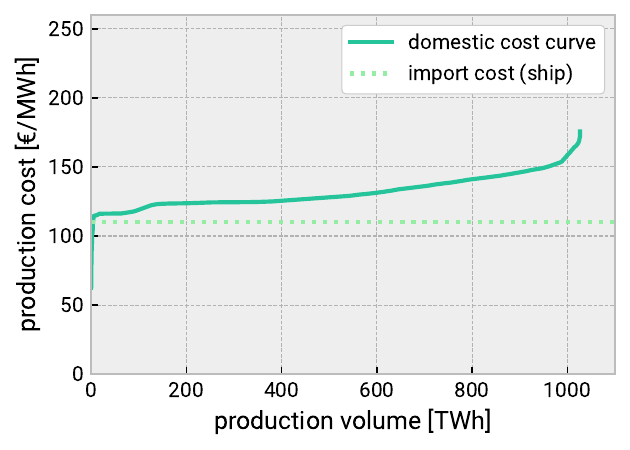} &
        \includegraphics[width=0.325\textwidth]{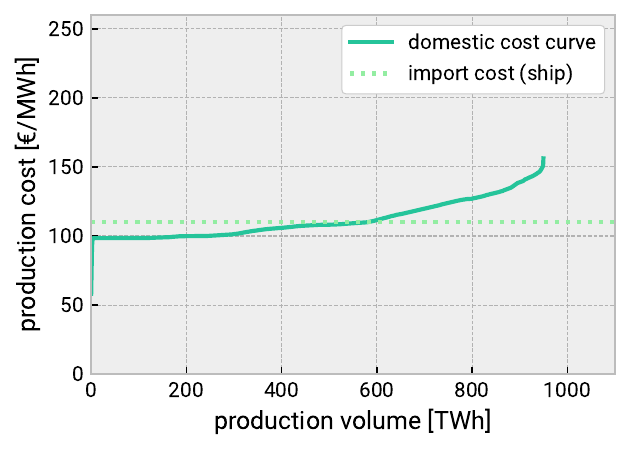} \\
        (j) CH$_4$ / no imports allowed & (k) CH$_4$ / only hydrogen imports & (l) CH$_4$ / all imports allowed \\
        \includegraphics[width=0.325\textwidth]{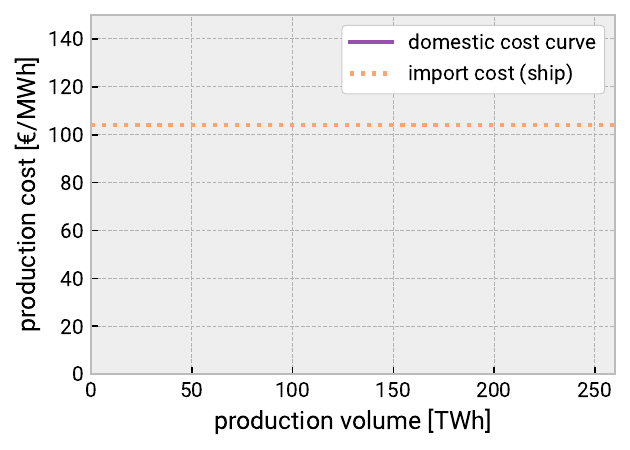} &
        \includegraphics[width=0.325\textwidth]{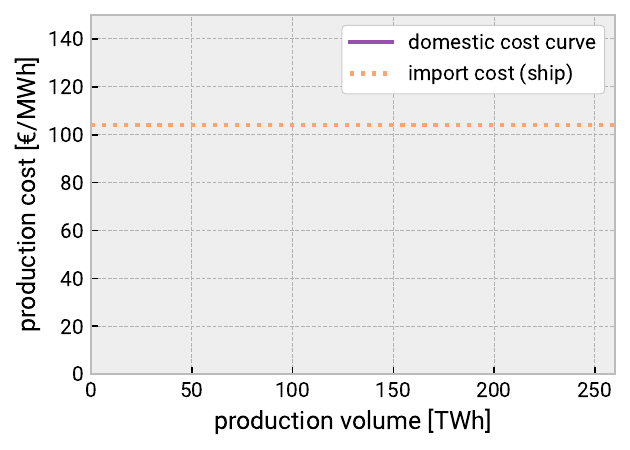} &
        \includegraphics[width=0.325\textwidth]{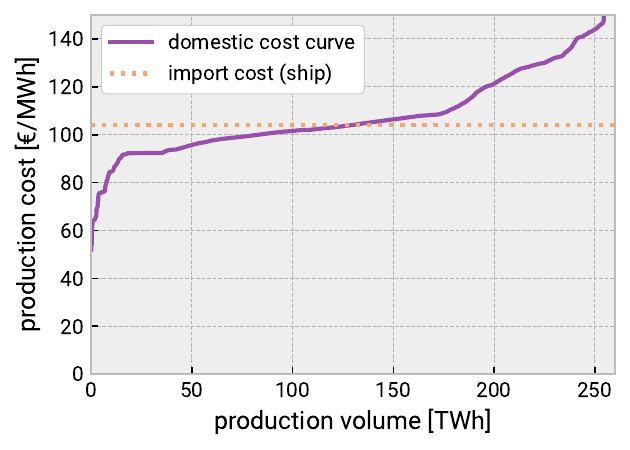} \\
    \end{tabular}
    \caption{\textbf{Domestic cost supply curves for different import scenarios and carriers.}
        The cost supply curves are built using sorted spatio-temporal market
        values with corresponding production volumes per region and snapshot. If
        the domestic supply curve is missing, no domestic production occured in
        the scenario. Shaded areas or dotted lines show import cost ranges of
        the respective carriers as reference.}
    \label{fig:si:cost-supply-curves}
\end{figure*}

\begin{figure*}
    \includegraphics[width=0.8\textwidth]{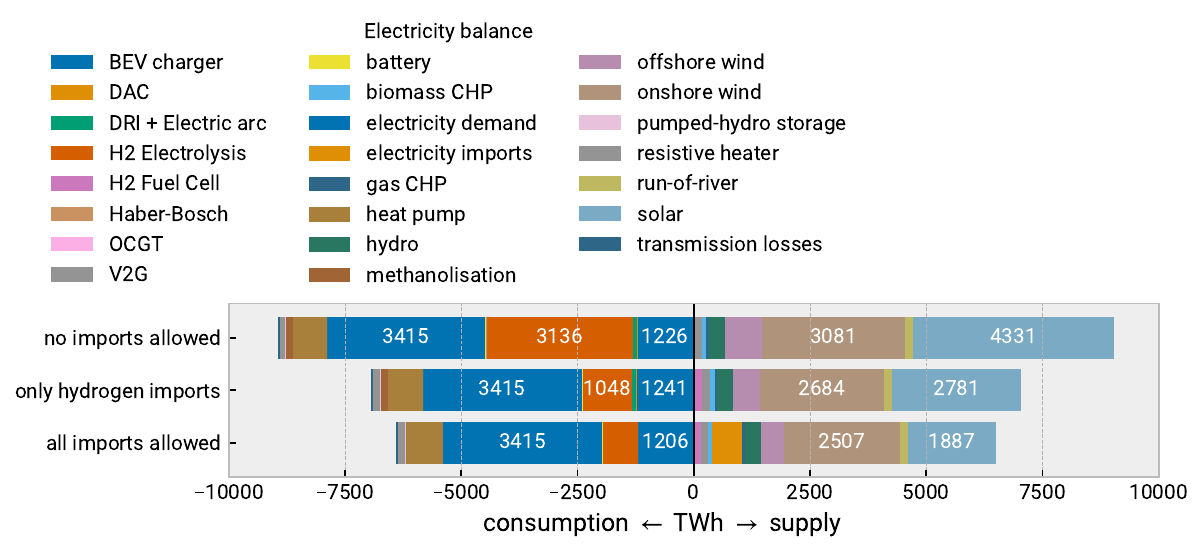}
    \includegraphics[width=0.8\textwidth]{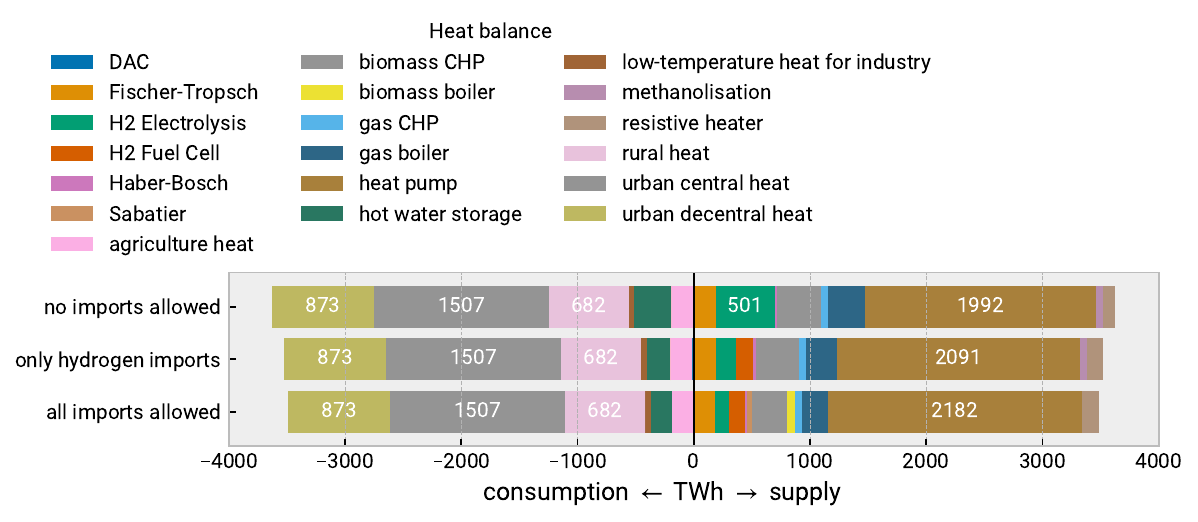}
    \includegraphics[width=0.8\textwidth]{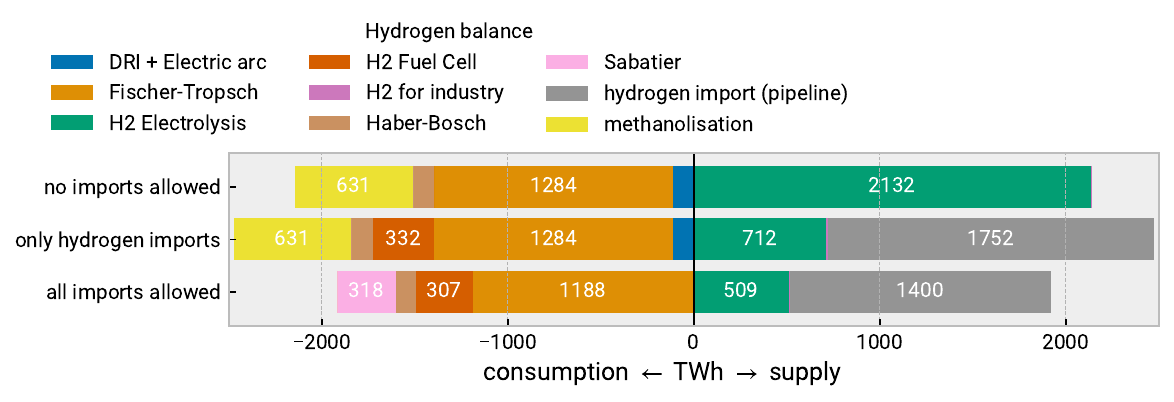}
    \includegraphics[width=0.8\textwidth]{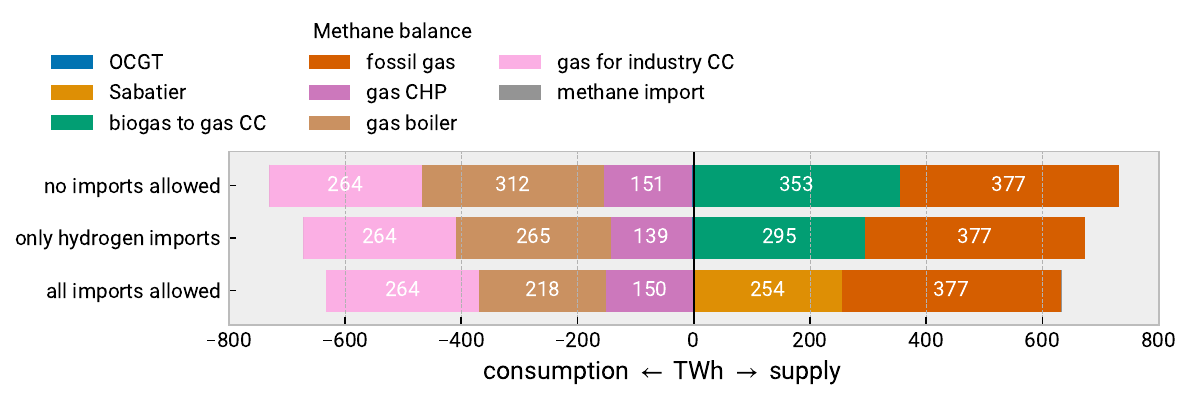}
    \caption{\textbf{Energy balances for three import scenarios for the carriers electricity, heat, hydrogen and gas.}
    }
    \label{fig:si:balances-a}
\end{figure*}

\begin{figure*}
    \includegraphics[width=0.8\textwidth]{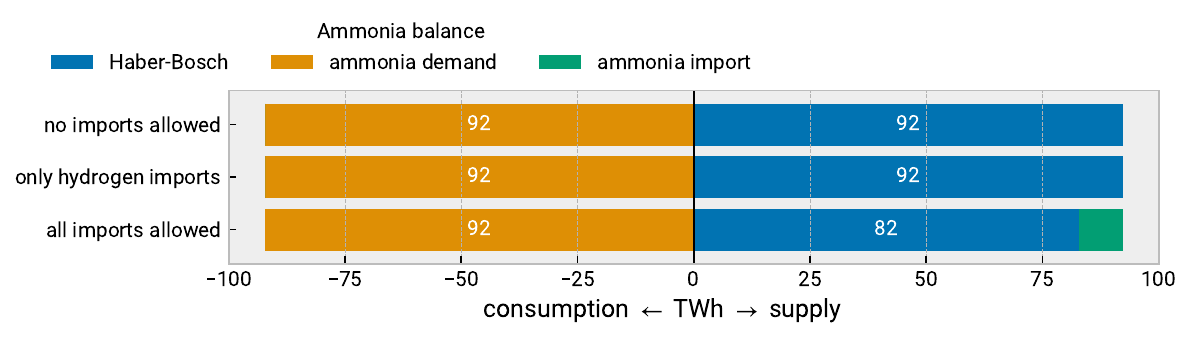}
    \includegraphics[width=0.8\textwidth]{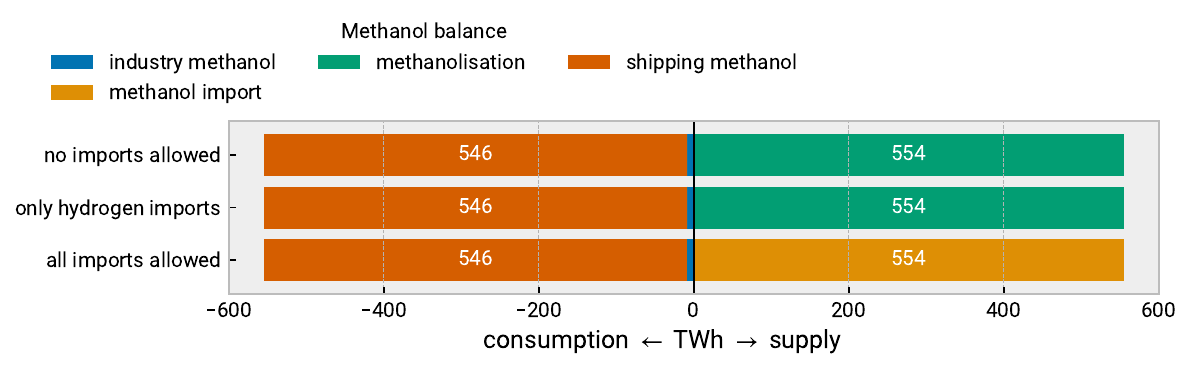}
    \includegraphics[width=0.8\textwidth]{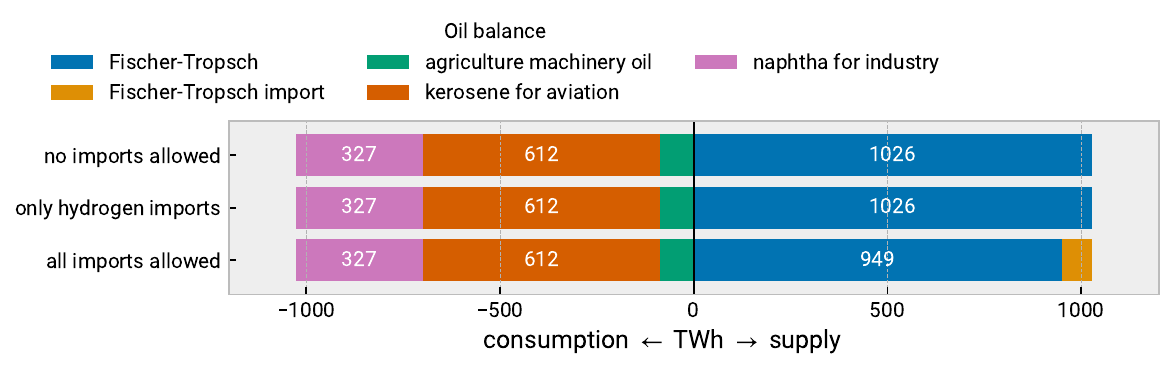}
    \includegraphics[width=0.8\textwidth]{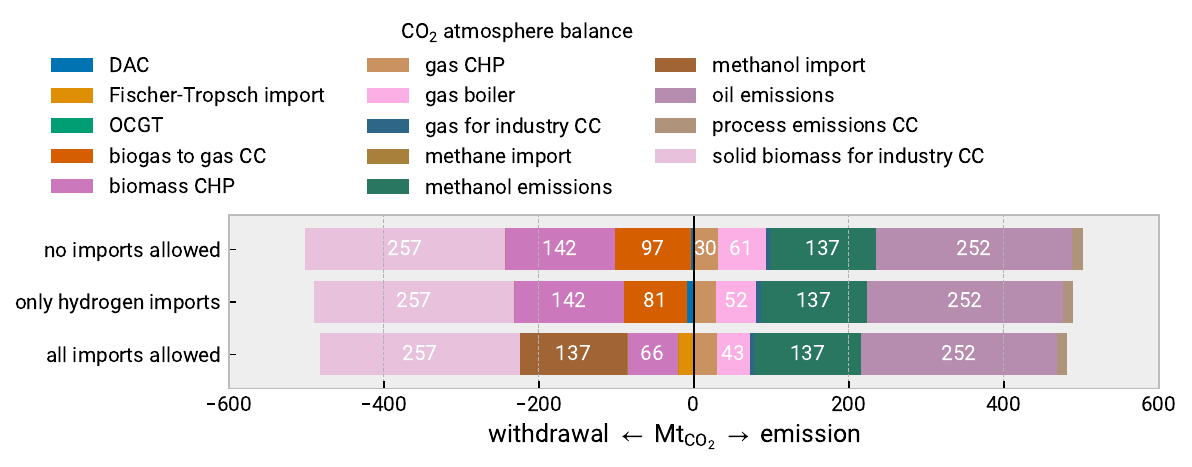}
    \includegraphics[width=0.8\textwidth]{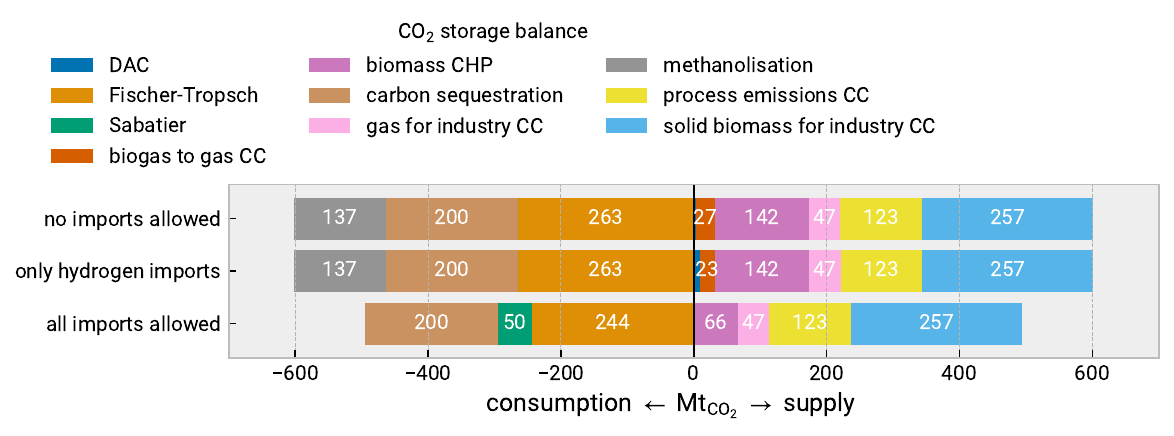}
    \caption{\textbf{
        Energy balances for three import scenarios for the carriers
        ammonia, methanol, and oil, as well as stored and atmospheric carbon dioxide.
    }
    }
    \label{fig:si:balances-b}
\end{figure*}

\begin{figure*}
    \includegraphics[width=\textwidth]{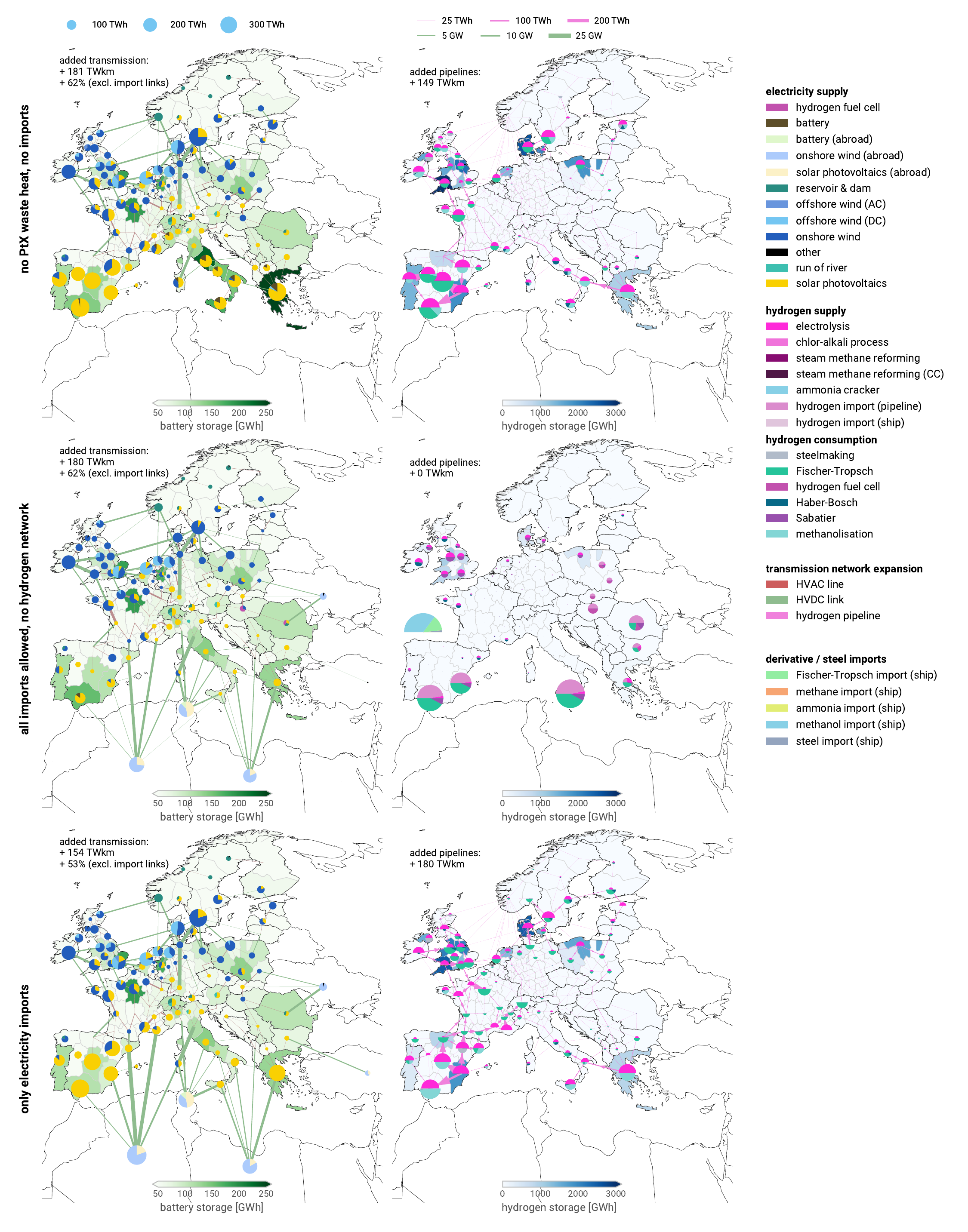}
    \caption{\textbf{Layout of European energy infrastructure for different import scenarios.} Role of PtX waste heat, hydrogen network, and electricity imports.
    Left column shows the regional electricity supply mix (pies), added HVDC and HVAC transmission capacity (lines), and the siting of battery storage (choropleth).
        Right column shows the hydrogen supply (top half of pies) and consumption (bottom half of pies), net flow and direction of hydrogen in newly built pipelines (lines), and the siting of hydrogen storage subject to geological potentials (choropleth).
        Total volumes of transmission expansion are given in TWkm, which is the sum product of the capacity and length of individual connections.
    }
    \label{fig:si:infra-b}
\end{figure*}

\begin{figure*}
    \includegraphics[width=\textwidth]{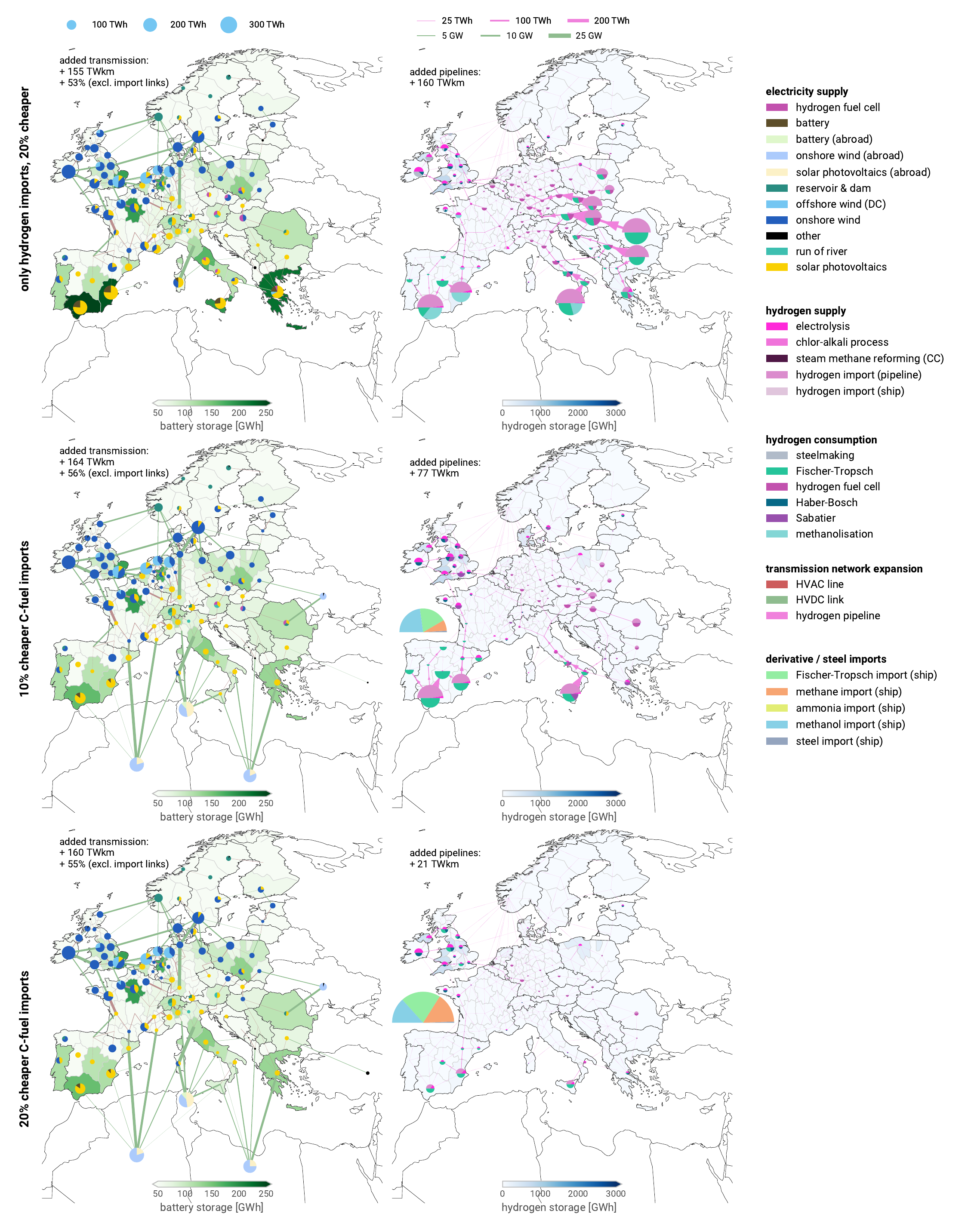}
    \caption{\textbf{Layout of European energy infrastructure for different import scenarios.} Sensitivities of infrastructure to import costs.
    Left column shows the regional electricity supply mix (pies), added HVDC and HVAC transmission capacity (lines), and the siting of battery storage (choropleth).
        Right column shows the hydrogen supply (top half of pies) and consumption (bottom half of pies), net flow and direction of hydrogen in newly built pipelines (lines), and the siting of hydrogen storage subject to geological potentials (choropleth).
        Total volumes of transmission expansion are given in TWkm, which is the sum product of the capacity and length of individual connections.
    }
    \label{fig:si:infra-c}
\end{figure*}

\begin{figure*}
    \includegraphics[width=\textwidth]{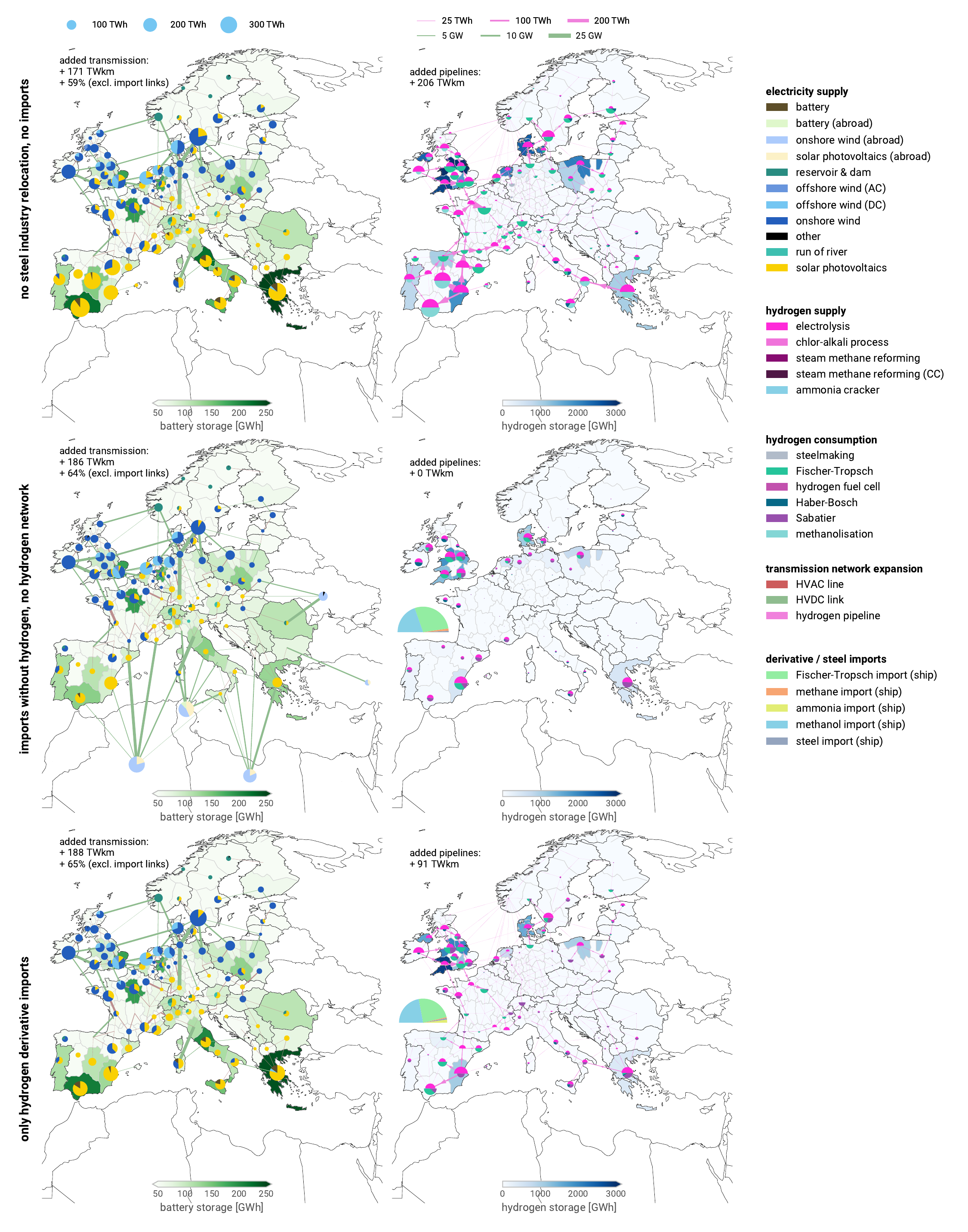}
    \caption{\textbf{Layout of European energy infrastructure for different import scenarios.} Role of industry relocation and focus on carbonaceous fuel imports.
    Left column shows the regional electricity supply mix (pies), added HVDC and HVAC transmission capacity (lines), and the siting of battery storage (choropleth).
        Right column shows the hydrogen supply (top half of pies) and consumption (bottom half of pies), net flow and direction of hydrogen in newly built pipelines (lines), and the siting of hydrogen storage subject to geological potentials (choropleth).
        Total volumes of transmission expansion are given in TWkm, which is the sum product of the capacity and length of individual connections.
    }
    \label{fig:si:infra-d}
\end{figure*}

\begin{figure*}
    \centering
    \footnotesize
    (a) average utilisation rate of import HVDC links \\
    \includegraphics[width=\textwidth]{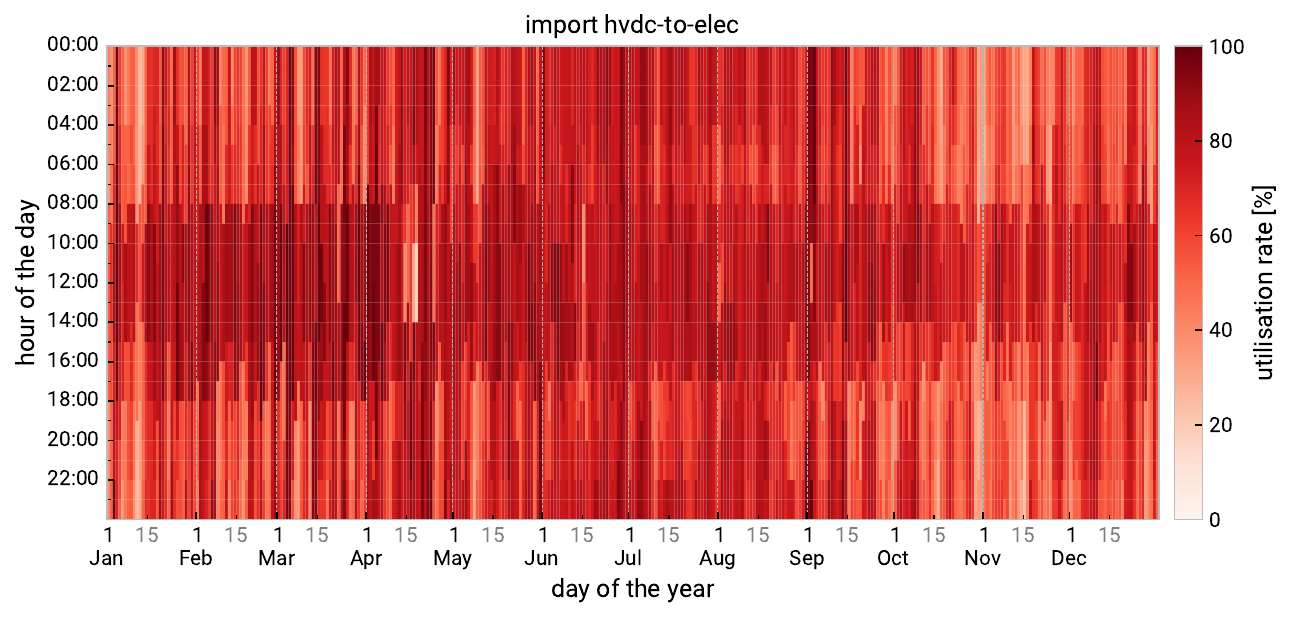} \\
    (b) average utilisation rate of import hydrogen pipelines \\
    \includegraphics[width=\textwidth]{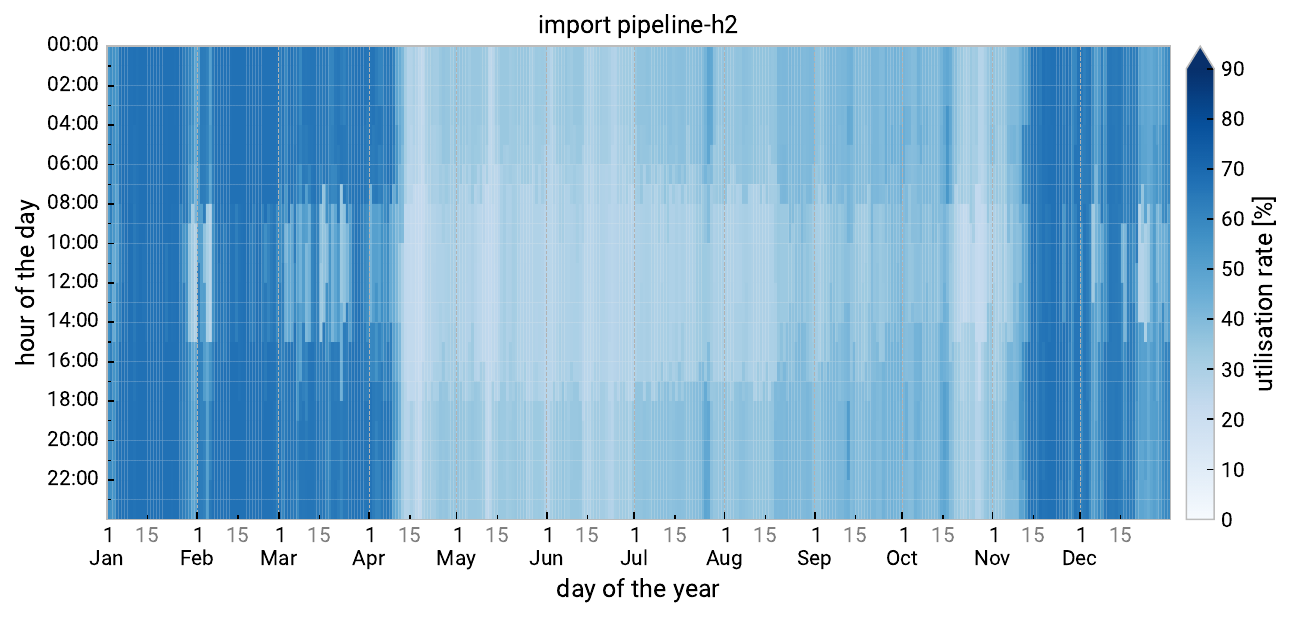}
    \caption{\textbf{Temporal usage pattern of electricity and hydrogen storage.}
    The capacity-weighted average utilisation rate is 72\% for import HVDC links and 45\% for
    hydrogen pipelines. For hydrogen import pipelines, a clear seasonal pattern
    with higher utilisation in winter is visible, demonstrated by an average utilisation rate of 56\% from November to April and 35\% from May to October.
    For other energy or material
    imports than hydrogen and electricity, the timing of imports is not
    informatively captured due to problem degeneracy caused by negligible
    storage costs of carbonaceous fuels and steel.}
    \label{fig:si:import-operation}
\end{figure*}

\begin{figure*}
    \centering
    \footnotesize
    (a) without imports \\
    \includegraphics[width=\textwidth]{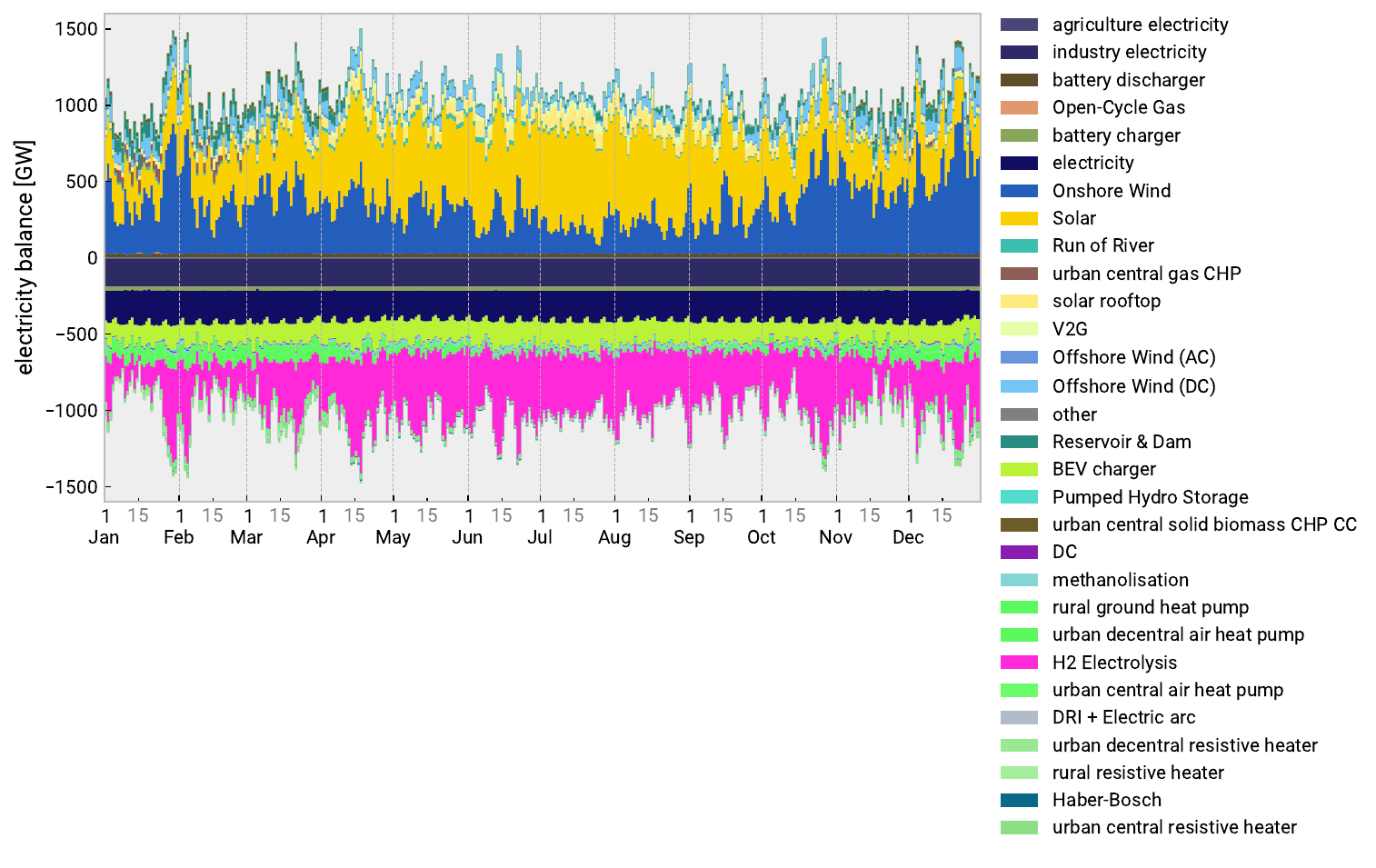} \\
    (b) with imports \\
    \includegraphics[width=\textwidth]{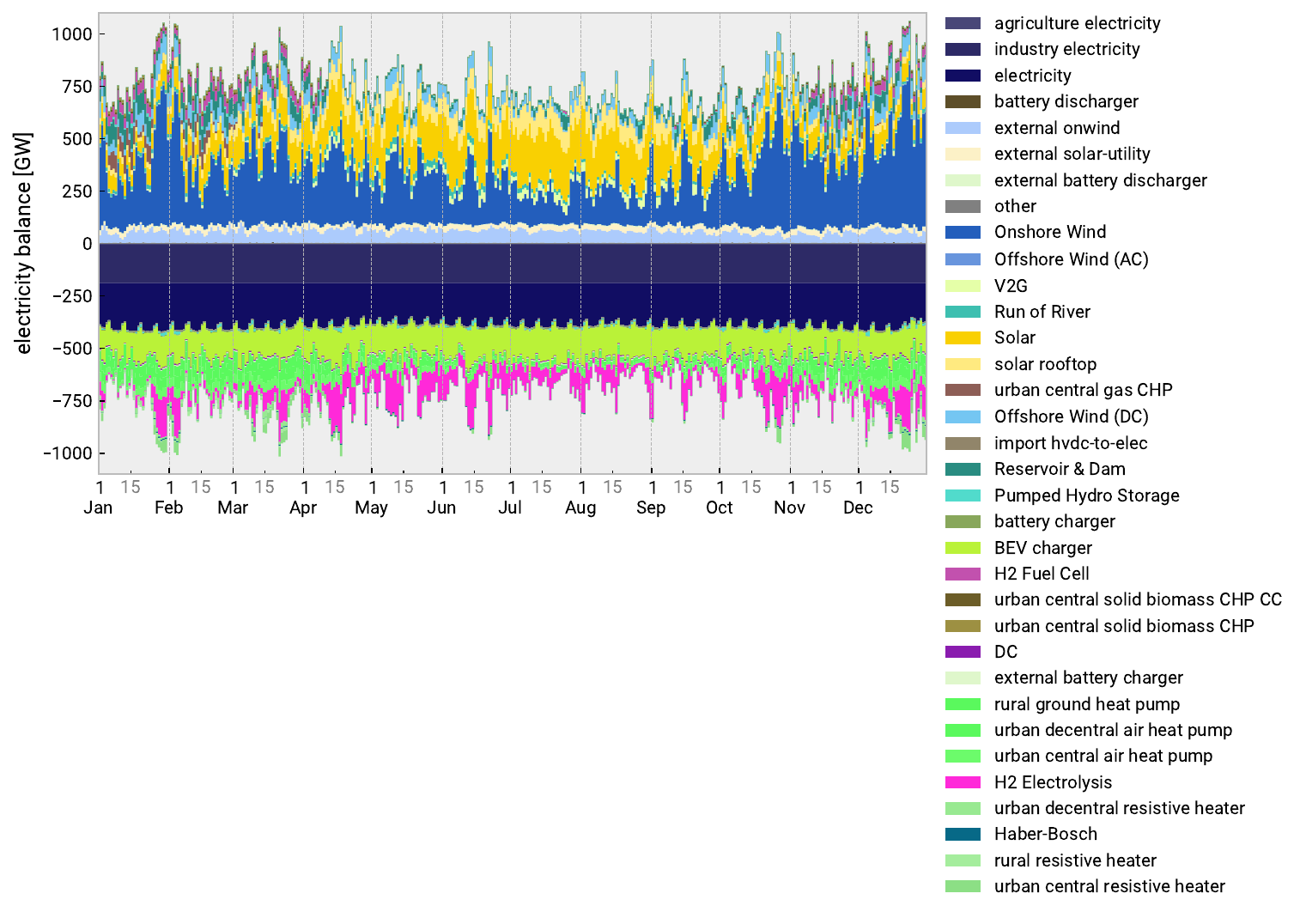}
    \caption{\textbf{Energy balance time series for electricity with and without imports.} Resampled to daily averages. Positive numbers indicate supply, negative numbers indicate consumption.}
    \label{fig:si:balance-elec}
\end{figure*}

\begin{figure*}
    \centering
    \footnotesize
    (a) without imports \\
    \includegraphics[width=\textwidth]{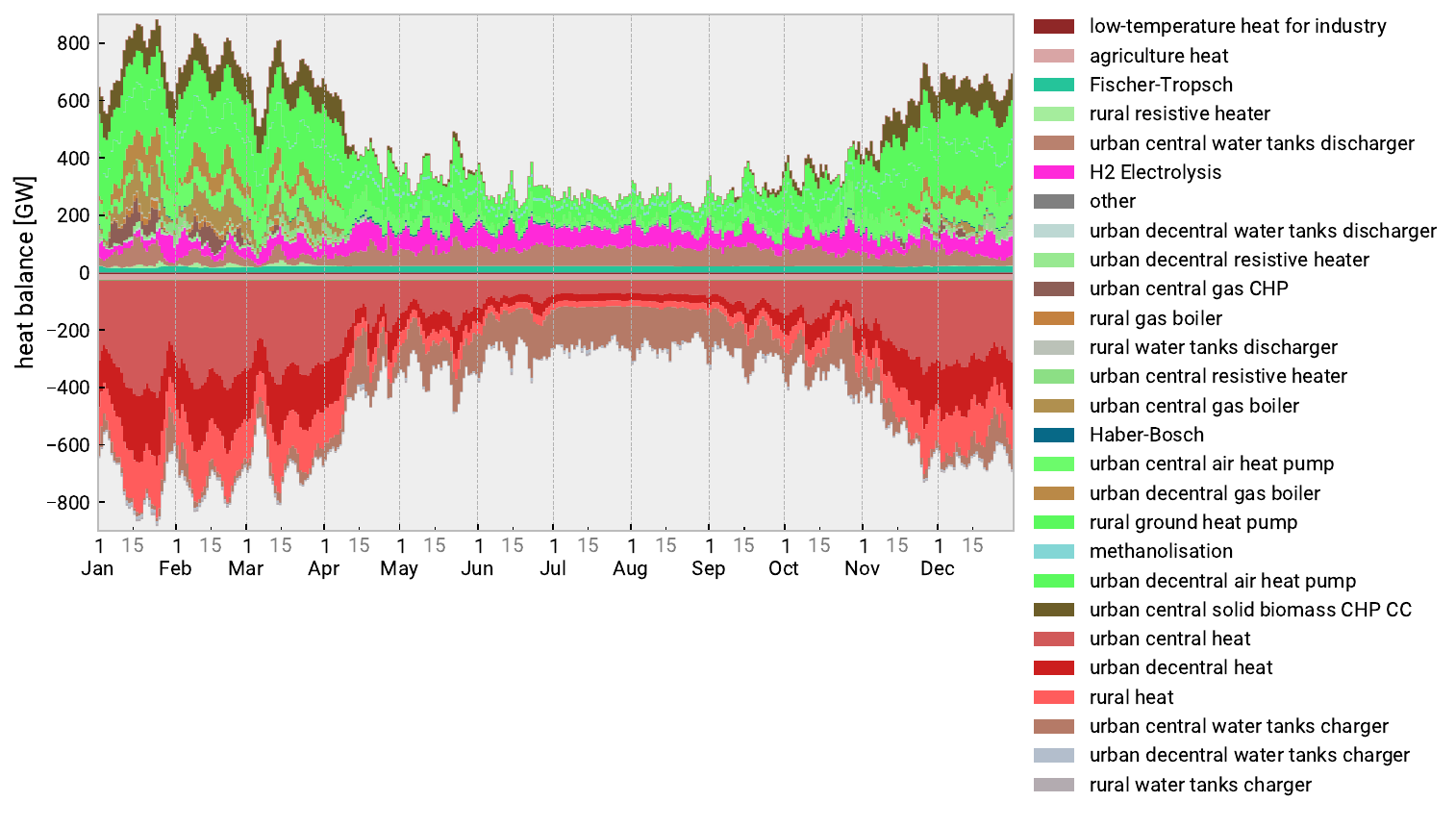} \\
    (b) with imports \\
    \includegraphics[width=\textwidth]{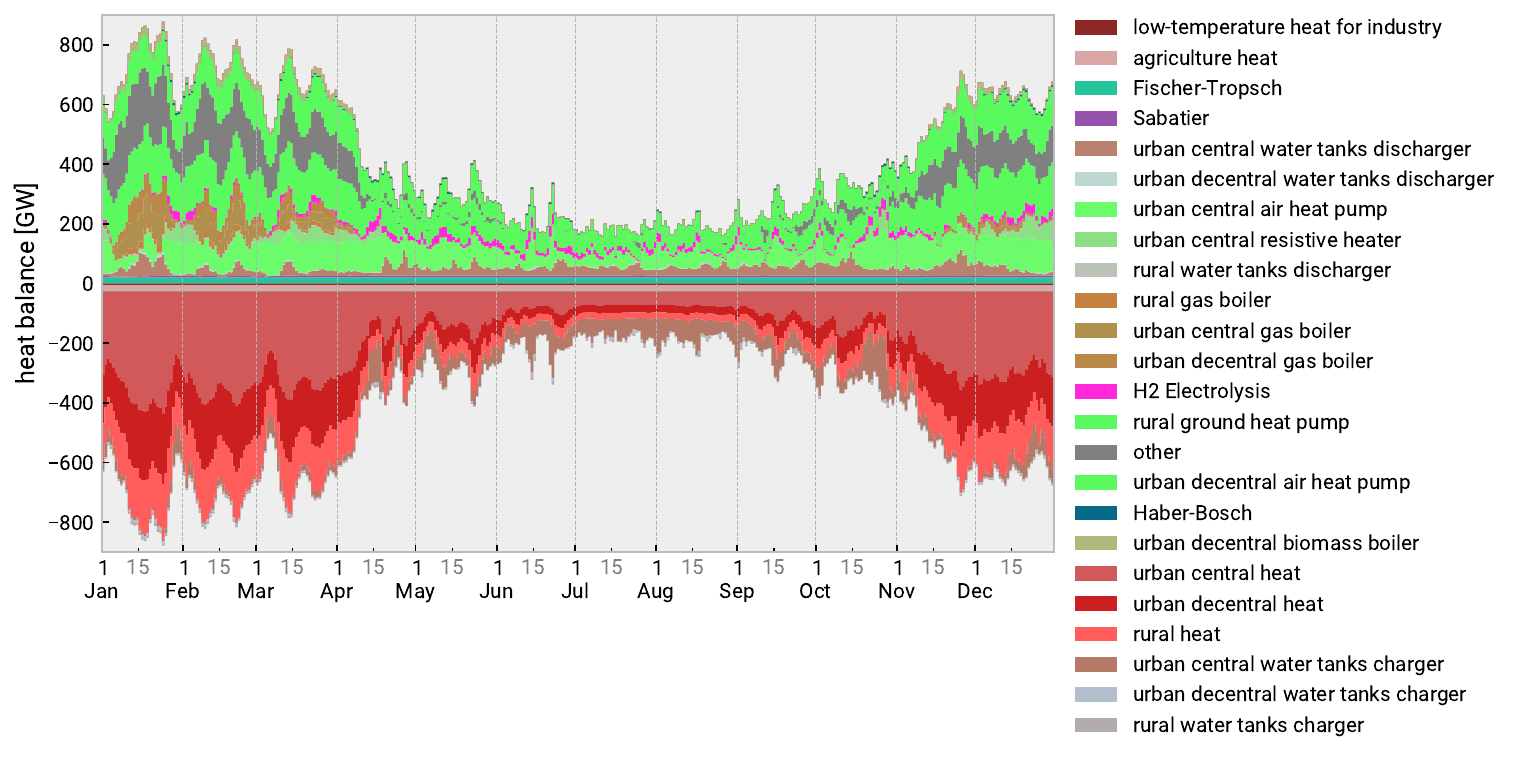}
    \caption{\textbf{Energy balance time series for heat with and without imports.} Resampled to daily averages. Positive numbers indicate supply, negative numbers indicate consumption.}
    \label{fig:si:balance-heat}
\end{figure*}

\begin{figure*}
    \centering
    \footnotesize
    (a) without imports \\
    \includegraphics[width=\textwidth]{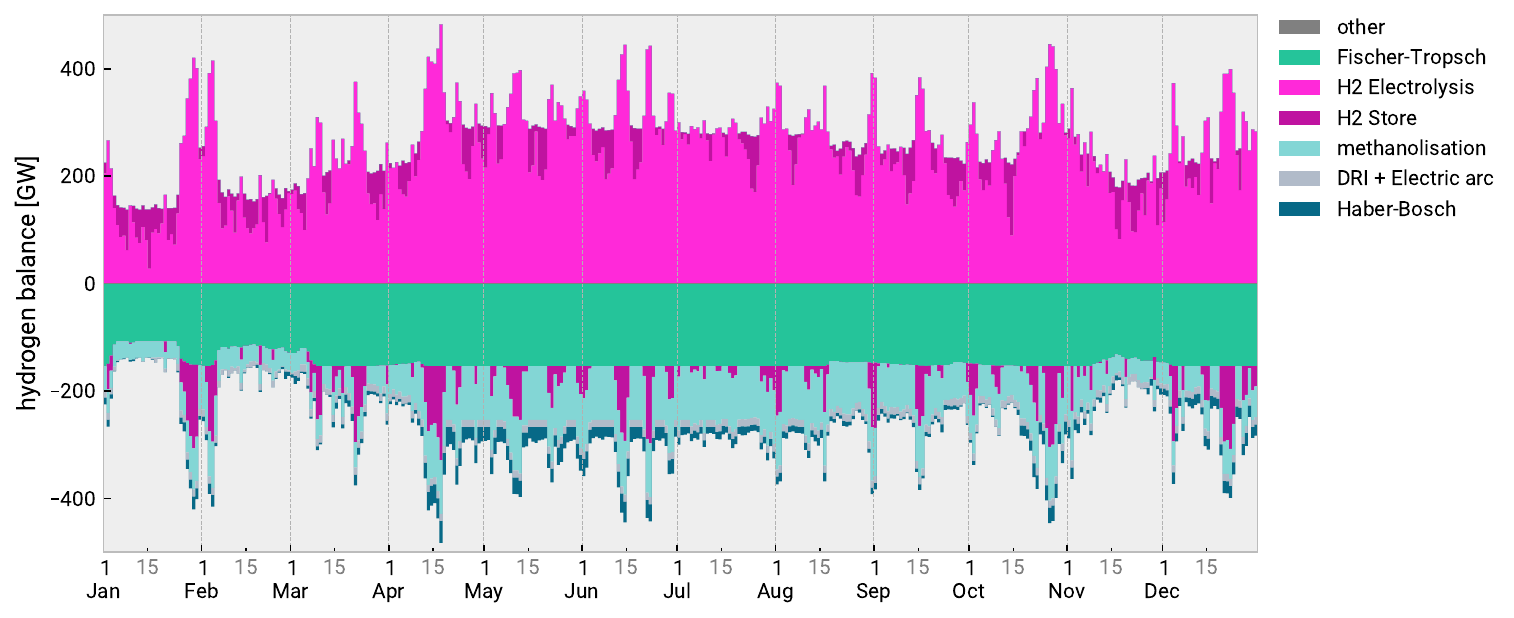} \\
    (b) with imports \\
    \includegraphics[width=\textwidth]{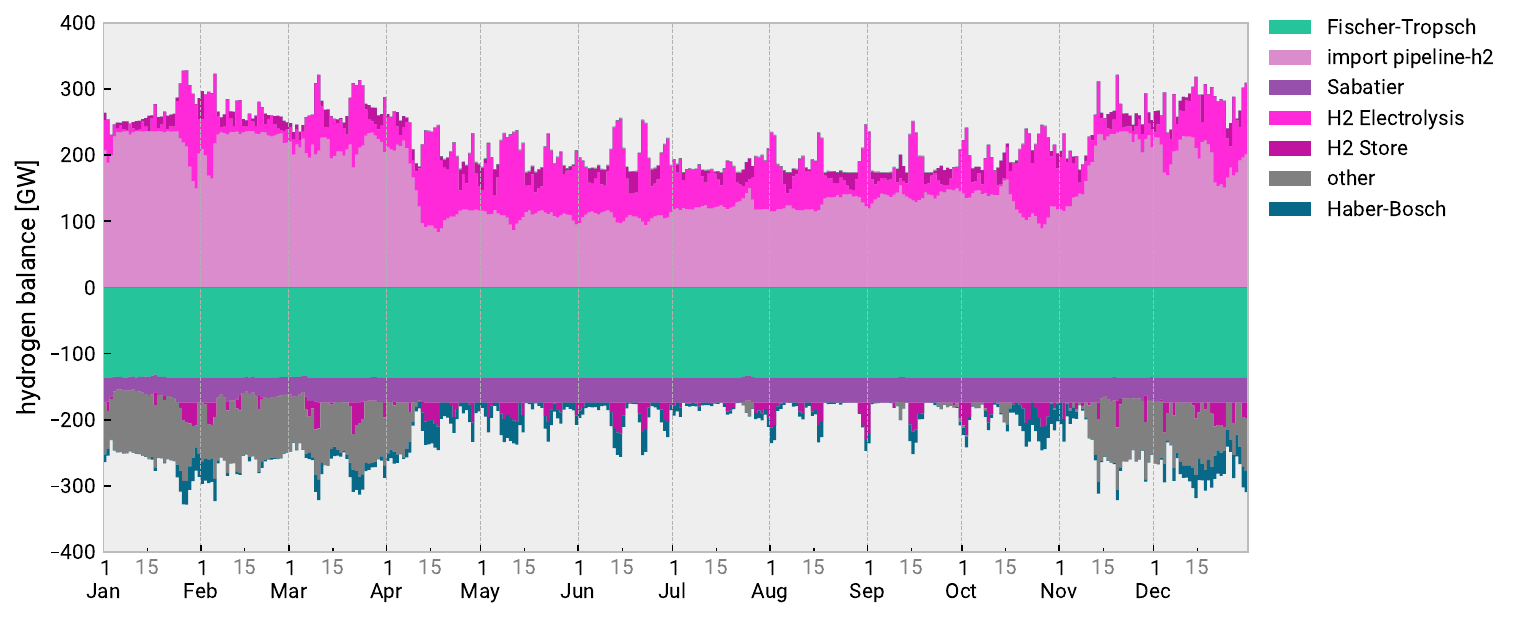}
    \caption{\textbf{Energy balance time series for hydrogen with and without imports.} Resampled to daily averages. Positive numbers indicate supply, negative numbers indicate consumption.}
    \label{fig:si:balance-h2}
\end{figure*}

\begin{figure*}
    \centering
    \footnotesize
    (a) gas transmission network \\
    \includegraphics[width=.8\textwidth]{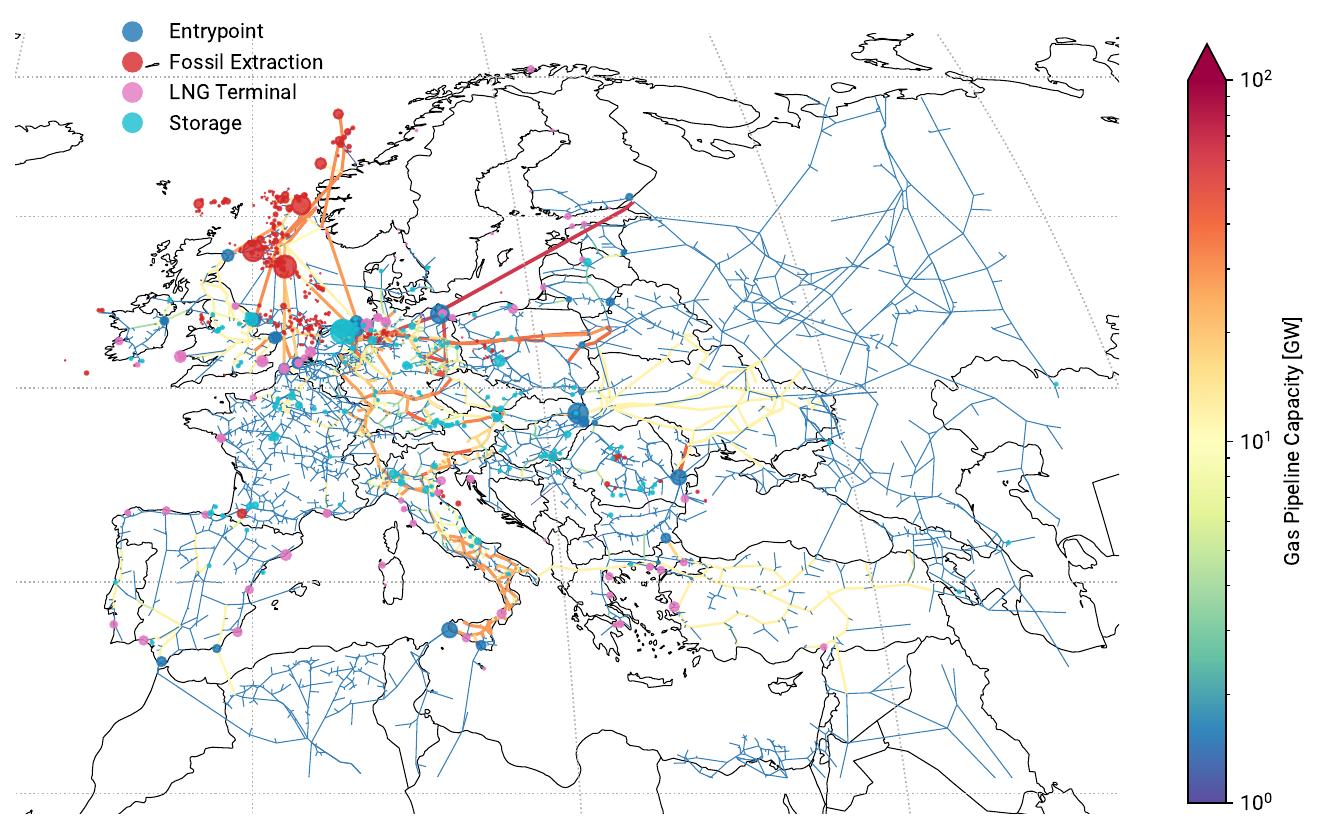} \\
    (b) electricity transmission network \\
    \includegraphics[width=.8\textwidth]{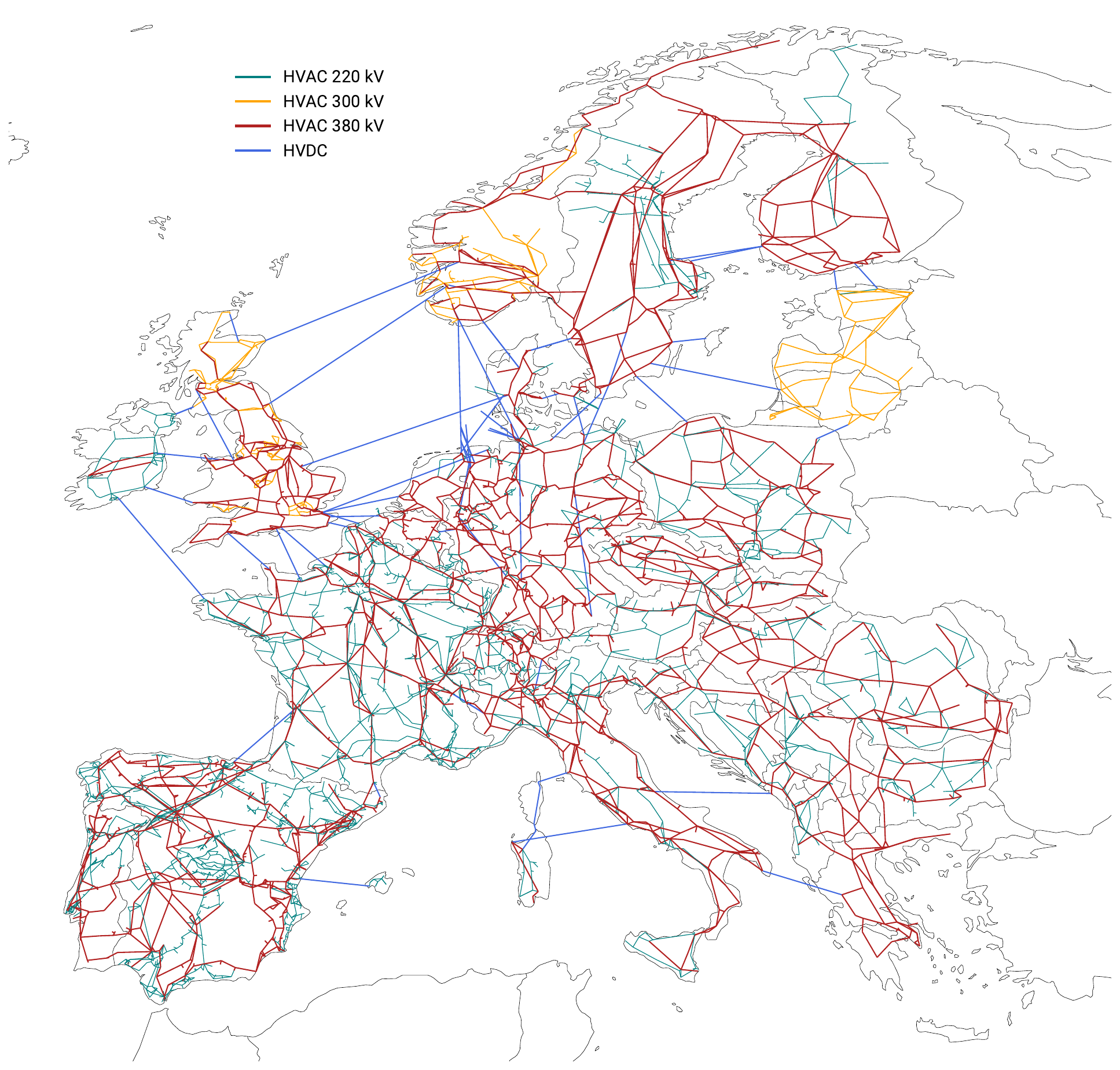}
    \caption{\textbf{Gas and electricity transmission network data.} For gas
    transmission, the map shows pipelines sized and colored by rated capacity,
    fossil gas extraction sites, storage locations, pipeline entrypoints, and
    LNG terminals. The data comes from SciGRID\_gas and is supplemented with
    data from Global Energy Monitor. For power transmission, the map shows
    existing transmission lines at and above 220~kV taken from the ENTSO-E
    Transmission System Map (\url{https://www.entsoe.eu/data/map/}),
    supplemented with planned TYNDP projects (\url{https://tyndp.entsoe.eu/}).}
    \label{fig:si:networks-raw}
\end{figure*}

\end{document}